\pdfoutput=1

\documentclass[12pt,twoside,vi,leftblank]{mitthesis}
\usepackage{cmap}
\usepackage[T1]{fontenc}
\usepackage[final]{listings}
\usepackage{multicol}
\usepackage{graphicx}
\usepackage{caption}
\usepackage{amsmath}
\usepackage{array}
\usepackage{parskip}
\usepackage[hyphens]{url}
\usepackage[hidelinks]{hyperref}
\usepackage[font=footnotesize,labelfont=bf]{caption}
\usepackage{float}

\usepackage[math]{cellspace}
\graphicspath{{./images/}}

\begin{document}

%
%
%
%
%
%
%
%
%
%
%

\title{Matching Startup Founders to Investors:\\a Tool and a Study}

\author{Yasyf Mohamedali}
\department{Department of Electrical Engineering and Computer Science}

\degree{Master of Engineering in Electrical Engineering and Computer Science}

\degreemonth{June}
\degreeyear{2018}
\thesisdate{May 24, 2018}

\copyrightnoticetext{\copyright\ \Mit\ 2018.  All rights reserved.}

\supervisor{John V. Guttag}{Dugald C. Jackson Professor}

\supervisor{Rei Wang}{First Round Capital}

\chairman{Katrina LaCurts}{Chair, Master of Engineering Thesis Committee}

\maketitle



\cleardoublepage
\setcounter{savepage}{\thepage}
\begin{abstractpage}
%
%
%
The process of matching startup founders with venture capital investors is a necessary first step for many modern technology companies, yet there have been few attempts to study the characteristics of the two parties and their interactions. Surprisingly little has been shown quantitatively about the process, and many of the common assumptions are based on anecdotal evidence. In this thesis, we aim to learn more about the matching component of the startup fundraising process. We begin with a tool (VCWiz), created from the current set of best-practices to help inexperienced founders navigate the founder-investor matching process. The goal of this tool is to increase efficiency and equitability, while collecting data to inform further studies. We use this data, combined with public data on venture investments in the USA, to draw conclusions about the characteristics of venture financing rounds. Finally, we explore the communication data contributed to the tool by founders who are actively fundraising, and use it to learn which social attributes are most beneficial for individuals to possess when soliciting investments.

\end{abstractpage}


\cleardoublepage

\section*{Acknowledgments}

First and foremost, I'd like to thank the team at First Round Capital (specifically Rei Wang and Phin Barnes) for sponsoring this thesis, and Professor John Guttag for supervising it. Without your patience, guidance, and support, I would have never finished.

Next comes my family. To my mother, Feloza, and my father, Zeid, for pushing me to come this far, and being with me every step of the way. To my brothers, Afyz and Izayah, for giving me a reason to keep striving forward. To my grandparents, aunts, uncles, and cousins, whose encouragement never ends. Thank you.

Miscellanea: The fine folks at the Crunchbase Venture Program, for providing me with my initial dataset. Anthony Zhang and the fine folks at KnowYourVC, for being an awesome partner. Thomas Stone of UCL, for giving me a solid academic foundation to build on. Professor Tomas Palacios of the MIT VI-A program, for making this thesis possible. Professors Anantha Chandrakasan and Ron Rivest, for helping bring all the pieces together. The partners and portfolio of Dorm Room Fund, for being constant guinea pigs. $\text{Nate}^2$ at HealthWiz, for inspiring the VCWiz name. Josh Lee, for the best design job I've ever seen. Haris Memon, Nick Abouzeid and the team at ProductHunt, for an incredible launch. Thank you.

And last but not least, to my friends. Alice, my best friend and partner, for having my back in whatever form was necessary, at any time. My cofounder Joe, for dealing with me at my worst and trudging on nonetheless. Bryan, for being my sounding board in the early days. Athena, for being my sounding board in the later days. Kimberli, for motivating me in the most difficult of moments. Divya, equally for helping me through tough times and for answering all my dumb questions. Will and Nikhil, for providing an endless stream of ideas to try. Deepti, Christina, Kiran, Sophia, and Michael, for being amazingly supportive friends. Thank you.

Bismillah.


\pagestyle{plain}
\tableofcontents
\newpage
\listoffigures

\chapter{Introduction}

In this thesis, we study the founder-investor matching process, as experienced by early-stage venture funds and the startup founders they invest in. We do this by exploring the creation of a tool (VCWiz) and completing a study on the data generated from this tool. The tool's goal is to capture existing best-practices in order to bring more efficiency and equitability to the matching process, while collecting data that allows for a further quantitative analysis. VCWiz was built collaboratively with input from best-in-class founders and investors, and has been actively used by thousands of founders. Communication data contributed to this tool lead to the creation of the VCWiz Email Graph, a social-professional graph of founders, investors, and their intermediaries. This data is used to examine how the fundraising process compares to commonly-held expectations. It is also used in our presentation of FounderRank, a methodology for ranking a founder's ability to fundraise based on their position in a social graph.

We begin in Chapter \ref{ch:ch2} by defining and detailing the inner workings of venture capital investments, and our chosen models of the behaviors of both founders and investors. We explore the opportunities in venture capital to leverage software, and justify the selection of VCWiz, a comprehensive fundraising tool for seed-stage founders, incorporating discovery, research, and outreach functionality. Chapter \ref{ch:ch3} documents the design, implementation, and launch of VCWiz. We document and analyze feedback gathered from founders across the spectrum of experience at various iterations of the tool. This feedback allows us to draw conclusions about the way founders use tools during fundraising, and how systems can be built to service those use cases. Chapter \ref{ch:ch4} presents a final VCWiz product: the end result of months of iteration and feedback. Following this, Chapter \ref{ch:ch5} presents a study that explores the data generated by founders on VCWiz. This study aims to discover the significance of social characteristics when fundraising. We explore how investors can rank a set of founders based on these characteristics, and what attributes of a founder set him or her up well to raise money from venture capitalists.

\section{Goals}

The goal of this thesis is to learn more about the process of matching founders to investors. To do this, we attempt to improve the existing process, and draw conclusions from the outcomes of this attempt. Improving the process entails making matching more efficient, in terms of the number and quality of successful matches that occur, and more equitable, by combating the latent biases found in venture capital today.

Our attempt to improve the process involves creating a tool that better equips a founder with the background knowledge, interfaces, and access necessary to find the right investors for their company. Within the context of fundraising, we wish to understand the social characteristics that predispose a founder for success. We want to show how existing information can be used to determine the likelihood of a founder's fundraising success, and how investors can better rank inbound investment opportunities as a result. We strive to create a tool and study that help both sides of the venture capital transaction achieve efficiency, without enforcing existing biases and stereotypes.

Specifically, our goals are to: determine the state of the tools and processes used for founder-investor matching today, build and evaluate a tool that helps founders match with their optimal investors, and use the data from the tool to educate both founders and investors on efficient matching.

\section{A Tool}

VCWiz is a comprehensive tool for seed-stage startup founders to raise their first round of financing. It is a holistic tool that covers discovering relevant venture firms, researching firms and investors of interest, and managing a structured outreach to those investors. This tool was developed over the span of eight months, with input from best-in-class founders and investors. This thesis documents the design and implementation of the tool, and evaluates its usage in production.

\section{A Study}

The communication data collected from VCWiz comprises the VCWiz Email Graph: a social-professional graph of founders, investors, their mutual connections, and the patterns of communication amongst all of them. This data is used in a study centered around FounderRank, a means of scoring founders based on the structure of the graph around them. The rankings generated by FounderRank serve to educate how social relationships can be used to evaluate a large set of founders, on the merit of their fundraising ability. Furthermore, these rankings demonstrate which social features of an individual are most crucial to successfully raising venture money.

\chapter{Background}
\label{ch:ch2}

\section{Venture Capital}

\subsection{Definitions}
\label{ch2:definitions}

A venture capital firm, or VC, is composed of a central pool of capital, contributed by individuals or organizations known as Limited Partners (LPs). This pool is managed by individuals known as General Partners (GPs), who are compensated for their work both with a fraction of the pool (the management fee) as well as a fraction of the returns on their investments (the carry).

In our simplified model, the sole goal of a VC is to trade capital from the pool for equity in companies that will later either enter public markets (via an Initial Public Offering, or IPO) or get acquired by another company. These liquidation events allow the VC to sell their equity for a profit. The success of a VC is measured by the realized capital gains that are accrued when equity is sold; the objective function of a VC is the expected value of this gain over all their investments. We will define efficiency for investors as the time investment per rate of return: the number of hours the GPs must work in order to achieve a given internal rate of return (IRR). With respect to matching investors with founders, we define efficiency as the aggregate number of hours spent by all involved parties on reaching a consensus on investment.

Founders are the other end of a venture transaction. We define founders as the individuals who start and incorporate new businesses (startups). In this thesis, we will focus on seed-stage founders, or founders who are raising their first round of money from institutional investors. Efficiency for founders with respect to fundraising is simply defined as the number of working hours it takes to receive venture funding above a given threshold, as defined by the size of the round. Equitability (in the context of fundraising) is defined as how easily a founder can establish a relationship with an investor with the intent of proposing an investment, regardless of their race, gender, socioeconomic background, or educational pedigree.

For more background on venture capital and the ongoing economic research in the field, we refer the reader to \cite{venture-survey}.

\subsection{Automation Opportunities}

In order to motivate our chosen problem of better matching between founders and investors, we first explore the set of opportunities for automation in venture capital as a whole. From these opportunities, we propose several potential products or tools to be built. A comprehensive overview of these proposals can be found in Appendix \ref{intro:products}. Finally, we identify our selected opportunity as particularly impactful.

The time GPs spend working is split between the activities of sourcing, analyzing, and supporting startups. One can imagine these forming a funnel-like pipeline:

\begin{enumerate}
  \item \textit{sourcing} fills the top of the funnel with high-quality companies,
  \item \textit{analyzing} filters these companies to only the investment-worthy ones, and
  \item \textit{supporting} increases the likelihood of a liquidity event in existing investments.
\end{enumerate}

While it is clear how sourcing additional companies and doing a better job of analyzing potential investments is beneficial to the bottom line of a firm, it is not self-evident that investing time into supporting portfolio companies leads to greater expected returns. To mitigate these concerns, we refer the reader to \cite{JOFI:JOFI12370}, which shows that supporting portfolio companies results in ``an increase in innovation and the likelihood of a successful exit''.

\subsubsection{Sourcing}

Sourcing entails GPs leveraging their networks and any available information (free or proprietary) to discover the optimal set of companies to consider. The stream of companies that are being considered is known as ``deal flow''. This is commonly split into outbound and inbound flow. Outbound flow is generated by the partners attending events and scouring their digital and analog networks for new investment opportunities. Inbound flow is generated by startup founders reaching out to the firm and requesting consideration for investment, or friends of the firm referring new companies for similar consideration.

Much of sourcing requires a human to aggregate large swaths of potentially-relevant signals, such as job changes, incorporations, and referrals, resulting in a few ``interesting'' highlights. The more data that can be ingested, the more potentially investment-worthy companies are surfaced. We can model this as an unsupervised graph problem, where nodes represent information accessible to a firm, and explore how we can learn to identify interesting nodes at a scale no human could manage.

When incorporating the founder perspective, the process of sourcing becomes an efficient matching problem. For any founder, there exists a set of investors who would be willing to invest in their company. For any investor, there exists a set of founders who would make for compelling investments. Facilitating these matches with minimal time burden on both parties is an exciting opportunity.

\subsubsection{Analyzing}

The process of analyzing and doing due diligence on startups is how the GPs of a firm decide whether or not to invest. This can include reviewing the product, financials, and traction of a startup, in addition to doing research on the founders and broader industry.

The lowest-hanging fruit in this process is automatically filtering, categorizing, and ranking companies in the pipeline. Investors are naturally limited to exploring a finite set of companies at any given time. As a result, they reject many companies on the basis of simple filters (such as a lack of academic pedigree) and pattern-matching historic successes. These filters are often sub-optimal and propagate unfair biases. The problem of clustering and ranking companies requires a semi-supervised, structured model that incorporates historic true positives and false negatives.

\subsubsection{Supporting}

Providing what is known as ``portfolio support'' is how venture firms attempt to ensure the companies they invest in survive long enough to realize a liquidation event (thereby allowing the VCs to cash out). This encompasses everything from advising the founders, to making key introductions, helping the company raise further funds, and helping publicize important announcements. There are several opportunities to build workflow automation tools that reduce the burden of time on the investor in carrying out these tasks.

\section{Evaluation of Opportunities}

To evaluate these opportunities for automation, we first justify and present our criteria.

The goal of our proposed solution is to increase efficiency and equitability, as defined in Section \ref{ch2:definitions}, for both founders and investors. In order to accomplish this, we decided that our solution must:

\begin{itemize}
  \item shorten the aggregate time spent by both parties of the venture equation,
  \item be free, open, and accessible to as many people as possible,
  \item not be tied to a specific institution, and
  \item combat existing biases in fundraising patterns.
\end{itemize}

There has been a great deal of attention recently on the issue of diversity and equality in the venture capital world, with many studies concluding that groups such as women and under-represented minorities are less likely to succeed in fundraising because of biases against them. For example, one study out of the Kauffman Centre for Entrepreneurial Leadership finds that it ``may be harder for female entrepreneurs to make the connection, to get in the door, or gain attention for their deal'', since ``women are outside the formal, predominately male venture capital network''~\cite{doi:10.1080/13691060118175}. In building this tool, we strive to use technology to fight these biases, rather than perpetuate them.

\section{Matching Founders to Investors}
\label{ch2:matching}

To balance the goals of equality and efficiency, we decided to build a solution to the founder-investor matching problem. This solution includes means of capturing data on the fundraising process that is useful for educating later analysis.

The problem of matching founders to investors in an efficient manner is crucial to the health of the venture ecosystem. Aside from the extremely well-known firms, there are countless venture firms in the country that are willing to invest in various niches and demographics. Often, the pain of seeking out these firms is what prevents a startup from raising money as expeditiously as possible. On the flip side, venture firms are always seeking out knowledge about new companies being started, particularly given the competitive nature of venture capital. Indeed, it has been shown that winning the competition to see new deals is vital to a venture fund's overall performance~\cite{doi:10.1111/j.1540-6261.2007.01207.x}.

We decided not to build a tool that explicitly aids in analyzing companies. Analyzing investments on the merit of the company is an invaluable aspect of the venture workflow at later stages of investment. However, when it comes to the seed-stage companies we are considering, data on a given company is often scarce. Thus, we instead opt to build a tool that, by aiding in more efficient and equitable matches, generates a dataset we can later use to better analyze the \textit{founders} behind these young companies.

Likewise, we avoided building workflow automation tools to aid investors in supporting their portfolio companies. While these tools can be extremely impactful for the startups affiliated with a given venture firm, it is difficult to build a solution that increases equitability for those founders struggling to be included in the inner circle of venture capital. These tools are often not public, and not shared, making them a poor fit for our goals. Furthermore, while impactful, these tools are often uninteresting to study, given their routine nature.

Though we seek to increase efficiency for both sides of the venture transaction, we decided to build a founder-facing tool, rather than one for investors. This is contrary to the majority of the technical work in the venture community today, which is focused on unilateral automated sourcing and triaging of new deals. Recent examples of this include Social Capital's Capital-As-A-Service\footnote{https://medium.com/social-capital/capital-as-a-service-a-new-operating-system-for-early-stage-investing-6d001416c0df} and the launch of Fly Ventures\footnote{https://techcrunch.com/2017/12/21/fly-ventures/}.

Automated sourcing clearly increases efficiency for investors, and helps them accomplish their goal of finding those rare companies that will exit. It also makes the process more efficient for \textit{some} founders, as they might be discovered by a firm they might not have otherwise interacted with. However, these solutions may often inadvertently harm the equality of the matching process: automated tools are trained on data sets of existing investment decisions that often contain biases against a given race, gender, or educational background. A recent study by venture analytics firm CB Insights claims that only 1\% of funded startup founders are black, and only 8\% are female.\footnote{https://www.cbinsights.com/research/team-blog/venture-capital-diversity-data/} Another study from the National Venture Capital Association shows that ``black employees comprise 3 percent of the venture workforce''~\cite{nvca-diversity}, an alarming statistic given that it has been shown that investors are more likely to invest in founders who share their ethnicity~\cite{BENGTSSON2015338}. We avoided building these investor-facing solutions for fear of exacerbating the disadvantages faced by minorities.

\section{Original Proposal}

The original proposal for the tool component of this thesis was to build a recommender system for early founders to find investors. This system would power a public-facing tool that collects relevant data from both founders and investors, and surfaces recommendations for each. This system would be bootstrapped with public data on venture investments, and augmented with attention-based data from the public tool. However, early attempts to build this system were unfruitful. The public data was too sparse to render any meaningful recommendations: there are simply not enough investments relative to the number of founders and investors. This corroborates the findings of Stone et al.~\cite{Stone:2013:EST:2541167.2507882}, who reported on the difficulty of building a recommendation system with hyper-sparse data sets such as the set of venture fundings in the US.

Instead of focusing our efforts on the recommender system, we refocused on the public tool itself, narrowing the scope of the tool to founders alone. Fortunately, we found that a rule-based system with custom sorting can provide sufficiently appropriate recommendations for the majority of early-stage founders.

\section{A Tool and a Study}

The next two chapters of this thesis focus on VCWiz, the public tool discussed above. This tool is now live at \url{https://vcwiz.co}, and aspires to be a comprehensive application for all the discovery, research, and outreach needs of a first-time founder. We will describe the design and implementation of VCWiz, which includes a graph-based interface that allows founders to explore their connections to investors. Through exposing this interface, we obtained email-based social graph data from $630$ founders actively raising their initial rounds of funding. This represents the subset of founders on VCWiz who volunteered their email data for this study. The remainder of this thesis details the characteristics of these founders, learnings from analyzing their fundraises, and the results of quantitative experiments on the aggregate graph. Each founder included in the study gave consent to have their data used for anonymous, aggregate purposes.

\section{Related Work}
\label{ch2:related}

Literature that explores the founder-investor matching process holistically is scarce. It is difficult to find any academic exploration on the tools and behaviour of founders when fundraising, since many of these efforts are often undertaken in private settings. Of course, this means there are many existing commercial or private solutions that help improve the founder-investor matching process. These products are detailed in Section \ref{vcwiz:existing}. There are not, however, any studies to date that examine the usage of these tools and draw conclusions from that data.

Our work stands out as unique in that it is the only public product that has been built as the result of an academic exploration. There have been private academic efforts to better discover and match with founders on the investor side, such as those undertaken by SignalFire (a data-driven platform that tracks patterns in consumer behavior and the movement of engineers to better inform investment decisions\footnote{https://techcrunch.com/2015/10/22/watch-out-vcs-chris-farmer-says-hes-about-to-massively-disrupt-the-industry/}), or Correlation Ventures (``one of the world's most complete databases of venture capital financings''~\footnote{http://correlationvc.com/approach/about} for use in predictive funding models).

On the specific topic of recommending investors to founders (and vice-versa), recent years have rendered a few studies, most involving the aforementioned Thomas Stone, whose thesis on Computational Analytics for Venture Finance~\cite{stone2014computational} delves into many of the problems with the publicly-available datasets.

When it comes to understanding and explaining the network characteristics of successful founders and investors, there is more work to build on. Existing studies show that better-networked investors, and the founders associated with them, are more successful in their endeavors~\cite{doi:10.1111/j.1540-6261.2007.01207.x}. There is evidence that being well-networked with these investors can impact one's probability of getting funded~\cite{doi:10.1287/mnsc.48.3.364.7731}. It has also been established that strong networks are crucial for entrepreneurs~\cite{BIRLEY1985107}, and specifically that network distance can strongly impact matching~\cite{pasquini2017matching} in venture finance. We build on these studies to demonstrate: the importance of various graph metrics for individuals who are fundraising, how well these metrics numerically correlate to fundraising success, and how investors can use these metrics to identify strong founders.

\subsection{Supporting Work}

There is a vast body of literature in financial and economic modeling of venture capital that we borrow from. For example, we use several findings from \cite{2017arXiv170604229H}, including the list of sector names to use as features for a company and the calculated features for both investors and founders. Another example is the set of economic models summarized in \cite{venture-survey}, which includes the problem of picking startups, matching founders to investors, and the interactions between venture firms and companies. While Rin et al. do not consider the practical ways we can improve these processes, they provide a background for the challenges at hand, and present useful mathematical abstractions.

\chapter{Initial Tool}
\label{ch:ch3}

\section{Motivation}

The culmination of our research and interviews with founders and investors was the proposal of VCWiz, a three-part platform geared towards helping first-time founders better research, discover, and reach out to their optimal seed investors. The focus on seed-stage companies reflected the industry opinion that early-stage venture was an insider's game, where as later rounds of funding were more dependent on quantitative metrics around the companies growth, traction, and success. In the spirit of making venture more accessible and efficient, we chose to focus on first-time founders, who do not necessarily have the connections and tribal knowledge that makes their more experienced counterparts so much more likely to succeed.

We conducted a series of user interviews (N = 21) to determine which aspects of the platform would be most crucial, and what functionality the first version of this application should contain. Appendix \ref{appf:survey:users} contains the questions that were used. The interviews were conducted across a spectrum of experience: from first-time student founders to industry veterans, based in New York, Boston, and San Francisco, working on everything from novel social networks to machine learning-powered drug discovery. We also consulted with several reputable investors at top-tier firms such as First Round Capital and General Catalyst.

We can summarize our user interview feedback into the following three feature buckets.

\subsection{Discovery}

Founders often complain that it is difficult to discover the set of investors that are applicable to their startup. Investors, especially at the seed stage, have a plethora of conditions imposed on the capital they distribute, including restrictions on location, industry, target market size, business model, amount of capital being raised, valuation of the startup, and terms of the deal. Frustratingly, these conditions are rarely published anywhere, meaning it is difficult to query a list of investors and reveal the ones that match your conditions as a founder. This leaves founders resorting to overwhelmingly large databases of investors, or boutique, curated lists that might miss less-well-known options for capital.

\subsection{Research}
\label{ch3:motivation:research}

Once an eligible set of venture firms has been found, the burden on the founder only increases. In addition to figuring out the specific constraints mentioned above, each firm has preferences that may or may not align with the founder's vision for their company. Furthermore, in today's markets, where capital is widely available and there are many similar sources, venture firms are fighting to differentiate themselves to founders. This adds another degree of freedom to the ranking function each founder must internally maintain. There is significant evidence that it is this ``extra-financial'' value of investors that dictates their helpfulness, given equivalent capital contributions~\cite{doi:10.1111/j.1540-6261.2004.00680.x}. However, determining the nature of this value for a given firm is often difficult without a meeting or phone call. A vast increase in the number of new seed-stage funds being started further exacerbates the problem: venture research firm CB Insights claims that ``the number of funds closed in 2014 was nearly 100\% more than 2013''~\cite{cbinsights-research-barbell}.

A separate concern from the selection of the venture firm is the selection of the General Partner within the firm. Our data shows that there are an average of $4.18$ partners per venture firm, with a standard deviation of $3.83$ (Figure \ref{vcwiz:fig:partners}). This does not include the several other associates and supporting staff on the investment team. We note that there has been a decline in the average partnership size over the years: a 1984 survey of venture firms saw $4.7$ partners per firm~\cite{GORMAN1989231}, while a 2008 survey of European venture capital firms saw $4.3$~\cite{BOTTAZZI2008488}. This downward trend can be attributed to the growing number of small, nimble seed-stage funds. Nonetheless, the task of selecting the correct entry point to a firm is daunting and vital to founders. Each investor often has an area of expertise and a type of company they prefer to consider, as well as a particular way to engage with the companies they support. However, as before, there is no easy way to tell which partners prefer which industries or business models, making the selection process for founders laborious at best, and arbitrary in the average case.

\subsection{Outreach}

The final major burden for founders who have identified and researched their ideal (seed) investors is to find a way to get connected to each one. It's commonly accepted in the industry that a so-called ``warm introduction'' (a direct introduction from someone who personally knows the investor) is the best way to start a conversation with a VC. Indeed, this is corroborated by the data. It has been shown that ``direct ties are strongly and positively related to the probability of investment''~\cite{doi:10.1287/mnsc.48.3.364.7731}, in support of the hypothesis that ``investors are more likely to invest in new ventures when they have a previously established direct tie to the entrepreneur than when they do not''~\cite{doi:10.1287/mnsc.48.3.364.7731}. The further removed a founder is in the global social graph from an investor, the lower the chance they will get a direct introduction, and the lower the probability of investment. For many first-time founders, it is simply impossible to get a direct introduction at all. The problem then shifts to finding the best ``intro path'' to a given investor. Barring any introduction, a founder endeavors to send the ``cold'' email that maximizes the chance the investor will consider taking a meeting. This process is often ad-hoc and confusing.

\section{Existing Solutions}
\label{vcwiz:existing}

Several solutions exist that solve one or many of the problems discussed above, but there has yet to be a comprehensive solution. As we will discuss later, the threshold at which a product is considered sufficiently feature-complete is very high. Though many of the founders we interviewed used pieces of these solutions, none of them were satisfied with the status quo, and each one thought that the state of existing tools could be improved.

\subsubsection{Crunchbase}

Crunchbase is an online database of companies, their founders, and the investors that back them. It was created ``to be the master record of data on the world's most innovative companies''~\cite{doi:10.1287/mnsc.48.3.364.7731}, and is largely used as a primary source to learn more about a startup's investors. While the database is very comprehensive (and indeed was use to seed the database for VCWiz), it has historically been cumbersome to navigate. The founders we interviewed found it to be a poor choice for discovery, though an excellent first step for research.

The database offered through the Crunchbase Venture Program \footnote{https://about.crunchbase.com/partners/venture-program/} was used to seed the VCWiz investor database.

\subsubsection{AngelList}

AngelList \footnote{https://angel.co} is ``a platform for startups'' that focuses on early-stage companies and investors (both angels and institutional seed investors). The core platform has social networking, and a directory of startups, their employees, and their early investors, akin to Crunchbase. AngelList's dataset is less comprehensive than Crunchbase's, and narrower in scope. Thus, the founders we interviewed found it less helpful for both research and discovery (though extremely helpful for recruiting employees).

\subsubsection{LinkedIn}

LinkedIn \footnote{https://linkedin.com} is a very popular professional networking platform that founders often use to find mutual connections to investors, so as to solicit introductions. The biggest complaint of founders using LinkedIn for this purpose was that it was not integrated into the rest of their workflow, though this was only expressed in a minority of those surveyed.

\subsubsection{NFX Signal}

Signal \footnote{https://signal.nfx.com} is a platform for founders to find introduction paths to VCs. Founders on Signal grant the application access to their Gmail inboxes, and in return can see the chain of people who comprise the shortest path to any given investor. The graph is built up based solely on email activity, and profile information for investors is self-reported. While this product successfully solves the problem of figuring out which individuals in one's network can provide the introduction to an investor, founders we interviewed often shied away from using it, citing privacy concerns. Signal is operated by NFX Guild, a venture firm, and founders are often unclear about how their email data is being used.

The methodology for displaying investors on Signal~\cite{signal-methodolody} was an inspiration for the VCWiz ranking algorithm.

\subsubsection{Streak}

The Streak CRM is a popular Gmail extension that embeds a spreadsheet-like customer relationship management (CRM) system in your mailbox. It tracks the progress of conversations, updating itself based on the emails being sent and received by the user. Streak offers a template set of headers and categories for fundraising \footnote{https://www.streak.com/startup-fundraising-management-inside-google-gmail} that is often used by founders. This method of tracking outreach and progress during a funding round was one of the most popular in the founders we interviewed: there were very few complaints, other than that this setup still requires substantial manual data entry.

The spreadsheet-like interface for tracking conversations with Streak was the inspiration for the conversation tracker in the first version of VCWiz.

\subsubsection{Affinity}

Affinity \footnote{https://affinity.vc} is a modern CRM solution that includes many features to make fundraising easier and more efficient. It is fully automated, presenting every conversation a founder has over email, along with pre-filled information about investors and firms. It solves many of the qualms founders have with simpler CRM systems such as Streak, and includes a social graph that suggests introduction paths, as Signal does. The platform solves many of the common complaints around fundraising tooling, though it comes at a premium. Our interviews showed that many founders consider it too expensive to use.

\subsubsection{Foundersuite}

Foundersuite \footnote{https://foundersuite.com} is a comprehensive set of tools for fundraising. It is as sophisticated as Affinity, but developed specifically for founders. The CRM component of Foundersuite features a card-based system that requires manual updating when the status of an investor in the fundraising pipeline changes. The software also includes a pre-populated database of investors that is used to autofill fields, and provide a search tool. Like Affinity, the common complaint with Foundersuite was the cost, and complexity.

The set of features and tools offered by Foundersuite inspired the starting feature set for VCWiz.

\section{Our Tool}
\label{chap3:tool}

The first step towards improving the founder-investor matching process is better tooling for both ends. Since founders are often the individuals initiating an interaction, it makes sense to first focus on founders. After reviewing the existing solutions, and the feedback of founders, it was clear that there is an opportunity for a product that is sufficiently comprehensive yet much more accessible (with respect to cost and usability).

Fundraising can look very different at different stages of a company's life. While raising a pre-seed or seed round can entail leveraging one's network to meet and impress sufficient investors until (at least) one decides to invest, later rounds of funding (e.g., Series A or B) are more predicated on the quantifiable traction of the company. Thus, many of the tools above are most impactful for seed-stage founders.

The process of fundraising also looks very different for the subset of founders that are so-called serial entrepreneurs: having raised money from institutional investors in the past means a founder no longer necessarily has the issue of discovering who they should take money from. Furthermore, their experience makes them more likely to succeed at fundraising and building a large business~\cite{gompers2010performance}. Research on retail businesses shows that even outside of technology, ``prior business experience increases the longevity of the next business opened''~\cite{doi:10.1086/683820}. The investing side of the equation also believes in the eminence of repeat founders---data from the First Round Capital 10 Year Project~\cite{first-round-10-years} indicates that ``repeat founders' initial valuations tended to be over 50\% higher'' than those of first-time founders. As a result, our tool is not focused on providing value to serial entrepreneurs.

Thus, in order to maximize the impact we have on the equitability and efficiency of the founder-investor matching process, we opted to design a tool geared towards first-time founders, who are raising their first (seed) rounds. The tool has discovery, research, and outreach components, borrowing interfaces and functionality from the best of the aforementioned tools. We call our new tool VCWiz. VCWiz went through three iterations, each substantially changing the functionality and interface according to feedback solicited from users.

The same three categories considered above (Discovery, Research, Outreach) are how we will partition the solution offered by VCWiz. We will discuss the theoretical solutions to be offered by the tool, and then dive into the implementation and lessons learned from the first two iterations. The final version, which is currently live, is described in the next chapter.

The current version of VCWiz can be found at \url{https://vcwiz.co/}.

\subsection{Discovery}

To solve the discovery problem, we identified several key characteristics that founders look for in a given venture capital firm. These include:

\begin{itemize}
  \item the location of the firm,
  \item the industries the firm has invested in,
  \item the average initial investment (``check size'') of the firm,
  \item the number of investments a firm makes annually, and
  \item the companies a firm has invested in.
\end{itemize}

The goal of the platform is to let the founder specify their preferences in any of these characteristics, and for the system to recommend relevant investors based on those preferences, and information collected about the founder (the stage, industry, location, and competitors of their startup).

\subsection{Research}

After surfacing recommendations to the founder, the platform strives to be the single location with all the relevant information about the partners of a given venture firm. The goal is that the founder never has to leave VCWiz (or a site linked from VCWiz) in order to make a decision about an investor. To this end, in addition to the characteristics necessary for discovery, we collect and report several other pieces of information:

\begin{itemize}
  \item the most recent investments of the firm,
  \item the most recent investments of a given partner at the firm,
  \item the firms that often invest alongside a firm (``co-investors''),
  \item the specific industries that a partner focuses on investing in,
  \item the topics a partner often discusses online,
  \item links to online profiles and content created by the firm or a given partner, and
  \item biographic and demographic information on each partner.
\end{itemize}

\subsection{Outreach}

Finally, once a founder has filtered their recommendations using the research tools on the platform, the final job of VCWiz is to ensure that conversations can begin with the desired investors. To measure progress on this front, the platform contains a ``conversation tracker'': a CRM that auto-populates the profiles of investors marked as desirable by the founder, and auto-updates as the founder has email conversations with those investors. The goal of this CRM is to be as automatic as possible, making assumptions wherever it can.

In addition to simply tracking conversations, VCWiz offers two tools for initiating them. The first is a NFX Signal-style introduction path system that leverages the social graph of the founder to identify the optimal shortest path to any given investor, if one exists. The second tool is a structured system for automated introductions: the founder can request an introduction to an investor, who gets a consistently-formatted, auto-generated dossier about the startup. The investor then has the choice of accepting or rejecting the introduction request. This system is an experiment to see how structure and consistency can improve the process of cold outreach, and is discussed in Section \ref{chap4:introrequests}.

Investors are prompted (by email) for feedback on why they make a decision one way or another. Currently, this feedback is stored but not used. Future work could explore how to best use this feedback to categorize investors, or to educate founders.

\section{Tool Iterations}

\subsection{Version 1}

\subsubsection{Design}

The first step in building the initial version of VCWiz was to spent time talking to 21 teams of startup founders, each going through a well-known accelerator such as Y Combinator (``one of the oldest and top-rated incubator/accelerator programs in the country''~\cite{stross2013launch}). This occurred in June of 2017. The goal was to capture these founders right as they were about to begin raising their initial rounds of funding. They all identified a need for personalized suggestions of investors. Thus our solution was to collect information from each founder on their ideal investor (characterizing investors and firms with features such as industry, check size, and location), and generate suggestions from a cluster of similar investors (using an item-based k-nearest neighbors model, as in \cite{Stone:2013:EST:2541167.2507882}). With this in hand, we built the first iteration of the VCWiz application.

User authentication for the VCWiz platform is handled by Google (Figure \ref{screenshots:v1:login}). The first version of VCWiz collected a founder's ideal investor profile (Figure \ref{screenshots:v1:signup}), as well as basic company information, before taking them to a screen of recommendations. The founder had the option to add any of the recommended investors to a list of ``target investors'' to begin tracking them, before being taken to the main card-based view of the app (Figure \ref{screenshots:v1:conversations}). This view presented a series of stages:

\begin{multicols}{2}
\begin{itemize}
  \item Waiting for Intro
  \item Waiting for Response
  \item Need to Respond
  \item Interested
  \item Not Interested
\end{itemize}
\end{multicols}

Investor cards showed summaries of a partner at a firm, alongside community notes on both the partner and firm. These cards could be moved between stages with dynamic buttons that captured the transitions between stages. At this time, data on user attention patterns was not utilized, save for sorting new users' recommendations by popularity in the existing user base.

\subsubsection{Feedback}

We learned a few crucial insights through the launch and testing of this first iteration of the application.

With respect to discovery, we realized that founders do not find investors by looking at clusters of similar investors after manually identifying a few, as our model assumed. Instead, investors were found by examining the previous investors of similar (or competing) companies. This meant that our investor-based kNN approach performed poorly for users, and that a rules-based system or user-based recommender system would perform better.

With respect to research, the biggest mistake we made was to include only a subset of the information we identified as useful for the founder. As a result, founders would end up leaving the platform to do further research, which was disruptive to their workflow.

With respect to outreach, the first major finding was that it was very difficult to convince founders to trust us with their investor conversations, and that anything we could do to build credibility and legitimacy (for example, auto-filling signup form fields or leveraging the brands of venture firms we were working with) vastly increased willingness to share data.

The next finding was that our users were very familiar with a spreadsheet-based experience (either through Streak or with actual spreadsheets), and that trying to replace it was difficult and unnecessary. A very common piece of feedback was that a ``smart spreadsheet'' would be a far superior interface to the existing card-based workflow.

\subsection{Version 2}

\subsubsection{Design}

The second iteration of VCWiz was started in July of 2017. It featured a new recommendation engine that first asked founders to identify competitors (or similar companies) that are more established. These companies were used to generate recommendations based on a simple algorithm that takes the set of investors from the identified competitors, filters out the eligible ones, and sorts them based on their relevance, popularity, and whether or not they are featured. The popularity of investor $i$ is based on the number of founders who have added $i$ to their outreach list.

\begin{lstlisting}[frame=single,mathescape=true,language=Ruby,basicstyle=\footnotesize,columns=fullflexible]
def recommendations(founder):
  investors $\gets$ founder.company.competitors.flat_map(c => c.investors)
  eligible $\gets$ investors.filter(i => i.industries $\cap$ founder.company.industries $\neq$ $\emptyset$)
  sorted $\gets$ eligible.sort_by(i => [
    i.featured,
    |i.industries $\cap$ founder.company.industries|,
    i.popularity
  ])
  return sorted
\end{lstlisting}

The interface for these investor recommendations is shown in Figure \ref{screenshots:v2:recs} (p. \pageref{screenshots:v2:recs}).

Another addition to the second iteration was an augmented spreadsheet interface (shown in Figure \ref{screenshots:v2:conversations}) that looks and feels like a traditional spreadsheet, but auto-fills cells based on the VCWiz investor database. Once sufficient information had been entered into a row of the sheet to uniquely identify one record in the database, the remaining fields were filled. This gave founders a familiar input experience with a sufficiently powerful addition to justify switching tools.

To build the VCWiz investor database, we started with a direct import of the Crunchbase Venture Program's dataset and augmented it with several additional sources. We discuss our data pipeline in depth in Section \ref{ch4:data}.

The venture firms in our database were tagged with the stage of company they invest in. These tags are assigned based on which fundraising round the VC is expected to initially invest at. The tags are defined (and ordered) as follows:

\begin{multicols}{2}
\begin{itemize}
  \item Accelerator
  \item Angel
  \item Pre-Seed
  \item Seed
  \item Series A
  \item Series B
  \item Venture
\end{itemize}
\end{multicols}

N.B. These categories are not necessarily mutually exclusive. For example, an accelerator may contribute to the pre-seed round of a participating startup. The category for ``Venture'' captures all growth-stage firms (Series C and later).

Finally, we refined the categories that a tracked investor could fall into, based on feedback from founders who were currently fundraising. The list below indicates the ordering over stages that is used throughout the platform.

\begin{multicols}{2}
\begin{itemize}
  \item My Wishlist
  \item Asked for Intro
  \item In Talks
  \item Need to Respond
  \columnbreak
  \item Pitching
  \item Committed
  \item Passed
  \item Not Interested
\end{itemize}
\end{multicols}

\subsubsection{Feedback}

Following the completion of the second iteration of VCWiz, we did another series of user tests, asking founders to focus specifically on the improvements over the first version. The dominant complaints are summarized below.

On discovery, the recommendations were not granular enough, and it was unclear why a certain investor was being recommended. Founders expressed the desire to filter and sift manually through investors. They demanded an interface for queries (comprised of a subset of the characteristics being used for recommendation), instead of being blindly handed what appeared to be arbitrary investors. While there was still a desire for recommendations, it seemed that the place for this was after some degree of filtering, rather than in place of the filtering.

On research, it was felt that the platform still did not provide sufficient information to supplant other tools. Furthermore, it did not display the information in an easy-to-digest way. A common suggestion was to incorporate content from social media and blogging platforms, as investors often use these platforms to demonstrate their interests.

On outreach, the major feedback was that the platform was too rigid. Having pre-defined stages and fields made it difficult to customize the tool for the variance in each founder's workflow. Founders felt like they were fighting the platform, rather than being empowered by it. A common feature request was a way to understand and leverage mutual connections in their outreach.

We used Reichheld's Net Promoter Score (NPS)~\cite{reichheld2003one} to track the growth potential of the product with respect to founders. At this stage, the product had an NPS of -50, which is very weak. Only 25\% of founders said they would recommend the product to a friend.

\subsection{Version 3}

The third iteration of VCWiz was started September of 2017, and aimed to incorporate all previous feedback. The final product of this iteration was a production-ready system that launched publicly. The interface and interactions were redesigned from the ground up, this time with the help of a professional designer. The functionality is still split across the three categories of discovery, research, and outreach, though each is now as feature-complete as the competing products that solve a narrower need. We embraced the feedback from founders that the product needed to be comprehensive and holistic, and have improved upon many of the popular features from other platforms.

We will first give an overview of the features in the third and final iteration of the platform. The implementation and launch of this tool is detailed in the following chapter.

\subsubsection{Discovery}

VCWiz's initial screen contains an interface to filter and search for investors (Figure \ref{screenshots:v3:filter}). Founder can filter by all the characteristics discussed previously, as well as by name and more novel metrics, such as topics often discussed. There are also options to constrain and modify the filters, such as changing a filter from a logical \texttt{OR} to a logical \texttt{AND}. In addition to filtering and searching, there are curated lists of investors that meet specific criteria, ordered by popularity, and selectively shown to founders based on the attributes of their startup (Figure \ref{screenshots:v3:lists}).

\subsubsection{Research}

Clicking on any of the results in the filter view brings up a research screen (Figure \ref{screenshots:v3:research}) that displays comprehensive information on both a firm, as well as every partner at that firm. Every attribute mentioned in interviews by founders as being useful is included in this view, including but not limited to biographies, social media links, recent investments, press mentions, blog posts and tweets, favorite topics to talk about, industries often invested in, and common co-investors.

\subsubsection{Outreach}

The outreach functionality of VCWiz is embedded in each screen. There is also a dedicated dashboard for tracking conversations.

Each view for research and discovery has buttons to add investors and firms to the VCWiz Tracker, a CRM that is synced throughout the platform. Founders see a contextual dropdown, showing the status of an investor in their pipeline, at every instance where they are on a research screen including that investor.

The conversations screen is a standalone dashboard that shows a table-like view, containing the information that was previously contained in spreadsheets. We used a table to strike the balance between giving founders an interface they are familiar with, and displaying the information in a sufficiently detailed way. The data in this table is populated and updated through an integration with the founder's email provider.

In addition to merely tracking conversations, this version of VCWiz overlays research and discovery screens with a subset of the founder's email graph, showing them the shortest path(s) to any given investor in their network. This allows the founder to understand how likely they are to be able to reach an investor without a single extra click.

Shortly after the public launch of this third version of VCWiz on January 25th, 2018, we surveyed active founders on the platform to again calculate the NPS (see Appendix \ref{appf:survey:founders}). This time, we received 117 responses, with an NPS of 0, a significant jump from the last version. 63\% of users agreed that they would recommend VCWiz to a friend, up dramatically from 25\% previously.

\chapter{Final Tool and Launch}
\label{ch:ch4}

This chapter will detail the final iteration of the VCWiz platform, and our efforts to launch it to the public.

\begin{figure}[ht]
  \centering
  \begin{minipage}[t]{0.45\textwidth}
    \centering
    \includegraphics[width=\textwidth]{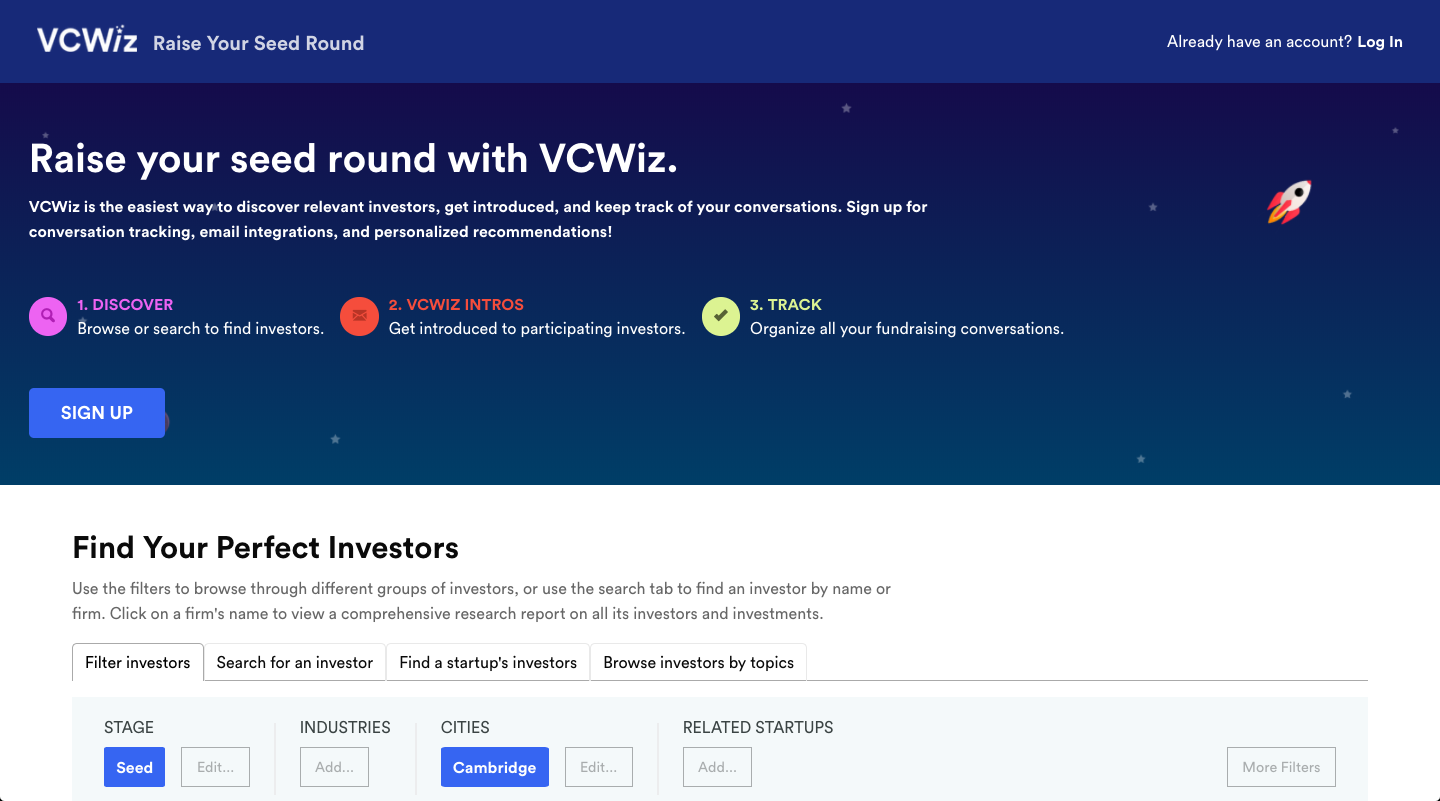}
    \caption*{VCWiz Landing Page}
  \end{minipage}\hfill
  \begin{minipage}[t]{0.45\textwidth}
    \centering
    \includegraphics[width=\textwidth]{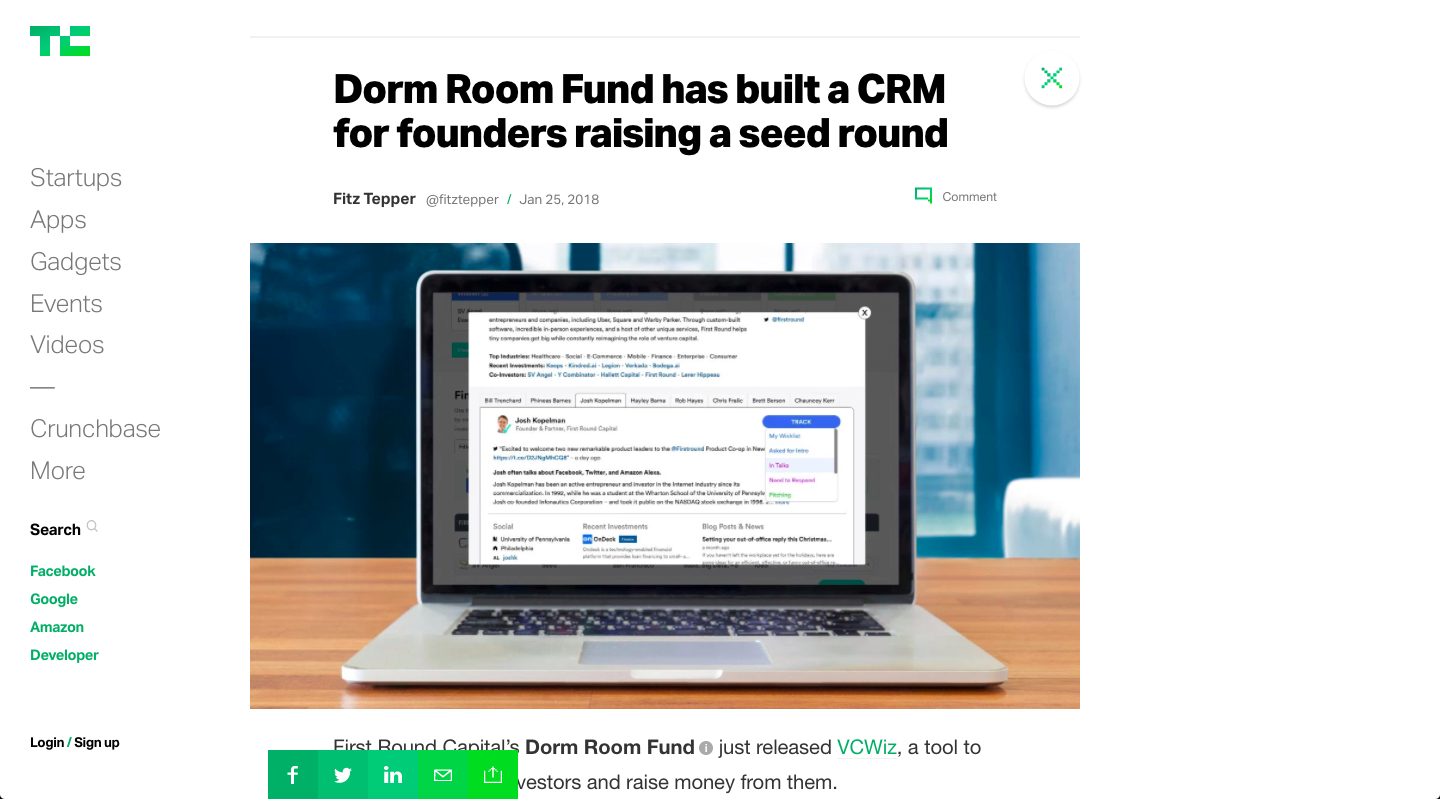}
    \caption*{Launch post on TechCrunch}
  \end{minipage}
\end{figure}

\section{Frontend}

The final platform incorporated all the feedback from previous iterations, and was built over a span of four months. Below, we expound the technical details of the platform's interface, exploring each aspect in the order a new user would.

\subsection{Onboarding}

Founders find the VCWiz platform through one of our launch partners (such as TechCrunch\footnote{\url{https://techcrunch.com/2018/01/25/dorm-room-fund-has-built-a-crm-for-founders-raising-a-seed-round/}}), or from search engines such as Google. We spent the months leading up to the launch generating research pages for every investor, firm, and company in our database (Figure \ref{screenshots:v3:onboarding}). These pages include comprehensive details on the entity in question, as well as an embedded view of all the VCs associated with that entity, and, if the user is signed in, all the personalization included in the platform.

After transitioning from their entry point to the main screen of the application, founders are able to filter, search, and explore lists of investors without creating an account. The site is fully functional from a discovery and research perspective, and about 80\% of users are content to peruse the content without signing up. If the founder decides to register themselves, we walk them through a series of questions to gather more information about their startup.

The signup flow begins by asking for the domain of the company. Using this as a unique identifier, we are able to query both our internal database, as well as external services (such as the Clearbit Logo API\footnote{https://clearbit.com/logo}) to gather as much information as possible on the founder's startup. The client browser makes a request to an API backend on the VCWiz server that initiates these requests in parallel, and returns a joined \texttt{Company} model within a given timeout threshold. This information is used pre-fill many of the following fields, including the name, description, industries, and competitors of the company. The founder is given a chance to verify this information, as well as provide mandatory information on their ideal investor profile. Finally, the founder is requested to log in with their Google account, in order to provide an authenticated email and social profile. We chose to use an OAuth2-based \cite{hardt2012oauth} login flow with an existing service provider to simplify the login experience, and to avoid having to store user credentials. Google was the platform of choice on account of it providing verified email address information for users, as well as to unify the authentication experience in the case that the founder also decides to provide API access to their email inbox (for the purpose of synchronizing their conversations with investors).

Providing access to one's email inbox is strictly an opt-in feature, and how the data will be used is explicitly described. As a result of our surveys to founders in previous iterations of the product (see Appendix \ref{appf:survey:users}), we found that it was necessary to have a plain-English description of our data use policy. We guarantee to founders that no human will ever read the individual messages of their inbox, that only aggregate data will be used for purposes other than their personal dashboard, and that we will only use metadata from their emails (headers, sentiment, etc.).  We allow ourselves to use features based on the body of the email, such as sentiment, provided they cannot be used to reconstruct a representation of the body.

After signing up, the founder is presented with a brief set of video clips that introduce the functionality to them (Figure \ref{screenshots:onboarding:intro}), including how to filter, search, and track investors. Following this, the site functions as it did before the founder signed up, with a few minor changes. Every screen with an investor has an integrated conversation tracker that shows the status of that investor, if any, in the founder's fundraise, as well as the email-based shortest intro path to that investor. The results of the filters are also personalized to the founder, based on the overlap in industries and location between each firm and the founder's startup. Signing up also unlocks the conversation tracker, with a preview of conversations on the main page (Figure \ref{screenshots:onboarding:summary}), and a dedicated screen for updating and viewing the status of each individual conversation (Figure \ref{screenshots:onboarding:conversations}).

\subsection{Ingesting User Data}
\label{vcwiz:ingesting}

One of the major insights from previous iterations of VCWiz is that founders have a variety of different ways they create and interact with data about their fundraising process, and they aren't often willing to change those. Thus, the tool we built had to meet founders wherever they currently were, in order to keep their conversation tracker on our platform updated. We built three independent tools for letting the system know about ongoing conversations, in addition to the integrations in the research and discovery sections.

The first (and easiest) way founders can import their conversations to the platform is to grant access to their Gmail inbox, either during the signup flow or when later prompted. This allows a regularly-scheduled job on our server to poll an API offered by Google\footnote{https://developers.google.com/gmail/api/}, and import new messages according to the pseudocode in Listing \ref{code:sync}. A \texttt{history\_id} parameter is cached in the \texttt{Founder} model to indicate the most recent thread fetched from Google, to avoid fetching duplicates in the future.

\begin{lstlisting}[frame=single,mathescape=true,language=Ruby,basicstyle=\footnotesize,columns=fullflexible,caption={Sync Inbox},label={code:sync}]
def sync_inbox(founder):
  for thread in fetch_threads(founder.address, founder.history_id):
    messages $\gets$ thread.fetch_messages()
    for message in messages:
      if message.from == founder.address:
        parse_outgoing(founder, message)
      else:
        parse_incoming(founder, message)
    founder.history_id $\gets$ thread.id
\end{lstlisting}

Parsing messages follows the algorithm in Listing \ref{code:parse}. This process also augments the founder's email-based graph with every email scanned.

As can be seen from the algorithm, when importing a user's emails and creating their email graph, we first started with the naive approach of scanning every email, creating a node (if one did not already exist) per address, and creating outgoing edges every time one node sent an email to another. While this works when only importing emails once, the APIs at our disposal were imperfect. The \texttt{history\_id} tracked from Google's API often expires, and imports must be repeated. Thus, we had to start tracking a unique message identifier in our own database to ensure emails are imported at most once.

\begin{minipage}{\linewidth}
\begin{lstlisting}[frame=single,mathescape=true,language=Ruby,basicstyle=\footnotesize,columns=fullflexible,caption={Parse Message},label={code:parse}]
def parse_message(message):
  if check_if_bulk(message):
    return
  founder.graph.connect(message.address)
  target_investor $\gets$ find_or_create_target_investor(founder, message)
  if !target_investor:
    return
  if !target_investor.email:
    target_investor.email $\gets$ message.address
  target_investor.stage $\gets$ guess_stage(message)
  create_new_email(founder, target_investor, message)
\end{lstlisting}
\end{minipage}

There are also several heuristics we use to skip messages that could be classified as bulk mail, as this adds unnecessary noise to the dataset. If the message meets any of the following criteria\footnote{The full algorithm can be found online at \url{https://git.io/vxunk}.}, it is logged and skipped:

\begin{itemize}
  \item There are more than 5 recipients
  \item The body contains a phrase often used in bulk mailings, such as ``unsubscribe'', ``terms of use'', or ``view in your browser''
  \item The headers contain one of several common vendor-specific listserv headers, such as \texttt{List-Unsubscribe}
  \item The return path of the message includes a popular bulk email vendor
  \item The local component of the from address is that of a commonly-automated inbox, such as ``noreply'' or ``info''
  \item The name of the sender includes common aliases, such as ``support'' or ``payroll''
  \item The domain of the sender or any recipient is common in transactional emails
\end{itemize}

N.B. In these criteria, the recipients are defined as the union of the TO, CC, and BCC fields, and body is defined as the concatenation of the text and HTML sections of the message.

The second way founders can inform the system about ongoing conversations is to CC (or BCC) a special email address. This address routes to a server that accepts the message and forwards the relevant metadata to an API endpoint on VCWiz. The metadata is parsed and the email is reconstructed before being run through the same algorithms as above. This alternate, manual way of updating VCWiz via email was added for the more privacy-conscious founders on the platform, who wished to have the convenience of updates based on emails without handing over access to their entire inbox.

The third and final ways founders can update the system in bulk is by uploading a existing spreadsheet of conversations. Our surveys revealed that the most commonly-used tool for tracking conversations with investors is a spreadsheet (or spreadsheet-like tool), so providing an easy way to migrate those onto the platform was essential. Founders can export a CSV file from any spreadsheet-based tool and import it on the conversation tracker page of VCWiz. The server parses the rows out of the CSV, and uses both the format of a column as well as it's header (based on the Levenshtein distance \cite{1966SPhD...10..707L} of a given header from a list of common choices) to guess which columns correspond to which internal database columns of a \texttt{TargetInvestor}. This mapping is presented to the founder for verification (Figure \ref{screenshots:import:columns}), and then is used to import the rows as a background job.

\subsection{Filter, Search, \& Sort}
\label{ch4:filtering}

\subsubsection{Filter}

The main filtering interface allows founders to display investors that match a set of criteria. We will describe each criterion before showing the algorithm used to filter. The logic behind the selection of criteria is to cover the majority of the ways founders describe their ideal investor: the stage the investor operates at, the industries that they invest in, the location they invest in, and their relationship to similar/competing companies (similar companies are generally a good sign, whereas directly competing companies might be prohibitive).

The first criteria is based on the stage at which a VC operates, as defined by the first funding round it typically participates in. This information can either be self-reported by a partner at the fund, or inferred from past investments. Note that both fund and funding rounds can be affiliated with multiple stages: when aggregating past investments, any stage that shows up at least half of the time is attributed to the fund. This criteria is always a logical \texttt{OR} when multiple are selected.

The next criteria is the set of industries that the VC commonly invests in. Once again, these are preferably self-reported, and there can be multiple associated with a fund (as well as an investor or a company). If this set needs to be calculated, it is done in the same way as the stage. The universe of industries is fixed: there is no free-form option when filtering. This universe (listed in Figure \ref{vcwiz:fig:industries}) was selected using the algorithm in Listing \ref{code:industries}, run over all the companies in the Crunchbase data set. The goal was to select the set of industries that cover the entire set, with minimal overlap.

\begin{minipage}{\linewidth}
\begin{lstlisting}[frame=single,mathescape=true,language=Ruby,basicstyle=\footnotesize,columns=fullflexible,caption={Display Industries},label={code:industries}]
def covering_industries(companies):
  all_industries $\gets$ companies.flat_map(c $\to$ c.industries)
  industry_options $\gets$ all_industries.unique()
  sorted_options $\gets$ industry.sort_by(i $\to$ all_industries.count(i))

  selected $\gets$ set()
  while companies.filter(c $\to$ c.industries $\cap$ selected $\neq \emptyset$).count() > 0:
    selected $\gets$ selected $\cup$ {sorted_options.pop()}

  return selected
\end{lstlisting}
\end{minipage}

By default, when a founder selects multiple industries, the filter is a logical \texttt{OR} of these industries. However, there is an option that can be toggled to make this filter an \texttt{AND}, such that returned VCs invest in \textit{all} the specified industries.

The next criteria is based on a set of cities. By default, this filter returns firms that are based in the cities specified (firms have a single headquarters, and an array of office locations, both of which are matched against). There is an option, however, to change this filter to instead return firms that have invested in \textit{startups} based in the specified city (each startup is affiliated with a single city).

Another criteria to be matched against is a set of relevant startups. The founder can select this set from the database of companies VCWiz tracks internally. By default, this set restricts the returned firms to those who have invested in at least one of the specified companies. An option can be toggled that changes this filter to restrict to the set of firms that have invested in \textit{similar} companies, based on the industries of each company in the set.

The final criteria to match investors against is a set of topics. In this case, a topic is anything found in the VCWiz entity database, which is built by extracting entities from the various data sources discussed in Section \ref{ch4:data}. At the time of writing, this database contains $98,000$ records. The filter based on these entities is a logical \texttt{OR}, and will return investors who often mention or discuss any of the given topics. We associate a topic with an investor if that topic is mentioned at least 5\% of the time in content created by or mentioning the investor.

Finally, there is a lone option to restrict the returned set of investors to those that operate solely in the US. This was a popular criteria for many founders on our platform.

\subsubsection{Search}

In addition to filtering against any combination of the above criteria, founders can also filter investors based on the name of the firm or individual. This is implemented as a simple fuzzy string match on the \texttt{name} field of \texttt{Firm}, and the \texttt{first\_name} and \texttt{last\_name} fields of \texttt{Investor}.

\subsubsection{Sort}

Once VCWiz has generated the set of investors that match a given filter-and-search query, it must decide the order in which to display the results. We devised a custom ranking function that sorts the results, using the following metrics to break ties:

\begin{enumerate}
  \item The number of investors in the firm that match the set of topics in the query
  \item The number of ``featured'' investors in the firm
  \item The number of intersecting industries between the firm and the query
  \item The number of intersecting cities between the firm and the query
  \item The number of founders who have initiated a conversation with the firm
  \item The number of ``verified'' investors in the firm
\end{enumerate}

N.B. Any metric that is missing the relevant filter (e.g., the topic filter for the first metric) in the query is simply ignored. ``Featured'' investors have been hand-picked as high-quality investors. ``Verified'' investors have completed their investor profile on VCWiz and self-reported their investment criteria.

This sorting function allows for a degree of personalization by using information from the founder's profile when the filter relevant to a metric is missing. Examples include the industries of a founder's startup and current location of a founder when the third and fourth metrics respectively are missing a filter. Note this profile information is only incorporated for ranking, \textbf{not} for filtering. Thus, if a founder specifies all possible filters, he or she will get the same results as another founder with the same query. However, if specific filters are omitted, information from the founder's profile will be used, rendering a unique ranking.

The interface by which these results are displayed to the founder allows for further sorting, based on the natural ordering of a given column. When this overriding sort is used, the custom rank is ignored.

\subsubsection{Implementation}

The algorithm for collecting the results of this filter, search, and sort process is shown in Listing \ref{code:filter}.

We start with every firm in the database, and filter out any that do not match the search string(s) provided by the founder. Often, the search string provided is a single word. In this case, we treat the string as the query for both the firm name and the individual investor name. The results then include the firms that directly match the query, and the firms that include an investor who matches the query.

We next apply the query's filter to the remaining set of firms, narrowing down the result set each time. In other words, the filters are aggregated with a logical \texttt{AND}.

\begin{lstlisting}[float,frame=single,mathescape=true,language=Ruby,basicstyle=\footnotesize,columns=fullflexible,caption={Filter and Search},label={code:filter}]
def filter_and_search(all_firms, founder, filters, search):
  firms $\gets$ all_firms
  investors $\gets$ firms.flat_map(f $\to$ f.investors)

  if search:
    first_name, last_name = extract_name_components(search)
    investor_by_name $\gets$ investors.filter(i $\to$
      i.first_name.contains(first_name) || i.last_name.contains(last_name)
    )
    firms_by_investor_name $\gets$ investor_by_name.map(i $\to$ i.firm)
    firms_by_name $\gets$ firms.filter(f $\to$ f.name.contains(search))
    firms $\gets$ firms_by_name $\cup$ firms_by_investor_name

  for filter in filters:
    firms $\gets$ apply_filter(firms, filter)

  sorted $\gets$ apply_ordering(firms, founder, filters)
  return sorted
\end{lstlisting}

\subsubsection{Display}

The result set is displayed in an infinitely-scrollable table to the founder (Figure \ref{screenshots:filtering:results}), with the following columns:

\begin{enumerate}
  \item The name and photo of the firm
  \item The company stages the firm invests at
  \item The headquarters of the firm
  \item The number of investments the firm has made in the last calendar year (``pace'')
  \item The top three industries that the firm invests in
  \item A drop-down to add or update the firm in the conversation tracker
\end{enumerate}

N.B. What founders really desire to see in the ``stage'' column is the average investment size of the firm. However, this number is difficult, if not impossible, to calculate given the limited public data on investor contributions to a given fundraising round. Thus, the company stage is used as a proxy.

The ``pace'' column is included to give the founder a sense of how active a given firm is. This was added by popular request, after many founders found it difficult to determine whether or not a firm was still actively investing.

Each column (other than ``industries'') also provides a button for overriding the custom ranking function. This allows the founder to sort the results by the data in a particular column. The ``firm'' and ``location'' columns are sorted lexicographically, the ``pace'' column numerically, and the ``stage'' and ``track'' columns by their inherently-defined orderings.

When a search string is provided, or the topic filter is used, it is valuable to not only surface not only the resulting firms, but the best-matching investor at each firm (e.g., if the search query matches the first name of a partner). In these cases, there is an additional, non-sortable column titled ``Partner'' that displays the name and photo of that best match (Figure \ref{screenshots:filtering:partner}).

\subsubsection{Initialization}

After completing the signup flow for VCWiz, the first screen a founder is presented with is the filter and search screen. To ensure a positive first experience when viewing the results, we initialize the filters to personalized values based on common assumptions. The funding round filter is set to ``Seed'', to reflect the target user of the platform. The industry and relevant startups filters are pre-filled with the industries and competitors of the founder's startup, respectively. Finally, the location filter is set to the nearest ``hub city'' to the founder's current location (based on their IP address).

A ``hub city'' is defined as a city that has at least 50 venture firm offices (as reported by our database) within it. This number was chosen as it is a natural threshold for cities with significant number of venture firms (see Figure \ref{vcwiz:fig:hubscuttof}). Figure \ref{vcwiz:fig:hubs} contains a list of the 69 hub cities on VCWiz at the time of writing.

\subsection{Introduction Requests}
\label{chap4:introrequests}

An experimental component of the final VCWiz platform is the ability for founders to request automated introductions to out-of-network investors on the platform.

The motivation behind this was to standardize the format and medium of ``cold'' intro requests in the venture community. As discussed in Section \ref{ch2:related}, there is both qualitative and quantitative data supporting the use of mutual connections to make introductions when reaching out to investors. This is corroborated by the results of our experiments on the VCWiz graph in Chapter \ref{ch:ch5}: centrality in the global social graph of startups and venture capital is highly correlated with how easily a founder will raise money. However, sometimes an introduction is simply not an option. In this case, founders resort to ad-hoc, unsolicited emails to investors, leveraging myriad folklore tactics\footnote{https://hbr.org/2016/09/a-guide-to-cold-emailing} to increase the chances of a response. This is a frustrating experience for both parties: investors are deluged with a stream of unwanted pitches, mixed haphazardly into their daily business, while founders are disappointed that their carefully-crafted custom email gets lumped in with the bulk email another founder sent to $1000$ investors.

We attempted to solve this problem by providing a tool for founders to send a templated introduction request that looks identical each time, save for a customizable blurb (Figure \ref{screenshots:intro:request}). The investor's email address is never initially revealed to the founder. Instead, a standardized request email is sent by the VCWiz platform to the investor, containing an automatically-generated dossier on the founder and their startup (from the information provided at signup). The investor can respond to this automated email with a simple ``yes'' or ``no'', and only in the case of an affirmative response is a second email sent from the platform, connecting the founder and the investor.

While many investors agreed that using an automated third-party such as VCWiz was preferable to founders directly sending countless emails and followups, there was significant doubt that such a platform would be adopted to a degree that could be considered a success. As it turns out, these concerns were well-founded.

When we launched the feature, we tracked a funnel of four success metrics:

\begin{enumerate}
  \item The number of introductions requested
  \item The number of requests that receive a response
  \item The number of successful connections
  \item The number of investments made as a result of an introduction
\end{enumerate}

N.B. a ``successful connection'' is defined as any introduction that results in at least one additional email from each party.

A few months after launching the feature, we saw very poor results. Out of 301 introductions requested, only 19 garnered a response from the investor, of which five were affirmative. One of these resulted in a successful connection, and none resulted in an investment.

Our hypothesis is that this experiment failed for two reasons. The first is that the founders who resorted to using this tool were inexperienced at fundraising or were not very well-connected, which presents an adverse selection problem: as we will demonstrate later, founders who are not well-connected will struggle to raise relative to those who are. The second reason is that investors have such a low response rate to cold emails of any kind that any improvement is negligible.

We tested this hypothesis by surveying every founder on the platform (questions shown in Appendix \ref{appf:survey:founders}), and every one of the 247 investors who had received an introduction request (Appendix \ref{appf:survey:investors}).

We asked the founders why they had or had not used the introduction request feature. $118$ founders responded, with $10\%$ saying they had tried the feature, $23\%$ saying they would never get a cold introduction to an investor of any form, and $16\%$ not understanding the value of the feature. The long tail of remaining responses ranged from not currently needing any intros, to hitting bugs when trying to use the feature. This data supports that many founders are skeptical of using the feature, or any cold introduction, because of how ineffective they are. A majority of the founders who have not used the feature ($65\%$) have exchanged emails with investors before, indicating they are somewhat experienced.

Only 13 investors responded to the investor survey. The overwhelming response was that the founders who had reached out were simply not high quality, or a good fit for their fund. This supports our earlier hypothesis on founders, and the very fact that there were such few responses corroborates our hypothesis about investor response rates to unsolicited emails. An interesting finding is that our solution still inconvenienced investors more than they were comfortable with. Their ideal solution would involve a centralized dashboard of requests that could be checked for interesting prospects by designated individuals at the firm, often not the investors themselves.

\subsection{Intro Paths}
\label{chap4:intropaths}

Founders who have shared their email graph with VCWiz can see an ``intro path'' to any given investor on the platform (Figure \ref{screenshots:intro:path}). The goal of displaying these paths (and the length of that path as investor's distance from the founder) is to assist founders in planning who can make an introduction for them (so as to avoid the problem faced in Section \ref{chap4:introrequests}).

The path is calculated by running a standard single-pair shortest-path algorithm between the founder's node and the investor's node. If multiple paths are found, the paths are ranked by the strength of the connections they represent (based on the sum of the frequencies of emails between nodes on the path), and the top three are returned. The Cypher script run on our Neo4j database instance to accomplish this has been reproduced in Listing \ref{vcwiz:cypher:intro}.

Intro paths are also available between founders and venture funds. In this case, the same algorithm as above is run, for every investor within the fund. We take the union of the resulting shortest paths, and rank them in the same way.

\section{Backend}

The backend architecture of the final VCWiz application is a Ruby on Rails\footnote{http://rubyonrails.org/} application that serves both the frontend React\footnote{https://reactjs.org} application and an internal API. Data on firms, investors, companies, and founders is ingested from many sources on a regular basis, using Sidekiq\footnote{https://sidekiq.org/}, a job scheduler, to update specific shards of the database in each job.


The main persistent store for data is a PostgreSQL\footnote{https://www.postgresql.org/} database running on Amazon Web Services (AWS). There are also instances of Redis\footnote{https://redis.io/} (for caching external API responses), Memcached\footnote{https://memcached.org/} (for caching internal intermediate data for rendering), and Neo4j\footnote{https://neo4j.com/} (for calculating introduction paths).

The application servers are deployed on Heroku\footnote{https://www.heroku.com/}, a platform-as-a-service that also runs on AWS.

In this section, we will detail the various aspects of the backend services, and how data flows through the system.

\subsection{Data Models}

Our main data models are the \texttt{Company}, \texttt{Founder}, \texttt{Investor}, \texttt{Firm}, and \texttt{Investment}. These, along with auxiliary models, are diagrammed in Figures \ref{vcwiz:model:hierarchy} and \ref{vcwiz:model:content}. Each of these models is backed by a similarly-named database table.

The decision was made to have many \texttt{Company}s per \texttt{Founder}, as founders on the platform very often have started a company before. This necessitates a denormalized \texttt{PrimaryCompany} model that keeps track of which \texttt{Company} is the one a founder is currently leading. One current issue with the data model is that any founder can affiliate him or herself with a startup already in the system, whether or not their claim is true.

Each \texttt{Founder} has many \texttt{TargetInvestor}s, each of which represent a conversation between a founder and an investor (or an investor on a founder's wishlist). \texttt{IntroRequest}s and \texttt{Email}s are then affiliated with a \texttt{TargetInvestor}.

Tweets, news articles, and blog posts mentioning either an individual investor or entire firm are each tracked by their own model. An \texttt{Entity} model that can be associated with any of these tracks mentions of extracted entities, and is used for topic-based searching. In order to reduce noise in the selection of entities, we made the decision to only create an entity record if a given entity has an entry on Wikipedia\footnote{https://www.wikipedia.org}.

\subsection{Data Pipeline}
\label{ch4:data}

There are several sources of information used by our data pipeline, each of which is abstracted, normalized, and merged into the existing schema of the system.  Instead of attempting to mirror the structure of each API in the server code, a wrapper class (\texttt{ApiObject}\footnote{https://git.io/vpvib}) was created that abstracts away common structure in the external API endpoints accessed. This allows simple property-based access of the JSON objects returned, with automatic detection of arrays and types that need to be converted (such as dates). The general methodology for populating an object is to start with a base source of truth (often Crunchbase), then augment with a variety of information streams, some of which are documented below.

\subsubsection{Crunchbase}

Through the Crunchbase Venture Program, we received access to the entire database of investors, firms, and companies on Crunchbase. Each \texttt{Company}, \texttt{Founder}, \texttt{Investor}, and \texttt{Firm} on VCWiz stores a unique Crunchbase identifier (\texttt{crunchbase\_id} or \texttt{cb\_id}) that allows changes on Crunchbase to be reflected in our models (when appropriate). Whenever an object with associations that have Crunchbase identifiers is updated, background jobs are initiated that attempt to fetch updates for each association. Furthermore, approximately once a month, a complete dump of the Crunchbase database is downloaded and imported (skipping over existing records).

\subsubsection{AngelList}

Each \texttt{Company}, \texttt{Founder}, \texttt{Investor}, and \texttt{Firm} also has a field for storing an AngelList identifier. The AngelList API\footnote{https://api.angel.co} is used to augment information on these objects when Crunchbase is ambiguous or incomplete. Through manual inspection, we found that AngelList's dataset often contains more information for companies that have not yet raised money from institutional investors, whereas Crunchbase focuses on venture-backed startups.

\subsubsection{Bing News Search API}

The Bing News Search API\footnote{https://azure.microsoft.com/en-us/services/cognitive-services/bing-news-search-api/} is used to periodically check for previously-unseen news articles on a given investor. These news articles are imported and processed, which involves summarizing them, extracting entities from their bodies, and categorizing their sentiment. All of this information is saved to a \texttt{News} record that is displayed to users on the research page for a \texttt{Investor}.

\subsubsection{Newsriver}

Newsriver\footnote{https://newsriver.io/} is a similar API to Bing News Search, and is also used to monitor for new press on an investor.

\subsubsection{Clearbit}

Clearbit\footnote{https://clearbit.com/} is a service that provides access to a dense graph of human profile information, with nodes that can be identified with an email address or social media profile. We use it to auto-fill profiles for both founders and investors.

\subsubsection{Text Processing API}

We use the Text Processing API\footnote{http://text-processing.com/docs/} for entity recognition and sentiment analysis of many pieces of text, including news articles and emails.

\subsubsection{Google Cloud Natural Language}

We use the Google Cloud Natural Language API\footnote{https://cloud.google.com/natural-language/} for the same reasons as the Text Processing API.

\subsubsection{Hunter}

Hunter\footnote{https://hunter.io/} is a service that collects common email patterns on a per-domain basis to aid in guessing a person's email address given their name and domain. When a founder requests an introduction to an investor who has not yet signed up for the platform, we use Hunter to guess their email.

\subsubsection{Twitter}

We use the APIs provided by Twitter\footnote{https://developer.twitter.com/en/docs} combined with the social media usernames reported by Clearbit to log recent tweets of every individual investor on the VCWiz platform. These tweets are displayed on the research page for the investor. Entities are extracted from these tweets, and are used to build a profile of the topics affiliated with an investor.

\subsubsection{Medium}

Medium\footnote{https://medium.com/} is a popular platform for blogging. When an investor has a profile on Medium, it is scraped regularly to identify new blog posts to display on the research page. Like tweets, entities are also extracted from blog posts for analysis.

\subsubsection{Homepages}

The homepages of investors, firms, and founders are all scraped for entity extraction, similar to the blog posts and news articles above.

\subsection{Inferring Partners}
\label{ch4:partners}

One of the most useful pieces of information about a given venture fund is the mapping of partners to investments; each investment generally has one partner as the champion. There is significant variance in the industry, business model, and founder background that each partner prefers, and selecting the most appropriate one can be crucial to securing an investment. Unfortunately, it is not common practice to make public which investor is the point partners on each deal, and founders are often left in the dark.

One of the key insights we had while building VCWiz is that there are often sufficient signals online to infer which partner at a firm was responsible for a given investment. These signals include the partner mentioning a portfolio company in their biography, frequently tweeting about a company, or often commenting to the press on behalf of the firm on matters regarding a company. While aggregating these signals manually would be tedious and difficult, it is a relatively easy process to automate. Our backend periodically queries for press and social media mentions of the portfolio companies of each firm, and scans those mentions for the names of the partners at the firm. If a partner appears more often than others, we assume they are the partner responsible for the investment.

While this method is not perfectly accurate, it has empirically shown to be sufficient.

\subsection{Security}

The internal API exposed by VCWiz presents an opportunity for abuse. The resources required to build and maintain our database are considerable; sites that expose a similar dataset go to great lengths to discourage web scraping and other illegitimate access. Crunchbase, for example, employs the services of Distil Networks\footnote{https://www.distilnetworks.com/}, which uses a variety of Javascript-based methods to prevent programmatic scraping. VCWiz exposes a JSON API to the public internet, so we sought to ensure that no client other than the VCWiz frontend could access internal resources. Additional measures were put in place to ensure the security and integrity of the founder-contributed data on the platform.

The first concern with respect to security is the storage of the data. All data is stored in a single database, with credentials in an environment variable on the web server. These credentials are rotated automatically on a regular basis. Access keys for third-party APIs are also stored in environment variables, and never recorded in code. User sessions are encrypted with a private key held only by the web server before being serialized into cookies. These sessions contain the primary key of the currently logged-in \texttt{Founder} (if any) to ensure that no one else can access a founder's data.

Protecting the internal APIs required preventing both unauthorized reads (of user or bulk data) and writes (that a user did not intend).

Unauthorized writes across founders are defended against with the measures described above. A common attack vector for an unauthorized write across sites is a cross-site request forgery (CSRF). CSRF is when ``a malicious site instructs a victim's browser to send a request to an honest site, as if the request were part of the victim's interaction with the honest site''~\cite{Barth:2008:RDC:1455770.1455782}. In our case, a malicious site could send a request to an internal API, impersonating the currently logged-in founder to, for example, request an introduction from an investor with arbitrary text. This presents a risk to the founder, and is mitigated by embedding a request-specific key in the meta tags of each page. This key is parsed by the frontend application, and sent in a header to the API with every request. If the key is present, the request must be from a legitimate source. If it is absent or incorrect, the request is illegitimate and is rejected before being routed.

Preventing read abuse of internal resources is accomplished through a combination of expiring server grants and rate-limiting. The encrypted user session maintains a timestamp, which is refreshed on every non-API page load. Each time an API request is made, this session timestamp is compared against the current server time. If more than one hour has elapsed, the API request is rejected with a \texttt{401 Unauthorized} status code. This ensures that only clients representing active users on the website can make API requests. Of course, there are instances where this mechanism results in a legitimate user being denied (because, for example, they left the page open and came back over an hour later). This rejection is handled transparently by legitimate clients, which contain an API abstraction service that triggers a page refresh upon receiving a \texttt{401}, thereby refreshing the server grant. Since it would be possible to obtain a grant for malicious purposes, the API is also rate-limited by session identifier and IP address.

\subsection{Performance and Caching}

In order to avoid rate-limits and slow response times in external APIs, a caching layer transparently caches every call. The query parameters and form data of the request are hashed with the domain and endpoint, and forming a key that is queried in a database before the request is made. If the cached value is not found, the request is made, and the raw result is stored in the database, with a default expiration time of one week.

Requests to internal APIs are similarly cached, by the fronted API abstraction service.

Performance at the application layer is not a concern, since all the major computation is done at the database layer. Our optimization efforts were spent on crafting efficient SQL queries, such as the one reproduced in Listing \ref{vcwiz:sql:query}. We instrumented our runtime to detect bottleneck queries, and denormalized data as necessary. This involved caching data that would have otherwise required a large table join, or an expensive aggregation. An example has been reproduced in Listing \ref{vcwiz:sql:view}.

\subsection{Routes}

The routing of VCWiz is split into the frontend application, resource paths that serve pre-compiled Javascript and CSS, and an internal API.

The endpoints in Figure \ref{vcwiz:routes:frontend} serve the pages for the frontend VCWiz application, which is a React app. The last section contains the pages that are auto-generated for search engines.

The endpoints in Figure \ref{vcwiz:routes:investors} serve the pages that investors interact with on VCWiz. The first group is a React app that allows investors to claim their profile on the platform, and make edits to the information that is displayed about them. The second group are the pages that investors land on when accepting or rejecting an introduction request from a founder.

The endpoints in Figure \ref{vcwiz:routes:api} comprise the internal API that largely serves to allow the frontend React apps to create, read, update, and destroy resources on the server.

\section{Launch}

\subsection{Marketing}

In the months leading up to the launch of VCWiz in January 2018, we ran a mass email campaign to every investor in the database, asking them to verify and amend pre-populated profiles. Investors had a strong incentive to verify their profiles: founders would be using the information to decide who to reach out to. Additionally, we awarded participating investors with a badge, viewable by all founders, indicating their profile was verified. Our initial email also requested the support of these investors in spreading the news about the new tool.

In the weeks leading up to the launch, we partnered with Product Hunt\footnote{https://www.producthunt.com/posts/vcwiz}, a popular website for launching technology products. They featured us in their weekly newsletter, and helped us reach a broad audience that includes many startup founders. Thanks to this partnership, the VCWiz homepage received $11,550$ views across $4876$ unique users within a week of launching.

A blog post detailing the full marketing efforts to launch VCWiz can be found online\footnote{https://medium.com/@dormroomfund/how-we-generated-1k-high-quality-leads-through-product-hunts-ship-ee8f1bebe6f6}.

\subsection{Metrics}

\begin{figure}[ht]
  \centering
  \begin{minipage}[t]{0.5\textwidth}
    \centering
    \includegraphics[width=\textwidth]{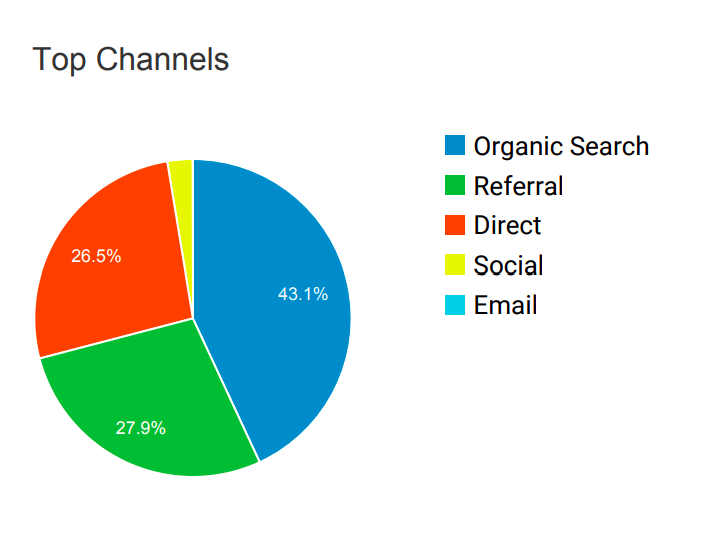}
    \caption{Popular user acquisition channels}
    \label{fig:acquisition}
  \end{minipage}\hfill
  \begin{minipage}[t]{0.5\textwidth}
    \centering
    \includegraphics[width=\textwidth]{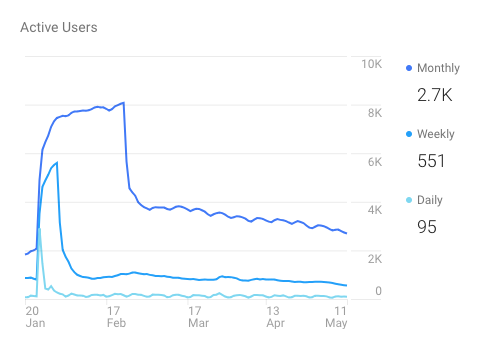}
    \caption{Active users over time}
    \label{fig:actives}
  \end{minipage}
\end{figure}

At the time of writing, there are $421,946$ VCWiz research pages indexed on Google. From these, there are roughly 7000 impressions per day, resulting in about $100$ clicks to the site. Similar stats are seen on other major search engines. Figure \ref{fig:acquisition} shows the top sources of new users. There are between $200$ and $300$ founders that actively use the site on a monthly basis, with around 1200 founders that have used the platform actively at least once since the launch. There are an additional $2700$ users who visit the site monthly, without signing up. Figure \ref{fig:actives} shows the trend of active users over time.

Registered founders visit $1000$ investor profiles monthly, for an average of four investors per founder per month. Of these, just over $50\%$ of them have granted access to their email inboxes for the purpose of tracking conversations with investors, and contributing anonymous, aggregate graph data for the VCWiz platform. These founders send and receive an average of 33 emails with investors per month.

To evaluate how deeply founders are engaging with the research component of the platform, we examine the pattern of session lengths. After filtering out sessions that end within $10$ seconds, we see that $20\%$ of the sessions since launch have lasted for at least $10$ minutes, with $5\%$ lasting over 30 minutes. The majority of this time is spent on the research pages generated for investors and firms. Interestingly, while founders tend to focus on the pages related to investors, they often come in through the pages focusing on a given company: $37\%$ of incoming search traffic is for a company page, with $30\%$ going to an investor page, and $20\%$ going to firm pages.

\subsection{Feedback}

Over the course of the months following the launch of VCWiz, we have done several surveys polling both founders and investors about the platform. The results of many of these surveys have already been discussed above. A few additional datapoints to highlight are that 38\% of founders surveyed spent some amount of time researching investors on VCWiz for the purpose of fundraising, and that 54\% of the founder who have not yet used the platform for research would do so if a single feature is added (the scope of these requested features ranges from simple tweaks to entirely new tools). This feedback, combined with the aforementioned metrics, lead us to the conclusion that though the platform is far from finished, the current iteration has indeed provided significant value to hundreds of founders.

One piece of feedback from a founder stands out in particular, and has been reproduced below.

\begin{quote}
VCWiz was very helpful in our fundraising journey. Through it we discovered several relevant investors that weren't on our radar, and we ended up building a robust target investor list that expedited our process.
\end{quote}

\section{Evaluation}

The goal of VCWiz is to be a tool that increases the efficiency and equitability of the founder-investor matching process, while generating data that facilitates further study. This chapter detailed the qualitative findings from the development of the tool, and founder reactions to the tool. The next chapter will cover a quantitative approach to the data collected.

\subsection{Equitability}

To evaluate equitability, we consider whether we made the fundraising process open and transparent. Our attempts have informed the research and discovery of over $18,000$ users across hundreds of cities, and have exposed thousands of founder reviews on the actions of investors. This information has been available to anyone, regardless of their background or experience in venture. Specifically, we have revealed information to these individuals that was previously only available to those ``in the know'': details such as the investment patterns of specific general partners, or which topics an investor often enjoys discussing. We believe, at the scale described here, this information dissemination has been impactful.

The true test of whether or not we increased equitability in the matching process would be to evaluate how VCWiz has impacted the fundraising ability of demographic groups who have previously struggled. Unfortunately, we did not collect sufficient data to determine this, nor has enough time elapsed to properly evaluate it. While there is anecdotal evidence that underrepresented founders feel more informed and comfortable after using VCWiz, there is no conclusive finding to report on.

\subsection{Efficiency}

To evaluate efficiency, we must consider both the time taken to complete the matching process, as well as the success rate of the matching process. Within the time to complete the process, we have two metrics: hours spent, and time elapsed.

Based on our research on the status quo for fundraising tools, it's clear that VCWiz would decrease the number of hours spent manually researching and discovering investors. This is corroborated by the fact that $47\%$ of surveyed founders indicated they were likely to recommend VCWiz to a friend\footnote{a score of 9 or higher on an 11-point poll}---an indication of the perceived value of the platform. With respect to time elapsed, we see that founders on VCWiz who successfully raised their seed round of financing took an average of $15$ weeks from the first investor conversation. This is significantly lower than the commonly-held average of four to five months (as seen in \cite{BRUNO198561}), though there is not sufficient information available to prove this is exclusively due to the use of VCWiz. A more detailed discussion can be found in Section \ref{ch5:timeline}.

Evaluating the success rate of founders on VCWiz is difficult as we lack a baseline to compare against. Just under $35\%$ of the founders who began raising seed rounds on the platform heard back from at least one investor, of which $66\%$ received investment from at least one investor. We believe this is in part due to founders' usage of the intro path functionality discussed in Section \ref{chap4:intropaths}, which aids founders in making connections to potential investors. $52\%$ of the founders on VCWiz used this functionality at least once.

While these numbers seem to imply that a large percentage of founders on VCWiz end up making a successful match (of those who find investors that are willing to engage), sufficient time has not yet passed to validate this. Venture funding volume waxes and wanes throughout the calendar year, and without a full year's worth of data, our results are as of yet inconclusive. Nonetheless, the success of our founders to find matches is encouraging, and motivates further work on the platform.

In a sense, the most ambitious tool for increasing the match rate of founders and investors was the intro request functionality discussed in Section \ref{chap4:introrequests}. If successful, this functionality would have increased efficiency (and to a certain degree, equitability through access) for both founders and investors. Unfortunately, this experiment failed miserably. There were effectively no successful matches made through this feature, with very little engagement shown from investors. Hopefully, future work can help incentivize investors to respond to promising founders.

\chapter{Graph Experiments}
\label{ch:ch5}

As part of our research into the founder-investor matching process, we sought to quantitatively demonstrate the importance of various commonly-accepted characteristics of fundraising. Furthermore, we wish to explore how we can leverage automated ranking systems to more appropriately match companies with a source of funding. This requires having a standard way to rank founders based on available information, including social graph data. To do this, we ran various experiments on a social graph of founders, investors, and their mutual connections. This graph is built from the information provided by founders on the VCWiz platform.

As discussed in Chapter \ref{ch:ch4}, one of the features of VCWiz is a CRM for founders that integrates with their inbox and scans (the headers of) all their emails with investors. As part of this optional integration, founders gave permission to have their aggregate email data used for research. As we scan these emails (filtering out any irrelevant ones, as described in Listing \ref{code:parse}), we build a graph, where each node represents an individual (using their email as a unique key), and each edge represents an email connection between two nodes. Edges are directed: there exists an edge from node $i$ to node $j$ if and only if an email has been sent from address $e_i$ to address $e_j$. Edges have weights equal to the total such number of emails sent. In order to comply with the privacy provisions made to founders, we avoided more sophisticated weighting schemes involving features extracted from the email body. The weights are ignored when doing manipulations and calculating graph metrics, unless explicitly specified.

Our motivation for modeling the underlying graph in this way follows from the fact that most fundraising-related communication happens over email. The majority of pre-pitch and post-pitch communication when fundraising happens over email, and almost every introduction made to an investor on behalf of a founder is done by email as well. Furthermore, emails are often sent as follow-ups to in-person meetings (at networking events, etc.). Finally, email is the preferred medium for ongoing communication between and founder and their investors. Thus, by capturing the entirety of the email graph for the subset of founders and investors on our platform, we get an accurate picture of the relationships at play.

\section{Experiments}

The remainder of this chapter will explore the findings of several experiments. The goal of these experiments is to better identify the social attributes that correlate with a successful fundraise, and how we can use this information to better match founders and investors. For the purposes of these experiments, we will define fundraising success as: raising at least as much money as planned, from a founder's top pick of investors, in as short a time as possible.

It has long been supposed that the characteristics of a founder in his or her professional network can impact, and indeed predict, how successful a fundraising attempt will be. There is lots of anecdotal evidence to support these claims\footnote{https://about.crunchbase.com/blog/fundraising-dos-and-donts/}, and recently there has been statistical evidence as well. A 2017 study uses AngelList data ``to estimate the effects of network distance in the matches resulting from Series A financing rounds'', and concludes that ``distance drives matching value and moderates preferences for experience and education''~\cite{pasquini2017matching}. We sought to verify and further elucidate this point with our graph data from VCWiz.

We would like to explore whether or not linear combinations of simple graph metrics can predict fundraising success. We first will define our metrics, and their intuitive meaning within the context of fundraising. The hypothesis is that commonly-accepted key metrics that correspond to a founder's ability to fundraise will be highly correlated with metrics indicating our definition of success. We will attempt to validate this hypothesis with our email graph, as well as analyze which factors are indeed the most important.

\section{Preprocessing}
\label{ch5:preprocessing}

A prerequisite to analyzing the graph is preprocessing the raw data from VCWiz. The first step was of course to build the graph. We followed the steps in Section \ref{vcwiz:ingesting} to import each founder's emails, adding nodes and edges to a global graph as described. Any messages skipped in the preprocessing phase are omitted from the graph. We added several additional rules for omission based on an analysis of intermediate graphs for outliers (for example, nodes that had significantly higher than average in-degrees or out-degrees).

Once the graph has been constructed, there is one last preprocessing step that must occur. Often, there are founders who sign up for the platform, but never interact with any of the email-related features. Their nodes are still added to the graph, but are orphans that have no neighbors. These nodes can slow down metric calculations unnecessarily, and make analysis harder, so we first filter them out using the Cypher query in Listing \ref{vcwiz:cypher:orphans}.

The final step is to label each node in the graph. Each node is labeled as a \texttt{Person}, with known investors and founders being labeled as \texttt{Investor} and \texttt{Founder} respectively. These two labels are mutually exclusive. In the case where a node could be labeled as both an \texttt{Investor} and a \texttt{Founder}, it is treated as an \texttt{Investor} if the person is currently employed by an institutional investment firm, and a \texttt{Founder} otherwise.

\section{Graph Analysis}

Before diving into our main experiments, we did some analysis of the graph. At the time of writing, the VCWiz email graph has $414,081$ individual nodes, including email relationships between $7679$ verified founders and $2134$ verified investors.

\subsection{Connectivity}

Looking at just the subgraph of founders who have signed up for VCWiz, we see that each node has a mean of $5.5$ neighbors, indicating that the founders on the platform often know other founders on the platform. This is consistent with the real-world behaviour of early-stage founders, who often communicate with a clique of other similar-stage founders and share resources such as tools. It is not, however, necessarily representative of the connectivity of real-world founders. It is possible that this connectivity is the result of founders spreading the word about the tool to their peers. An interesting observation is that though these founders are not as professionally isolated as those who would benefit most from using the tool, they are perfectly set up to use the network-based functionality of VCWiz, such as intro paths.

\subsection{Communities}

In order to determine how much of the connectivity present is the result of founders sharing the tool with peers, we performed a connectivity analysis. If it turns out that the graph is comprised of weakly connected cliques that are strongly connected internally, it would support our hypothesis. Using Label Propagation (LPA), we can section the graph into partitions by flooding the nodes with labels: assigning each node an initial label and propagating these labels with a set of rules until distinct communities evolve. We refer the reader to \cite{2007PhRvE..76c6106R} for a comprehensive description of the algorithm.

Upon partitioning the founder graph, we find that there are $584$ communities, with an average of $1.2$ founders per community. Our model of the founder community on the platform was not accurate; it is not the case that there are isolated pockets of founders who are spreading news of the product to each other. Indeed, it seems that the majority of the founders on the platform are all part of a larger, loosely-defined community that cannot easily be partitioned.

An interesting finding is that the few most-populous communities are easily recognizable, after which there is a long tail of independent communities with only one or two founders. The three top communities found by LPA are Dorm Room Fund Partners, Dorm Room Fund Portfolio Companies, and YCombinator Portfolio Companies. Given that both of these organizations helped influence this work, it is not surprising to find these communities.

We also ran an alternative connectivity analysis, using the Louvain Method~\cite{2008JSMTE..10..008B}. This revealed another significant community of founders: Student Founders at UC Berkeley. However, the remainder of the communities still appear to be insignificant. We note that this is not necessarily representative of early-stage founders as a whole, and instead reflects the lack of communities in the subset of founders who use VCWiz.

\subsection{Propensity to Investors}

We sought to answer the question of whether or not the community and neighborhood of a founder's node can predict their propensity to engage with certain investors. However, using graph structure alone, we lack sufficient signal to predict anything. We revisit this question in describing future work, taking into account additional node metrics and founder characteristics.

\subsection{Patterns}

The profile data collected by VCWiz, when joined with the founder's email history, offers an opportunity to validate commonly-held assumptions in venture. We test a few of these hypotheses below.

\subsubsection{Email Volume}

The classic pattern of communication when fundraising is as follows. At the start, the founder has a very high frequency of outgoing emails, as they reach out to and/or get introductions to investors. At this stage, the founder is fighting to stay relevant, and will be following up often. Once the investors show interest and begin engaging the founder, the majority of communication happens over phone calls and in-person pitches, resulting in a decreased email volume. Finally, as the round begins to close and investors begin to commit, we expect email volume to increase again, reaching a peak as the founder pushes for final decisions and coordinates the transfer of capital.

We can evaluate the accuracy of this pattern by plotting weekly email volume over the percentage of ``committed investors'' (investors who have decided to make an investment). For email volume, we specifically use the percentage of total emails sent during the fundraise that were sent in a given week. Fitting a third-degree polynomial to this data should show a relatively high volume to start, a sharp decline as a few investors commit, and a gradual increase as the fraction of decided investors approaches $1$.

As shown in Figure \ref{fig:patterns:email}, we see a similar pattern, but not exactly what was expected. The volume of emails sent in the early stages of the round are not as high as we believed them to be, indicating that there are minimal emails back-and-forth while the first investors are diligencing and deliberating a company. Additionally, the email volume peaks when around $70\%$ of the investors are committed. Our best explanation for this is that the majority of the coordination around closing the round happens with the lead investor(s), including clarification and negotiation of the fine points. Following this, the remaining investors largely fall in line without much delay. We believe founders are engaging in these discussion with the lead(s) before all the investors have committed.

\subsubsection{Fundraising Timeframe}
\label{ch5:timeline}

Fundraising is a notoriously time-consuming activity. The oft-quoted timeline for raising a seed round is $90$ to $150$ days of engaging investors in direct solicitation for investment, with several months prior of relationship-building. One study that surveyed founders in Northern California showed a median fundraising period of $4$ to $5$ months~\cite{BRUNO198561}, with $20\%$ of founders reporting the process taking over $8$ months. It's difficult to find quantitative reports of fundraising timeframes that don't involve surveys, so we analyzed the average length of a seed fundraising attempt on VCWiz.

\begin{figure}[H]
  \centering
  \begin{minipage}[t]{0.45\textwidth}
    \centering
    \includegraphics[width=\textwidth]{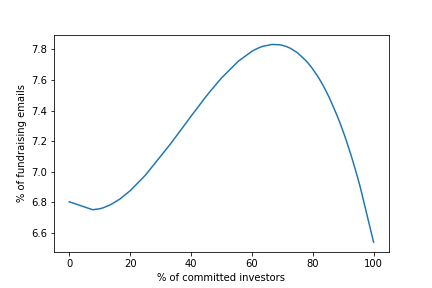}
    \caption{Weekly email volume versus percent of eventual investors committed}
    \label{fig:patterns:email}
  \end{minipage}\hfill
  \begin{minipage}[t]{0.45\textwidth}
    \centering
    \includegraphics[width=\textwidth]{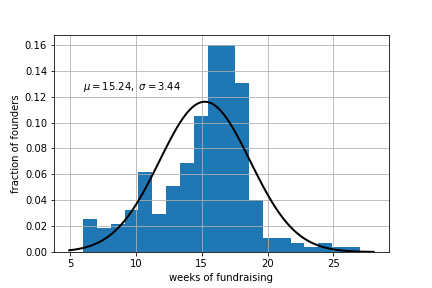}
    \caption{Histogram of weeks spent fundraising}
    \label{fig:patterns:fundraising}
  \end{minipage}
\end{figure}

We measure the period of fundraising by defining the start as the first time an investor is added to a founder's wishlist, and the end as the last time a wishlist investor's status is updated. Based on this definition, founders on VCWiz spend an average of 107 days actively fundraising. Figure \ref{fig:patterns:fundraising} shows the distribution of weeks spent fundraising across all founders on the platform. This timeline is significantly shorter the expected fundraising period identified above, which might indicate that founders who use VCWiz are more qualified than the average founder, or that the tool helps shorten the time elapsed while fundraising.

It is important to note that this method of calculating the fundraising period is not perfect. As shown in Figure \ref{fig:patterns:email}, much of the communication near the end of a round happens over non-email channels. Therefore, VCWiz might have a delayed or inaccurate view of the status of the round in the tail end of fundraising. This can skew the calculation of the period send date.

\section{Baseline Ranking}

We now shift to an attempt to rank founders based on their propensity for fundraising. In order to evaluate any future scoring functions, we need a baseline to compare against. In the absence of a quantitative baseline scoring function, we hand-crafted a baseline ranking, incorporating our knowledge of venture capital, and the results of the research cited thus far. The goal of this baseline is to rank founders based on how likely they are to raise the largest round in the shortest period of time. We will briefly document the process of exploring features for this baseline before defining the actual scoring function.

\subsection{Potential Features}

Below are the features that were considered for the function, and our evaluation of each one for inclusion.

\subsubsection{Total funding of current round}

When available, the total amount of money raised in the current round is an excellent indicator of fundraising success, as it is normally the goal of the founder to raise as much money as possible, up to some internal maximum. This metric should definitely be included in the scoring function. Unfortunately, for many founders on the platform who are currently attempting to raise their seed rounds of funding, there might not yet be any money raised.

\subsubsection{Total funding of previous rounds}

As referenced earlier in Section \ref{chap3:tool}, the First Round Capital 10 Year Project \cite{first-round-10-years} indicated much higher fundraising success rates for repeat founders. In this case, having raised money either in a previous round for the same company, or for a previous company, would give a founder the credibility and experience necessary to improve their chances in their current fundraise.

We experimented including both this number, when available, as well as a binary feature indicating that this number is nonzero. Ultimately, we decided to use the numeric feature, as variations in this metric are significant. Instead of using the raw number, we scale the amount raised for a given startup by the average round size of the industry the company operates in. This accounts for different capital requirements across industries: having raised \$5M for a social media company conveys much more credibility than the same amount for a biotechnology company.

\subsubsection{Number of intro requests accepted}

While it would be useful to include a metric capturing a founder's success rate when requesting cold introductions on the VCWiz platform, there is simply not enough data for the metric we report to be indicative of anything. This follows from our earlier analysis (Section \ref{chap4:introrequests}) on why this feature saw very little utilization.

\subsubsection{Number of interested investors}

One metric that is indicative of global investor interest is the number of investors who have exchanged multiple emails with the founder during the timeframe of the raise. While these investors may or may not end up investing, the fact that they were interested enough to email several times is a strong signal that the founder will have options to select from when it comes time to close the round.

\subsubsection{Fraction of investors who respond}

A similar metric to the last is the percentage of investors who have emailed a founder back after the founder has initiated contact with them (either directly or through an introduction). A high value here indicates that the founder and the startup are compelling enough to garner investor interest. It's also indicative of a founder's ability to reach out to investors who are a good fit for the startup.

\subsubsection{Average response time per investor}

The average time an investor takes to respond to a founder's email might be indicative of the investor's excitement for the founder, if all such communication was captured by the platform. However, many founders are manually including VCWiz on specific emails, rather than allowing access to their inbox. This can prohibit accurate calculation of response times, as investors often do not include VCWiz in their responses. This can create outliers that heavily skew the average and add a burdensome level of noise. Furthermore, many urgent conversations occur by phone or in-person, which is also not factored into the calculation. Thus, this feature will not be included.

\subsubsection{Length of fundraising period}

A strong fundraise (based on our earlier criteria) is one that results in sufficient dollars being raised, in the shortest amount of time possible. Thus, the amount of time spent fundraising should be a high-signal feature: a short fundraise alone is not indicative of success, but assuming success, shorter times are better. However, our method of calculating the fundraising period is error-prone: if a founder stops using the platform, or is unsuccessful in closing their round, we will not account for this. For this reason, we omit this feature.

\subsubsection{Average sentiment of investors}

Intuitively, a high positive sentiment from an investor in an email indicates an affinity, and a strong relationship. Looking at the data on VCWiz, strong sentiment is often associated with a pre-existing relationship, often from a previous company. However, it is a very noisy source of data, so it should not be relied upon exclusively.

\begin{figure}[H]
  \centering
  \begin{minipage}[t]{0.48\textwidth}
    \centering
    \includegraphics[width=\textwidth]{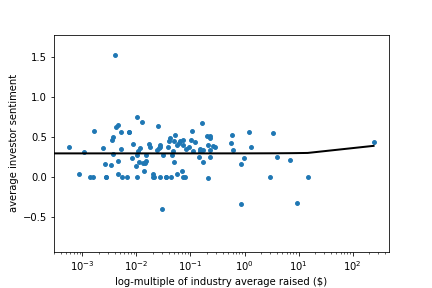}
    \caption{Incoming investor sentiment versus log-multiple of industry average round size raised}
    \label{fig:sentiment:raised}
  \end{minipage}\hfill
  \begin{minipage}[t]{0.48\textwidth}
    \centering
    \includegraphics[width=\textwidth]{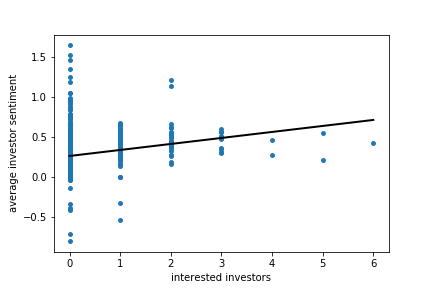}
    \caption{Incoming investor sentiment versus number of interested featured investors}
    \label{fig:sentiment:interested}
  \end{minipage}
\end{figure}

Using this metric adds a new dimension to our ranking function that is not highly correlated with any other. Figures \ref{fig:sentiment:raised} shows the average sentiment of an incoming email during fundraising versus the total money eventually raised (adjusted for the average round size of the startup's industry), across all founders on the platform. Figure \ref{fig:sentiment:interested} shows the average sentiment versus the number of featured investors who expressed interest in a startup during its fundraise. In both cases, we see that sentiment doesn't follow a strong pattern.

\subsection{Evaluation}

After evaluating each of the above metrics, we drew the following conclusions.

The best metric, when available, is the aggregate funding the founder has raised to date, across any company. This is the sum of the two funding-related features.

Next, the number of interested investors is a good proxy for success during the fundraise, as it represents absolute interest.

The email outreach success rate (fraction of investors who respond) tends to be too noisy, as it is skewed by high-profile founders who send a very small number of emails, to already-established connections. However, the number of investors who respond after being added to a wishlist on VCWiz is a very strong signal, as it indicates success with no prior relationship.

Finally, we include the average investor sentiment (as indicated by incoming emails), as it provides a dimension not captured by the other metrics.

\subsection{Ranking Function}

To create a ranking function from these features, we calculate a weighted sum of sort indexes for each founder (one per feature), then sort the list based on this aggregate index. The position in this sorted list of a founder is that his or her rank. We use ranks, not absolute values, across features to account for differences in scale.

We demonstrate the scoring function in Equation \ref{eq:baseline}, where $S_{m, f}$ gives the sort index for founder $f$ when the founders are sorted by metric $m$, and $W$ is the set of (metric, weight) pairs. The total ordering of the baseline is the set of founders, sorted by this score, in ascending order (a lower score represents a higher-ranked founder).

\begin{equation}
\label{eq:baseline}
  score(f) := \sum_{(m, w) \in W} S_{m, f} * w
\end{equation}

\noindent The weights we use for the baseline are found in Figure \ref{fig:nfr:baseline:weights}.

\begin{figure}[ht]
\begin{tabular}{c | c}
\textbf{Feature}           & \textbf{Weight} \\\hline
Aggregate Funding          & 4 \\\hline
\# Interested Investors    & 3 \\\hline
\# Responded From Waitlist & 2 \\\hline
Average Investor Sentiment & 1
\end{tabular}
\centering
\caption{Baseline Metric Weights}
\label{fig:nfr:baseline:weights}
\end{figure}

This baseline, while not defensible as a ground-truth ranking for rigorous statistics, is built on sane assumptions about fundraising and venture capital. Random sampling indicates that the results are aligned with expert expectations. We will use this baseline to compare and evaluate our numerical methods for ranking, but acknowledge that it is not a perfect ranking according to the criteria we defined.

\section{Evaluation Criteria}

The next step is to define how to evaluate other ranking functions against our baseline. There are many options for evaluating ranking functions. We will discuss the options before justifying our selection. For mathematical definitions of each evaluation metric independent of our use case, we refer the reader to Section 3.2 of \cite{DBLP:journals/corr/abs-0704-3359}. These ranking functions are often used to score a permutation $\pi$ of documents given a document-set, query tuple $(D, q)$. In this case, we assume $q$ is the fixed query of founders that are most likely to succeed at fundraising, and $D$ is our list of founders.

In selecting the evaluation criteria, we imagine two relevant use-cases for our ranking functions. The first is to display a sorted list of founders to an interested party, such as an investor. The second is to use this ranking as a feature in a later process, such as enhanced matching between founders and investors (which we will explore in the next chapter).

Winner Takes All (WTA) and Mean Reciprocal Rank (MRR) are often used when displaying results based on a ranking, but both assume there is only \textit{one} top-ranked document. This does not align with either of our uses cases, so we discard them.

Discounted Cumulative Gain (DCG) takes the sum of \textit{relevances} of each founder in the ranked list (where the relevance in our case is $N - r$, with $N$ being the total number of founders, and $r$ being the rank of that founder in the baseline), weighting each founder by how early it appears in the list. We therefore get a score that increases as we put the highest-ranked founders near the start of the list, but penalizes incorrect ordering of later founders less and less. In other words, this is a metric of relative regret. This aligns well with our first use case, and provides some useful, though imperfect, information for our second.

Normalized Discounted Cumulative Gain (NDCG) simply normalizes the DCG against the number of founders in the list, so the metric can be compared across lists of different size.

Precision@n is another metric that gives preference to the top results, in this case explicitly only considering the top $n$. This metric is simply the fraction of founders in the top $n$ results that are also in the top $n$ founders of the baseline. It is a quick, useful way of evaluating a ranking for our first use case.

Root-Mean-Square Error (RMSE) is a very common error function for recommender systems~\cite{Cremonesi:2010:PRA:1864708.1864721} that measures the differences (errors) in ranking for every founder in the list. It works very well for our second use case.

Mean Absolute Error (MAE) is another common error function, which measures the average absolute distance between a ranking and the ranking of the baseline.

Kendall's rank correlation coefficient ($\tau$) measures the ordinal association between the two lists of founders, and is often used to report on the correlation between to rankings. It roughly measures the agreement in rank over all pairs of items.

Spearman's rank correlation coefficient ($\rho$) is another measure of rank correlation. It measures the direction of association in rank between the two lists of founders.

We decided to use NDCG, Precision@n (for $n \in \{5, 10, 20\}$) for evaluating the quality of the rankings for the purposes of recommendation, $\tau$ and $\rho$ for calculating the rank correlation with the baseline, and RMSE and MAE for scoring the ranking as a feature to later processes. Figure \ref{fig:evaluation:formulas} shows the formulae used for each, where $N$ is the number of founders in the list, $B$ is the baseline, $X$ is the ranking being evaluated, $R_i$ is the founder in the $i$-th position of ranking $R$, $\text{rg}(f)$ gives the rank of founder $f$ in the baseline, and $\text{sc}(R, f)$ gives the \textit{score} of founder $f$ under ranking $R$. Each ranking that we will be evaluating has a scoring function over all founders in the range $[0, 1]$. For the baseline, we assign scores by simply scaling the rank to be in this range.

\begin{figure}[H]
\begin{equation}
  \begin{array}{c >{$\displaystyle}Sc<{$}}
    \text{\textbf{Metric}} & \text{\textbf{Formula}} \\
    \text{DCG}(X) & \sum_{i=1}^{N} \frac{2^{N - \text{rg}(X_i)} - 1}{\log_2{i + 1}} \\
    \text{NDCG}(X) & \frac{\text{DCG}(X)}{\text{DCG}(B)} \\
    \text{Precision}(X, n) & \frac{|X_{0:n} \bigcap B_{0:n}|}{n} \\
    \tau(X) & \frac{\sum_{i=1,j=1,j \neq i}^N \text{conc}(i, j, X)}{N (N - 1)} \\
    \rho(X) & 1 - \frac{6 \cdot \sum_{i=1}^N \text{drg}^2(X, B, i, i)}{N (N^2 - 1)} \\
    \text{RMSE}(X) & \sqrt{\frac{\sum_{i=1}^N \big(\text{dsc}(X, B, X_i)\big)^2}{N}} \\
    \text{MAE}(X) & \frac{\sum_{i=1}^N \big|\text{dsc}(X, B, X_i)\big|}{N}
  \end{array}
\end{equation}

\begin{align*}
 \text{drg}(X, Y, i, j) &:= \text{rg}(X_i) - \text{rg}(Y_j) \\
 \text{dsc}(X, Y, f) &:= \text{sc}(X, f) - \text{sc}(Y, f) \\
 \text{conc}(X, i, j) &:= \text{sign}\big(\text{drg}(X, X, i, j)\big) \cdot \text{sign}\big(\text{drg}(B, B, i, j)\big)
\end{align*}

\centering
\caption{Evaluation Criteria Formulae}
\label{fig:evaluation:formulas}
\end{figure}

Finally, we must define our null hypothesis, so that we can provide p-values for our rank correlations. In this case, the null hypothesis is that the correlation between the baseline and the ranking in question is 0. Thus, p gives the probability that the two uncorrelated rankings would give them same rank correlation metric value.

\section{FounderRank}

We will be ranking founders based on information collected from the VCWiz platform. We wish to score nodes based on graph metrics in an email graph of fundraising relationships. To do this, we first need to define a single scoring function that captures our goals. To this end, we introduce FounderRank.

FounderRank is a metric in the range $[0, 1]$ that quickly communicates the strength of a founder in the global fundraising graph of founders, investors, and their mutual connections (i.e. the VCWiz Email Graph). A strong node is able to rapidly spread the word about their startup, start conversations with relevant and desirable investors, and convince investors to invest in them. These abilities are crucial for fundraising, which is in turn crucial for the survival of a startup.

Note that we are currently evaluating how well a founder can fundraise conditioned on them knowing who they would like to fundraise from. We are not tackling the issue of discovering investors, which we have touched on in the previous chapter, and will explore further in the next.

\subsection{Core Characteristics}

A combination of existing studies and our own interviews with numerous seed-stage firms reveals three intuitive properties of a founder's node in a professional or social graph that are desirable when fundraising: \textbf{importance}, \textbf{influence}, and \textbf{access}. Note that these characteristics do not take into account factors such as domain expertise, personality, and pedigree, all of which will also contribute to a successful fundraise. We are focusing exclusively on network properties for this experiment.

The first characteristic is \textbf{importance}. Importance looks at how crucial a founder is in his or her own ecosystem. This property is important as it is indicative of the founder's degree of expertise. It has been shown that in efficient professional information networks, entrepreneurs are bounced from expert to expert until they have sufficient information to answer their query~\cite{BIRLEY1985107}. The more crucial a founder is to a domain, the more access to high-quality information he or she will have. Thus, founders who have more importance are likely to be seen as a less risky investment, increasing the chances an investor responds positively to a fundraising proposal.

The second characteristic is \textbf{influence}. Influence captures how effectively a founder can effect change in their ecosystem and solicit aid from their peers. In other words, how likely other founders are to help this founder. A study on interorganizational networks of young companies found supporting evidence to the fact that ``third parties rely on the prominence of the affiliates of those companies to make judgments about their quality''~\cite{10.2307/2666998}. Founders who have high influence can leverage this to convince investors to give them funds.

The third characteristic is \textbf{access}. This is the notion of how well a founder can get in front of the investors of their choice. The more directly connected a founder is to an arbitrary investor, the more likely that investor is to respond to an inbound request (either via an introduction or cold) for funding. Additionally, a high degree of access means a founder has many options to chose from when it comes to starting conversations with investors. This follows from the more general finding that proximity in a graph to providers of valuable resources gives a node access to many viable alternatives~\cite{10.2307/3069443}. This can be valuable when a founder's top choice of investor does not work out, which is often the case.

\subsection{Graph Metrics}

To quantitatively evaluate the impact of each of these characteristics, and measure their predictive capability, we need to find corresponding graph metrics.

For \textbf{importance}, we selected PageRank~\cite{page1999pagerank}, the canonical starting point for ranking nodes in graphs. PageRank recursively evaluates node importance by analyzing the importance of nodes that link to the node in question. It is a widely-accepted measure of node importance. We use a normalized PageRank~\cite{berberich2007comparing}, which accounts for the number of nodes and structure of the graph.

For \textbf{influence}, we selected Betweenness Centrality~\cite{10.2307/3033543}. Betweenness Centrality is the count of the number of shortest paths (over all pairs of nodes) that pass through the node in question. The rationale is that if a node is on the intro path for a pair of people, that node has influence over that pair, as it can control whether or not the introduction is made. This makes the assumption that every communication request goes along the shortest path, which is not perfectly accurate, but is sufficient for our purposes.

For \textbf{access}, we selected Closeness Centrality~\cite{FREEMAN1978215}. The Closeness Centrality of a node is inversely proportional to the node's distance from every other node. Thus, this metric measures how ``close'' a node is to the other nodes in the graph, based on shortest-path lengths, normalized for the number of nodes in the graph. The rationale is that if a node can access every other node in the graph, on average, via a short path, the node must have better access than a node that must use longer paths. Once again, this assumes that the length of the shortest path between nodes is the determiner of connection strength. Based on the data presented and cited thus fair, this is a fair assumption.

To use these metrics as features in our experiment, we take the pre-processed graph from Section \ref{ch5:preprocessing}, and calculate the raw metric value for each node. We then normalize as specified, and finally scale each metric to be in the range $[0, 1]$. We now have a number that represents each metric for a node, normalized relative to the other nodes in the graph, and on a standard scale. Note that we do not yet specify how to combine these metrics, we are simply calculating them.

\subsection{Random Model}

In order to have a model to compare against, we first start with a random model, which simply generates a score randomly and uniformly in $[0, 1]$ for each founder, then ranks them based on this score. The metrics for our random model are found in Figure \ref{fig:rand:results}.

While the random model would never be used for a serious application, it is worth noting that this way of sorting founders is not too far from the techniques used by many analysts in the real world today. The status quo at many firms involves haphazardly picking new companies to investigate based on ``gut feelings'': heuristics based on pattern-matching previous successes without any grounding in data. In one extreme case, an individual we interviewed claimed he triages companies by first sorting them alphabetically.

\begin{figure}[ht]
\begin{tabular}{c | c | c | c | c | c | c | c}
NDCG & P@5 & P@10 & P@20 & $\tau$ & $\rho$ & RMSE & MAE  \\
0.212 & 0.000 & 0.000 & 0.000 & -0.0550 & -0.0826 & 0.272 & 0.251 \\
\end{tabular}
\centering
\caption{Random Model Results}
\label{fig:rand:results}
\end{figure}

$\tau$ and $\rho$ have p-values of $0.0395$ and $0.0389$ respectively. As expected, there is essentially no correlation between the random model and the baseline.

\subsection{Naive FounderRank}
\label{ch5:nfr}

We now begin to explore ways to combine these three selected graph metric to produce a scoring function that we can compare against our baseline. The Naive FounderRank (NFR) method simply averages these three numbers, per node, to arrive at a score in $[0, 1]$.

The first observation we made when trying to rank the founders based on NFR is that our rankings totally missed founders on the platform who are not currently fundraising, but who are known to be repeat founders who have successfully fundraised in the past. This was a result of our email graph only spanning the last year. Relationships in the venture capital world take years to build up, and we were not evaluating them over a large enough timeframe. Thus, we re-built our graph to incorporate the last five years of data, which gave us a much more comprehensive view of the ecosystem.

The below observations and conclusions are all drawn from this new graph, with up to five years of email data from $630$ founders.

We sorted the founders in our original list by their NFR score, and then evaluated this ranking based on our earlier-established criteria. The results are shown in Figure \ref{fig:nfr:results}.

\begin{figure}[ht]
\begin{tabular}{c | c | c | c | c | c | c | c}
NDCG & P@5 & P@10 & P@20 & $\tau$ & $\rho$ & RMSE & MAE  \\
0.525 & 0.2 & 0.1 & 0.15 & 0.413 & 0.581 & 0.197 & 0.158 \\
\end{tabular}
\centering
\caption{NFR Results}
\label{fig:nfr:results}
\end{figure}

The p-values for $\tau$ and $\rho$ are respectively $10^{-54}$ and $10^{-57}$.

\subsubsection{Discussion}

As expected, the naive model easily and confidently beats the random model on every metric. While this is intuitive (there must be \textit{some} information in this graph), it is important to point out that even such a simple, naive model can best a standard seen often in venture today. If the goal is to use this ranking to surface interesting founders, NFR is an acceptable solution, albeit not great. The precision is still too low to be considered truly valuable. As a feature, the rank correlation observed in NFR is impressive. The degree to which naively combining graph metrics gives a score correlated to the baseline indicates how important the founder's position in their social-professional graph is.

We can interpret $\tau$ as saying that there is reasonable agreement between the rankings presented by NFR and the baseline, and $\rho$ as saying the direction of association in the rankings is positive, and significantly so: more often than not, the ranking of NFR tends to increase when the baseline does the same.

A final observation is that, as expected, all of the relative regret metrics improved after rebuilding the graph over a longer time period. It is easier to identify strong founders when relationships over a long period of time can be taken into account. Founders in the middle of the ranking largely stayed in the same position, as no new information was revealed about them.

\subsubsection{Edge Weights}

One interesting experiment we considered was using the frequency of emails exchanged between two nodes as an edge weight. Running the same metric algorithms on this new weighted graph should render even more accurate data about relationships, the position of a given node, and therefore the strength of the founder in the graph. Unfortunately, there is a significant amount of work involved in re-writing the metric algorithms to incorporate edge weights, and we did not have the time to explore this. We leave it as future high-potential work.

\subsection{Weighted FounderRank}

The next model we experimented with is a linear combination of the three graph metrics, as determined by a simple linear regression, using the hand-crafted baseline ranking as labels. We fit a standard linear regression model to the baseline ranking, using each of the three graph metrics as a feature. We get an $R^2$ value of $0.355$, with evaluation metrics as shown in Figure \ref{fig:wfr:results}. The change in metrics is largely positive: NDCG decreases by $16\%$, but every other metric is better. $\tau$, $\rho$, and precision all improve significantly, as do the error functions. We see that this weighted ranking function would be superior to the naive at predicting fundraising success, though it would fall short if used to rank and present the entire list of founders on their fundraising merit.

\begin{figure}[ht]
\begin{tabular}{c | c | c | c | c | c | c | c}
NDCG & P@5 & P@10 & P@20 & $\tau$ & $\rho$ & RMSE & MAE  \\
0.441 & 0.2 & 0.2 & 0.2 & 0.481 & 0.662 & 0.0995 & 0.0729 \\
\end{tabular}
\centering
\caption{WFR Results}
\label{fig:wfr:results}
\end{figure}

More interestingly, we can examine the effect on our evaluation metrics and $R^2$ by using different combinations of the three graph metrics, with the goal of determining their relative importance.

\subsubsection{Sole Metrics}

If we rank our founders based solely on their PageRank, we get uniformly poorer evaluation metrics than WFR (albeit not drastically), with the exception of NDCG, which jumps up to $0.525$. This implies that if our sole concern was ranking the entirety founders of the batch, sorting by PageRank would be sufficient. In other words, PageRank is a good, not great, filter for sorting a list of founders by their expected fundraising performance. This corroborates our conclusions from extending the timeframe over which the graph is built: the best fundraisers often are well-embedded and have built up strong relationships over years, giving them a high PageRank.

If we rank solely based on a founder's betweenness, every metric is strictly worse. The $R^2$ value of the regression actually goes negative, indicating that this metric alone is a very poor model of a founder's potential for fundraising success. Furthermore, a founder's betweenness is highly correlated with their PageRank (a Pearson correlation coefficient of $0.917$), making it unnecessary to even consider.

Finally, if we rank solely on closeness, we end up with a much lower NDCG (at $0.347$), and a marginally increased P@20 ($0.25$), $\rho$ ($0.669$) and $\tau$ ($0.485$). The RMSE increases to $0.133$. The conclusion here is that how much access a founder has in the graph can offer a better ranking in isolation, though at the cost of less accurate individual scores. For ranking, it appears that closeness explains more of the optimal ranking than any other metric, and this is supported by an $R^2$ value which is relatively close to that of WFR: $0.307$. While it is far from a perfect model, a node's closeness centrality in the email graph explains about 30\% of the variance of a hand-tuned ranking of the founders in the graph.

Our conclusion from testing these individual metrics is that access is most important aspect of a founder's node in the graph, but a sufficiently high importance (PageRank) can compensate for a lower closeness centrality.

\subsubsection{Optimal Combination}

Based on the above, the optimal combination of features is some linear combination of PageRank and Closeness Centrality. We ran another linear regression with these two features, resulting in a model that has similar evaluation metrics to WFR, save for the highest NDCG seen yet ($0.528$). Adding PageRank to Closeness Centrality in our model brings $R^2$ to $0.354$, a $15.3$\% increase.

\subsubsection{Conclusion}

The goal of these experiments on the VCWiz Email Graph was to explore whether the data stored in the graph structure and relationships could improve the process of predicting a founder's ability to fundraise, for use in the founder-investor matching process by both humans and other models. We have shown that there is significant signal in our selected graph metrics of founder nodes.

Through these experiments, we have determined that graph metrics alone can do a good job scoring founders. The rankings are not perfect, and can only explain about 35\% of the variance seen in the baseline they are compared against, but they significantly best the status-quo of an effectively random model, and have considerable rank correlation with the baseline ($\rho = 0.669$, p $\approx 0$). Upon examining the individual graph metrics, we see that Closeness Centrality, which corresponds to a founders ``access'' in the graph, is the driving factor, with exceptions in the case of very successful founders who have extraordinary PageRank. We also find that Betweenness Centrality, which indicates a founder's influence over introductions, is a near-meaningless metric, with information that is captures almost entirely by PageRank.

The takeaway of access being a driving factor to founder fundraising success further motivates our work on the VCWiz platform. We've demonstrated that founders who have a short average distance to investors tend to see success fundraising. A stated goal of the platform is to increase founders' access to investors, a demonstrably high-impact outcome. We accomplish this by helping founders discover and understand investors who are within reach of their network, and facilitate connections that add edges in the relationship graph. VCWiz enables these connections for all founders, enshrining equality in what is otherwise an insider's game.

\section{FounderRank with Investment Data (FR+I)}

The next experiment explores whether the conclusions of the FounderRank experiment hold true on the global graph of venture investments. If so, we wish to explore whether adding global investment data to the VCWiz Email Graph can augment it and provide even better rankings.

We begin with considering the public graph of venture fundings, as reported by VCWiz. At the time of writing, this database contains $305,033$ founders, $94,190$ investors, and $298,862$ investments. For an overview of how this data is collected, see Section \ref{ch4:data} (\pageref{ch4:data}). The graph is constructed by creating a node for every founder and investor on the platform, each tagged as such. For each instance of an individual investor investing in a founder's company (see Section \ref{ch4:partners} for an overview of how this information is inferred), two edges are added: one from the founder to the investor, and one from the investor to the founder.

\subsection{Baseline}

As before, there is no ground-truth ranking of the founders in this graph, so we need to generate a baseline. We will focus on public metrics involving the past experience of the founders, as well as the aggregate money they have raised. We rely on the work done in \cite{2017arXiv170604229H} as justification for selecting these metrics.

\begin{quote}
We see that the top non-sector features are related to the past experience of the leadership (executive acquisition, executive IPO, advisory IPO, leadership age). The investor feature maximum acquisition fraction is also one of the top non-sector features. This suggests that companies with experienced and successful leadership and investors have increased drift which results in a higher exit probability.
\end{quote}

The features we selected are Job IPO, Job Acquired, Executive IPO, Executive Acquired, Advisory IPO, and Advisory Acquired. These features track initial public offerings and acquisitions across any company the founder has worked at, started, or advised. The specific feature we use is the sum of all these numbers. We call this feature Affiliated Exits.

Using the same process as before (a combination of manual inspection, sane priors about venture capital, and consulting experts), we arrived at the baseline weights shown in Figure \ref{fig:fri:baseline:weights}. We use the same scoring function construction as in Section \ref{ch5:nfr}. We will also use the same evaluation metrics.

\begin{figure}[ht]
\begin{tabular}{c | c}
\textbf{Feature}   & \textbf{Weight} \\\hline
Affiliated Exits   & 4 \\\hline
Aggregate Funding  & 1
\end{tabular}
\centering
\caption{FR+I Baseline Weights}
\label{fig:fri:baseline:weights}
\end{figure}

The rationale for these weights follows from the metrics from \cite{2017arXiv170604229H} that have been shown to correlate with a founder's success rate, including ability to successfully fundraise. We additionally add the Aggregate Funding feature from the last experiment, to acknowledge that founders who have raised a large amount of capital in the past are more likely to be trusted as stewards of capital in the future.

One interesting observation about this baseline is that rewarding a large number of exists gives us a ranking that surfaces many founders who are now investors. This aligns well with career paths seen in the venture world: founders who take a company from birth to exit have experience that is highly sought-after in investors.

\subsection{Random Model}

Much as before, a random model shows the expected characteristics with all our evaluation criteria. There is no rank correlation with the baseline.

\subsection{Optimal FR+I Model}

The best linear combination of graph metrics for the global funding graph tells a similar story to that of the VCWiz Email Graph, albeit with diminished efficacy.

In this case betweenness, which we previously identified as a near-useless metric, has a large negative coefficient when included, with a model that is effectively the same as omitting it altogether. The remaining combination of PageRank and closeness gives us a largely poor result: an $R^2$ of $0.268$, with precisions of $0$. However, the rank correlation still somewhat exists. We see a $\rho$ of $0.494$ and a $\tau$ of $0.343$ (with a p-value of effectively $0$). Moreover, we see that closeness explains $94\%$ of the variance of the optimal weighted model, and alone gives rank correlations that almost fully capture the correlation observed with the optimal combination of metrics ($\rho = 0.482$).

We see that there is no way to reasonably surface the best fundraisers in this graph with just these metrics, though there is sufficient information to give a ranking function that would beat a random model. It is clear that the trends identified in the social-professional email graph hold for the broader venture ecosystem, but that the former has much more information than the latter.

\subsection{Additional Venture Relationships}

We further experimented with the graph of funding data, augmenting it with data about both co-founding relationships (founder-founder) and co-investing relationships (investor-investor). The co-investing relationships (an edge between a pair of investors if and only if those two investors have participated in the same funding round of a company) did not significantly improve our models, but the co-founding relationships did.

In this new graph of investment and co-founding relationships, the best model is the one that solely uses the closeness metric. This is consistent with all our previous conclusions. With this model, we get a $\rho$ of $0.496$. While this is not as strong a correlation as the email graph was, it still shows a similar result. Once again, we expect there to be less information in a graph of public investment and cofounding data than there is in a social-professional graph that reveals the many conversations which do not end in an investment or new company.

\subsection{FR+I with Email Data}

We have shown so far that the data in the global funding graph alone is not sufficient to show significantly meaningful correlation. However, we know that the private email graphs of founders \textit{do} have predictive power. To explore whether or not there is any value in the global funding graph's relationships in the context of ranking founders, we took the VCWiz Email Graph and overlaid the graph of investment and co-founding relationships from the previous section.

The email graph contains sufficient information to predict a founder's fundraising ability. Given this, the results confirm what we expected: the public funding graph adds useful information, which increases the correlation of the model's ranking with the baseline. Upon adding the public graph nodes and relationships, we see an increase in $\rho$ from $0.662$ to $0.696$, and a doubled P@5 of $0.4$.

\subsection{Conclusion}

Our conclusion from the above experiments is that the public graph of venture funding relationships does not provide adequate information on social-professional networks to draw conclusions about the strength of a founder (and his or her ability to fundraise). While private graphs, such as the VCWiz Email Graph, which capture \textit{all} communication between founders, investors, and their intermediaries can be useful in evaluating and characterizing founders, public data falls short. We believe this is in part due to private conversations including those which do not culminate in a fundraising relationship; while an investor may not invest in a founder, a strong relationship might still exist that can be leveraged at a later date. Furthermore, public data makes it difficult to infer social friendships, which are captured in emails and might be crucial to introductions.

Despite these shortcomings, the public-data graphs exhibit the same trends that the email-based graphs do: closeness (access) is the key to success while fundraising. Additionally, this public data can still be used to successfully augment an existing private graph with additional relationships from venture investments past and present. Since the public graph adds relevant relationships to those implied by emails, we see an increase in the efficacy of the email-based graph when augmenting it with the global funding graph.

\chapter{Conclusion}
\label{ch:ch6}

\section{Summary of Contributions}

In this thesis, our goal was to learn more about the state of founder-investor matching by attempting to improve the efficiency and equitability of this process. In doing this, we have presented two major contributions. The first is VCWiz, a tool that aspires to aid founders in finding and connecting with seed investors. The second is a series of experiments centered around FounderRank, a method of ranking founders based on their likelihood of a successful fundraise.

\subsection{VCWiz}

Through the process of designing and implementing VCWiz, we identified and enumerated three major areas, each with several opportunities, where software tools could improve the investor-founder matching process: sourcing, analyzing, and supporting. We surveyed a group of seed-stage founders and discovered that existing tools for investor discover, research, and outreach were isolated, lacking in functionality, or not practical to use. To address this, we built three iterations of a holistic fundraising tool.

The first iteration of the tool revealed the high bar for functionality and ease of use when it comes to replacing household tools like spreadsheets. The second iteration's feedback focused on the need for a platform to be comprehensive and customizable. The third iteration, which is currently live and in use by thousands of founders, has proven that there are opportunities to make early-stage investing easier and more manageable to a diverse range of startup founders. The feedback and data from this tool indicate that VCWiz does make progress towards its goals of making fundraising more efficient, and more accessible. Hundreds of founders, amongst them many underrepresented minorities, have successfully used the VCWiz platform to discover new investors, leverage their networks, and raise a round of financing. However, a lack of comprehensive data and a baseline to evaluate against leave us without conclusive evidence that VCWiz succeeds in its goals.

The difficulties we faced in evaluating the impact of VCWiz on founders with respect to efficiency and equitability are indicative of the need for further research. There are very few quantitative studies on the founder-investor matching process to reference as baselines, and we are in need of standard metrics and infrastructure to understand the impact tools have on underserved founders. It is the hope of the authors that we have opened the door to further study on this front.

The VCWiz Email Graph, collected from the founders on the platform, serves as the basis for the study described in this thesis.

\subsection{FounderRank}

Using the VCWiz Email Graph, we explored the structure and interactions of a graph of seed-stage founders from a variety of backgrounds and pedigrees. We identified that there are three important social characteristics of a founder when fundraising: importance, influence, and access. We identified three graph metrics that correspond to the characteristics: PageRank, Betweenness Centrality, and Closeness Centrality.

With a linear regression, we showed that the optimal linear combination of these three graph metrics can explain about 35\% of the variance found in a hand-crafted baseline ranking of founders, with a rank correlation of $0.66$. Furthermore, we identified access as the most important characteristic a founder can have when fundraising, followed by their importance. Influence, as proxied by Betweenness Centrality, has a near-perfect correlation with importance, and provides very little information when ranking founders.

By running an identical process over the graph of public funding data, we discovered that a similar trend applies: there is a correlation between Closeness Centrality and founders who have achieved many exists. Despite this trend, there is not sufficient information in the relationships of venture investments alone to predict fundraising success. However, this graph does contain information that can better inform rankings of founders when used to augment a social graph such as the VCWiz Email Graph. Doing so increases the rank correlation defined above non-trivially, to $0.70$.

The experiments detailed in this thesis indicate that there is an exciting opportunity to further study how the social characteristics of founders can impact and predict fundraising success. We hope to see additional research exploring how one might use this information to more efficiently match founders with investors, and to educate inexperienced founders on the relationships they should be building.

\section{Future Work}

\subsection{VCWiz}

\subsubsection{Requested Features}

On the VCWiz platform, there are several high-demand features that founders have requested repeatedly. In a survey of 118 founders, the most requested features were:

\begin{itemize}
  \item Importing and syncing intro paths from LinkedIn in addition to email
  \item Supporting Microsoft Outlook for email syncing
  \item Per-investor notes across communities of founders
  \item Shared accounts for co-founders to share
  \item A faster, more responsive filtering interface
  \item Incentives for investors to respond to intro requests
  \item Custom CRM columns
  \item More angel investors in the database
  \item Information on why an investor made a given investment
\end{itemize}

Future work would include evaluating and implementing these features, as well as continuing to find creative data sources for aggregating more information on investors. Performance is another opportunity for future improvement: the average API request involving a filter operation takes about one second.

\subsubsection{Cold Outreach}

While our data shows that VCWiz is effective at aiding founders in discovering investors, and researching them in-depth, there is still work to be done on supporting the outreach to investors. Many founders found the tool to be less useful than expected when leveraging the conversation tracker and associated features, as it was often inaccurate. Future work includes finding ways to better detect transitions in investor status during an active fundraise, and novel incentives for investors to engage with the intro request functionality on the platform.

\subsubsection{Community Support}

One major area of exploration to consider is adding community functionality to VCWiz. Currently, the platform has no direct way for founders to contribute back information on the investors they interact with. While their usage patterns and communication history are used to inform rankings, there is an opportunity to add founder-reported attributes to investor profiles. Doing this would allow founders to learn from the aggregate knowledge of their peers, without having to undertake a laborious set of meetings and phone calls. Examples of this include personality traits of investors, investment criteria, and evaluations of the extra-financial value discussed in Section \ref{ch3:motivation:research}.

We began exploring this set of features by forming a partnership with KnowYourVC\footnote{https://knowyourvc.com/}, a founder-oriented review site for venture capitalists. While founders cannot currently report their experiences directly on VCWiz, the can see reviews and tags pulled in from KnowYourVC's API.

\subsubsection{Ranking \& Filtering}

The results of our study show that there is merit to ranking founders and investors based on the information collected on the VCWiz platform. There is further opportunity to use this information in surfacing investors to founders during the discovery phase of their fundraise.

The ranking function discussed in Section \ref{ch4:filtering} is currently very similar for every founder on the platform. Future work could explore further customization of this ranking, based on preferences collected from the founder. For example, finding an investor that is in the same physical location might be much more important to a founder than finding one who has deep expertise in a specific industry, or vice-versa.

\subsection{FounderRank}

While we have examined the relationship between founder and investors in social graphs, we have yet to rank and score the two sets of nodes jointly. There are several opportunities to explore the efficacy of this technique, including explicitly exploring how the scores of the founders that an investor has funded correlate with the rank of the investor, and vice-versa.

Additionally, we have yet to explore the rankings calculated by combining social graph metrics with the baseline metrics used to evaluate FounderRank. Considering characteristics of the company and the founder jointly should result in holistic rankings which can be evaluated over time. We believe these joint rankings would be superior to anything discussed thus far.

\subsubsection{Clustering}

The rank assigned to a founder (or investor) is an indicator of their social characteristics, which might make it a useful feature when clustering individuals. One hypothesis to test in the future is that investors prefer to invest in founders who have a similar relative rank to themselves. If this is true, and investors have an affinity to similarly-ranked founders, then it would be prudent to show founders investors of a comparable rank in a tool such as VCWiz, so as to maximize their chances of an investment being made.

\subsubsection{Recommender Systems}

The experiments to date with data generated from founders on VCWiz has focused on scoring and ranking founders and investors. However, there is another, related application of this data in recommender systems. We have the property that similar users (founders/companies) are positive about similar products (investors). Therefore, the discovery problem can be reduced to a version of the very popular Netflix Prize~\cite{netflixpize} problem. This allows us to apply an entire body of recommendation-system research.

The crucial hypothesis to test is that incorporating data from the platform increases the quality of recommendations over classic techniques. This information includes attention-based features, such as which investors are clicked on, reached out to, and interacted with. If this hypothesis is proven, a recommender system would increase the efficiency of investment matches, with founders discovering investors they would not have otherwise. However, we must be careful to not sacrifice the equitability of our system: as discussed in Section \ref{ch2:matching}, models trained on existing data can incorporate the dangerous biases that are pervasive in venture capital.

One could test this hypothesis by beginning with a classic item-based recommender system, trained on features extracted from the investor's profile. This model could be evaluated against a simple baseline, which suggests the co-investors of a startup's competitors as the recommended investors. The performance of this model could then be compared to a hybrid model, which adds a user-based layer that is trained on the click and outreach data of founders on the VCWiz platform.

\appendix
\chapter{Potential Products}
\label{intro:products}

We have spent time exploring possible tools that could be built to aid in various stages of the venture pipeline. Each tool is identified below, along with the motivation and a brief summary of the technical challenges involved.

\section{Sourcing}

We have identified two opportunities to do with sourcing, both on the outbound flow side.

\subsubsection{A system to aggregate signals from founders and predict the intent to start a company}

Founders often emit signals that indicate they are starting a new company, often long before they officially announce their new endeavor. These signals can be explicit (changing a job title on LinkedIn, or biography line on Twitter) or implicit (leaving a job, moving cities, or attending entrepreneurial events). In isolation, these signals are not strong, but in aggregate they can be strongly correlated with the intent to start a company.

We see an opportunity for a system that monitors the social networks of a GP, identifying and aggregating potential signals. The technical challenges include linking seemingly-unrelated signals across networks and schemas, and inventing a ranking algorithm which can present the most likely potential founders given a set of signals. We would likely use these signals as machine learning features.

\subsubsection{A system to discover and monitor promising out-of-network individuals and organizations}

While there are a plethora of announcements and releases online which would indicate an investment-worth company has formed, humans are not capable of monitoring and filtering the wealth of information generated on the internet on an ongoing basis. Thus, a GP's sourcing abilities are largely limited to the founders they can discover in their network.

We propose a system which treats the relevant information on the Internet as a connected, directed graph, which can be monitored and have its nodes ranked (as PageRank does for search engines). Every interesting community (such as educational institutions) could have its own independent graph, and the top-ranked nodes of each graph could be surfaced for easy human review. The technical challenges around this system include a lack of labeled training data (what constitutes an ``interesting'' node?) and the noisiness of the web (there are many sites linked from a community that contain irrelevant or even misleading information).

\section{Analyzing}

When is comes to analyzing, there are two major project proposal we considered.

\subsubsection{A system to filter, categorize, and rank the companies in a venture pipeline}

Many seed-stage funds suffer today from an overwhelming pipeline of startup companies to consider. There is considerable data available on these companies which seems to be correlated to how investment-worthy the company is at first glance. At the very least, the cheap filters applied by investors are mimicable through existing data (alma maters of founders, size of initial market, sentiment of partners after first meeting).

We propose a system which uses the information associated with pipeline companies to categorize each company into buckets that predict how far in the pipeline the company will move, using these buckets to filter and prioritize the pipeline. This will be an online, semi-supervised clustering problem which receives constant feedback from partners. The technical challenges include identifying and extracting the relevant features (which may include leveraging NLP techniques on descriptions, pitch decks, and meeting notes), and finding a way to incorporate user feedback in a meaningful way. Evaluation methods are also difficult to formulate a priori.

\subsubsection{A system to surface and summarize key trends and news in a given industry}

Many hours of time is wasted at venture firms serially researching and identifying key facts and risks about both a company and its broader industry. This act of information extraction and summarization is well-suited for classic Natural Language Processing.

We propose a system which ingests both internal data on the company at hand, as well as recent news and evergreen data sources (such as Wikipedia) and delivers a digest of key risks identified in the company (based on pitch decks and partner notes), as well as a one-pager on the given industry.

\section{Supporting}

Finally, with regards to portfolio support, there are two tools we considered building.

\subsubsection{A system for the discovery of and supporting outreach to the optimal set of seed-stage investors}

It is widely accepted in the venture industry that there is significant merit to a founder finding the ``right'' set of investors when raising money. Not only does the strategic focus of a firm and its network impact said firm's ability to help a company, but the particular focus of a partner within a firm can also influence whether or not a company even gets funded. There is strong empirical evidence that partners at venture firms do indeed specialize and focus on a very specific subset of companies~\cite{Stone:2013:EST:2541167.2507882}.

Matching a founder to the most relevant partner at each firm, and the most realistic and appropriate firms at each funding stage, is a challenging problem for humans to tackle alone.

We propose a hybrid recommender system which suggests relevant and strategic investors to founders, based on their company and ideal investor profile. This would follow the models laid out in recent literature on recommender systems~\cite{Burke2002}.

Our tool would also provide an interface for planning and tracking the process of reaching out to these investors, as a way to collect structured training data for future iterations. Technical challenges here include building a sufficiently strong user experience so as to inspire trust in the tool, determining how to identify a user as features, and building a labeled database of investors and the founders they have backed.

\subsubsection{A system to predict and propagate viral company news}

As the number of companies in a seed-stage venture firm's portfolio grows, it becomes increasingly difficult for partners to keep track of the movements of each company. This makes it difficult to identify when a company is in the process of making a big press release (which the VC could support). Furthermore, there is no easy way for a VC to know the latest public change in each of their companies.

We propose a tool to monitor the social media accounts of portfolio companies, summarizing news and sharing the posts that are estimated to be the most popular or viral. Text summarization is an open research problem that has several standardized solutions~\cite{textsummarization}, each of which can be tuned for the domain with manual feature engineering and additional rule-based systems. Estimating social media popularity and virality can be done with linear point-process models such as SEISMIC~\cite{seismic}, or more complex Bayesian models like the one presented in \cite{bayesiantweets}, which uses more features from the graph generated by the post and its shares. The biggest technical challenges here are around coaxing and tuning these algorithms to give sufficiently good results for our domain.

\clearpage
\newpage

\chapter{Feedback Surveys}
\label{appf:survey}

At various times during the design and iteration of VCWiz, we solicited feedback from founders and investors. This appendix documents the questions that were asked, and the set of answers permitted, when applicable.

\section{VCWiz Founder Survey}
\label{appf:survey:founders}

\begin{enumerate}
  \item How likely are you to recommend VCWiz to a friend?
    \subitem \textit{The founder was presented with a scale from 0 to 10}
  \item Have you used VCWiz?
    \subitem No
    \subitem Yes, I used it for my fundraise!
    \subitem Yes, I used it to research investors
    \subitem Yes, I clicked around a bit
    \subitem Yes, I used it to fill out my investor profile
    \subitem Other \textit{(free-form)}
  \item What's preventing you from using VCWiz more often?
    \subitem No
    \subitem Yes, I used it for my fundraise!
    \subitem Yes, I used it to research investors
    \subitem Yes, I clicked around a bit
    \subitem Yes, I used it to fill out my investor profile
    \subitem Other \textit{(free-form)}
  \item VCWiz has an intro request functionality, that helps you get connected to investors. Have you tried this out?
    \subitem Yes
    \subitem I didn't know I could do that
    \subitem It didn't work when I tried
    \subitem I couldn't figure out how to use it
    \subitem I don't need any intros right now
    \subitem I prefer to get my intros warm
    \subitem Other \textit{(free-form)}
  \item VCWiz also has a feature called Link, which scans your inbox and syncs your VCWiz dashboard with ongoing conversations with investors. Have you tried this out?
    \subitem Yes
    \subitem No, because I have privacy concerns
    \subitem No, because I don't need it
    \subitem No, because I use a different CRM
    \subitem No, because I don't know how to enable it
    \subitem Other \textit{(free-form)}
  \item Any other feedback for us? What's your \#1 feature request?
    \subitem \textit{The founder was presented with a free-form response box}
\end{enumerate}

\section{VCWiz Investor Survey}
\label{appf:survey:investors}

\begin{enumerate}
  \item Why did you end up not engaging with the founder(s) that reached out to you?
    \subitem I didn't see the intro request email
    \subitem I didn't know what it was, so I ignored it
    \subitem I was annoyed about receiving the email
    \subitem I didn't have time to look at the email
    \subitem I don't respond to cold emails
    \subitem Other \textit{(free-form)}
\end{enumerate}

\section{VCWiz User Testing}
\label{appf:survey:users}

All of the following questions were presented with boxes for free-form text responses, unless otherwise specified.

\begin{enumerate}
  \item What's your name?
  \item Were you able to test out VCWiz?
    \subitem Yes
    \subitem No
  \item What went wrong?
  \item What did you like about the product?
  \item What did you dislike?
  \item How do you discover new investors currently? What are your biggest complains with this process?
  \item What tools have you used for discovering investors in the past?
  \item How do you organize your outreach to new investors currently? What are your biggest complains with this process?
  \item What tools have you used for organizing your outreach to investors in the past?
  \item Have you used dedicated software for managing investor relationships?
    \subitem Yes
    \subitem No
  \item What have your biggest complaints been with those tools?
  \item What has prevented you from using dedicated tools?
  \item Would you use VCWiz as it exists today?
    \subitem Yes
    \subitem No
  \item What would have to change for you to use VCWiz?
  \item How likely are you to recommend VCWiz to a fellow founder?
    \subitem \textit{The founder was presented with a scale from 1 to 5}
  \item What is the one most important feature to add to VCWiz?
  \item Anything else we should know?
\end{enumerate}

\clearpage
\newpage

\chapter{VCWiz Queries}

\begin{lstlisting}[frame=single,language=SQL,basicstyle=\tiny,columns=fullflexible,label={vcwiz:sql:query},caption={Firm Filter Query}]
SELECT
  subquery.*,
  firm_partners.partners AS partners,
  firm_recent_investments.recent_investments AS recent_investments,
  firm_coinvestors.coinvestors AS coinvestors
FROM (
  SELECT
    distincted.*,
    firm_velocities.velocity
  FROM (
    SELECT
      DISTINCT ON (fullquery.id) fullquery.*
    FROM (
      SELECT
        firms.*,
        wo.rn
      FROM (
        SELECT
          subquery.id,
          row_number()
          OVER (
          ORDER BY
            subquery.ti_sum DESC,
            subquery.c_cnt DESC) AS rn
        FROM (
          SELECT
            firms.id,
            SUM(COALESCE(firm_investor_aggs.target_count, $1)) AS ti_sum,
            COUNT(DISTINCT companies.id) FILTER (
              WHERE companies.location = $2
            ) AS c_cnt
          FROM
            firms
            INNER JOIN investments ON investments.firm_id = firms.id
            INNER JOIN companies ON companies.id = investments.company_id
            INNER JOIN firm_investor_aggs ON firm_investor_aggs.firm_id = firms.id
          WHERE (
            firms.id IN (
              (
                  SELECT
                    firms.id
                  FROM
                    firms
                  WHERE (firms.location && $3)
              )
              UNION
              ::~ (
                SELECT
                  firms.id
                FROM
                  firms
                  INNER JOIN investments ON investments.firm_id = firms.id
                  INNER JOIN companies ON companies.id = investments.company_id
                WHERE (companies.location = $4)
              )
            )
          )
          GROUP BY
            firms.id
          ORDER BY
            ti_sum DESC,
            c_cnt DESC
          LIMIT $5
        ) AS subquery
        LIMIT $6
      ) AS wo
      INNER JOIN firms
      USING (id)
    ) AS fullquery
  ) AS distincted
  LEFT OUTER JOIN firm_velocities ON firm_velocities.firm_id = distincted.id
ORDER BY
  rn OFFSET $7
LIMIT $8) AS subquery
  INNER JOIN firm_recent_investments ON firm_recent_investments.firm_id = subquery.id
  INNER JOIN firm_coinvestors ON firm_coinvestors.firm_id = subquery.id
  INNER JOIN firm_partners ON firm_partners.firm_id = subquery.id
\end{lstlisting}

\newpage

\begin{lstlisting}[frame=single,language=SQL,basicstyle=\footnotesize,columns=fullflexible,label={vcwiz:sql:view},caption={Denormalization of Individual Investor Properties}]
SELECT
  firms.id AS firm_id,
  COALESCE(SUM(investors.target_investors_count), 0) AS target_count,
  bool_or(COALESCE(investors.featured, false)) AS featured,
  bool_or(COALESCE(investors.verified, false)) AS verified
FROM firms
  INNER JOIN investors ON investors.firm_id = firms.id
GROUP BY firms.id
\end{lstlisting}

\begin{lstlisting}[frame=single,basicstyle=\footnotesize,columns=fullflexible,caption={Finding Intro Paths},label={vcwiz:cypher:intro}]
MATCH (other:Investor), path = shortestPath((me)-[*1..4]-(other))
WHERE id(me) = {founder_neo_id} AND id(other) = {investor_neo_id}
RETURN path, reduce(count = 0, r IN relationships(path) | count + coalesce(r.count, 0)) AS total
ORDER BY total DESC
LIMIT 3;
\end{lstlisting}

\begin{lstlisting}[frame=single,basicstyle=\footnotesize,columns=fullflexible,caption={Removing Orphans},label={vcwiz:cypher:orphans}]
MATCH (n:Person)
WHERE NOT (n)--()
DELETE n
\end{lstlisting}

\clearpage
\newpage

\chapter{VCWiz Data Models}

These Entity-Relationship diagrams are generated with rails-erd\footnote{https://github.com/voormedia/rails-erd}. In the diagrams, \texttt{Firm} is referred to as \texttt{Competitor} for legacy reasons.

\begin{figure}
  \includegraphics[width=\textwidth]{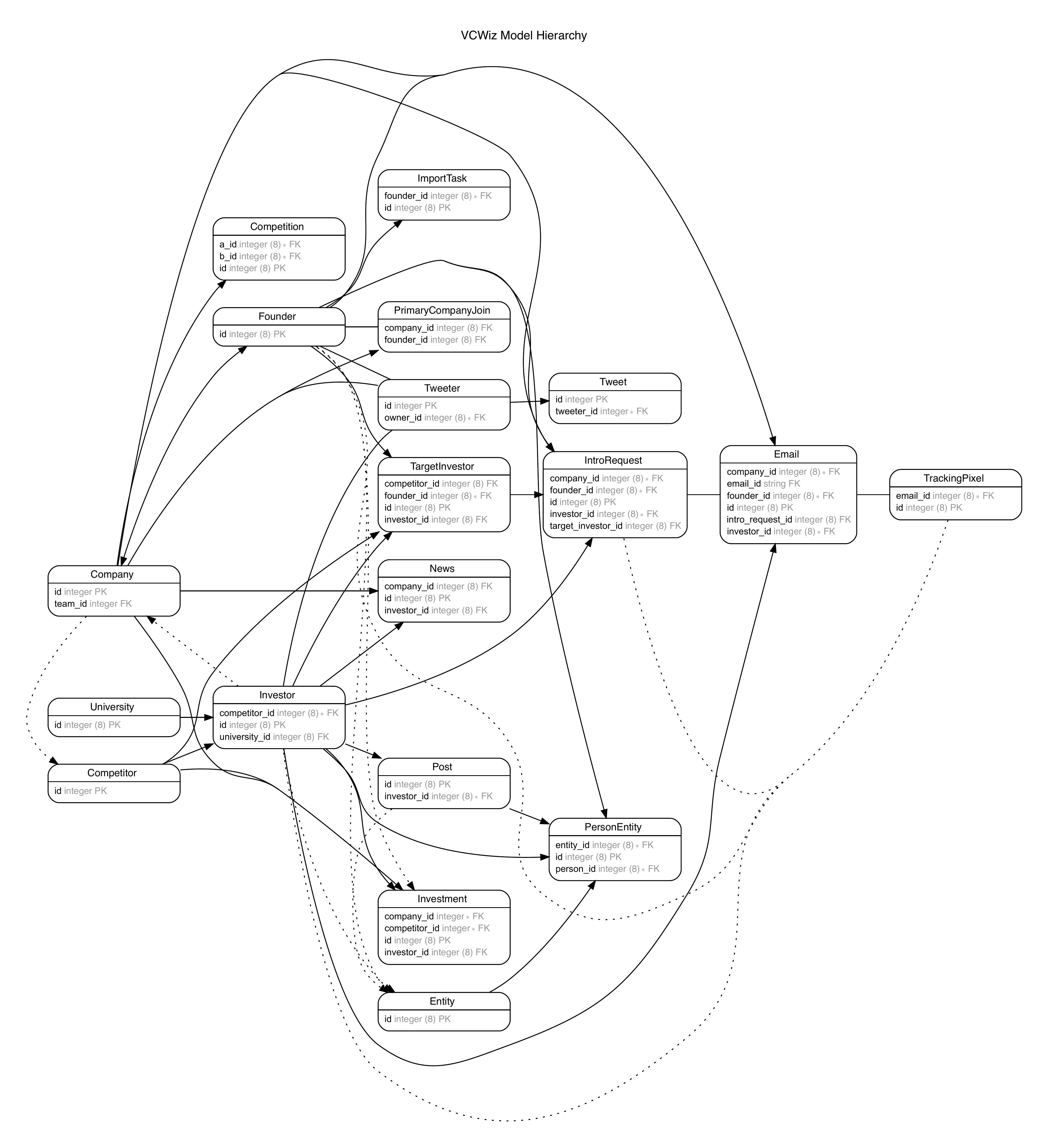}
  \caption{VCWiz Model Hierarchy}
  \label{vcwiz:model:hierarchy}
  \centering
\end{figure}

\begin{figure}
  \includegraphics[width=\textwidth]{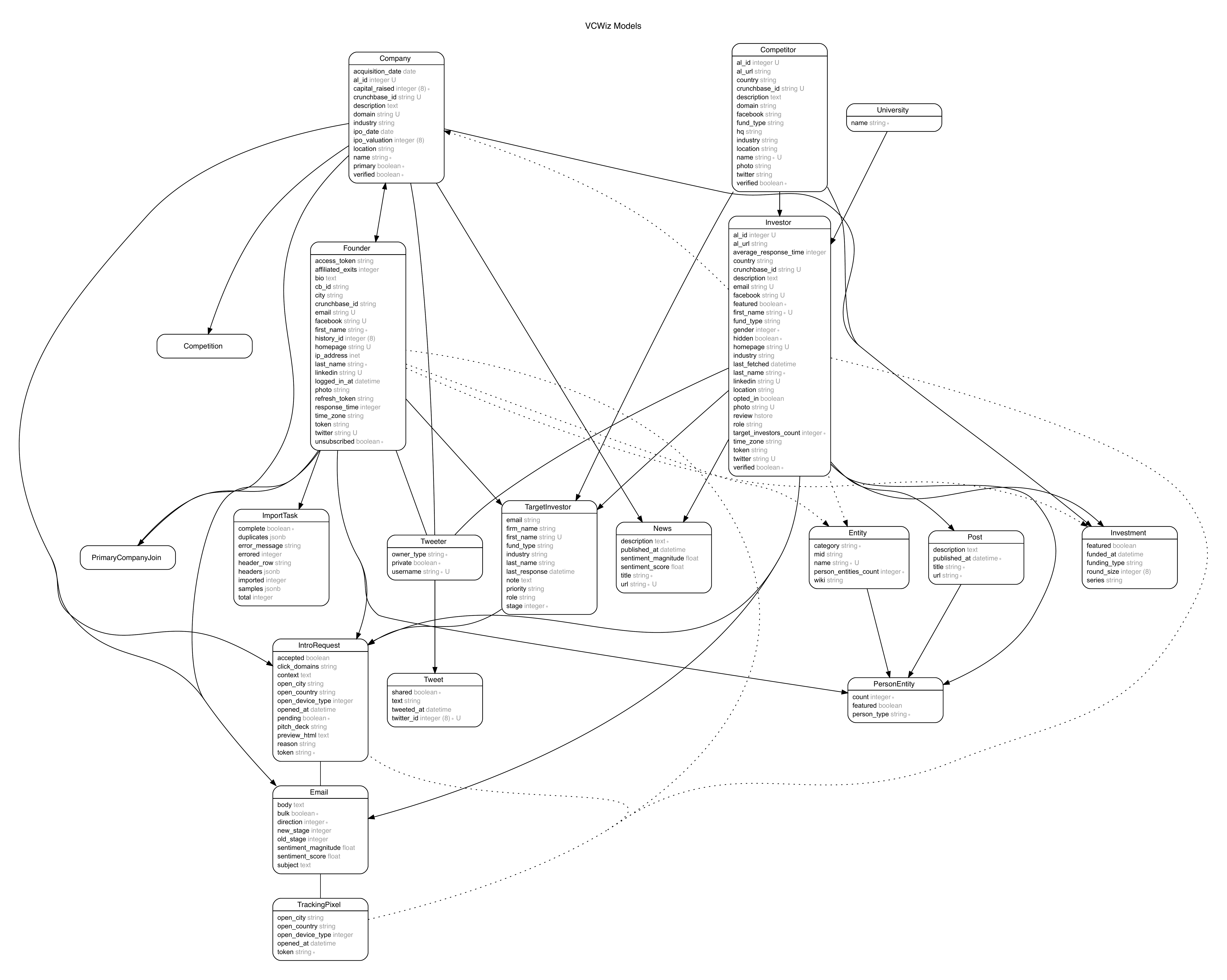}
  \caption{VCWiz Models}
  \label{vcwiz:model:content}
  \centering
\end{figure}

\clearpage
\newpage

\chapter{VCWiz Screenshots}

\begin{figure}[ht]
  \centering
  \includegraphics[width=\textwidth]{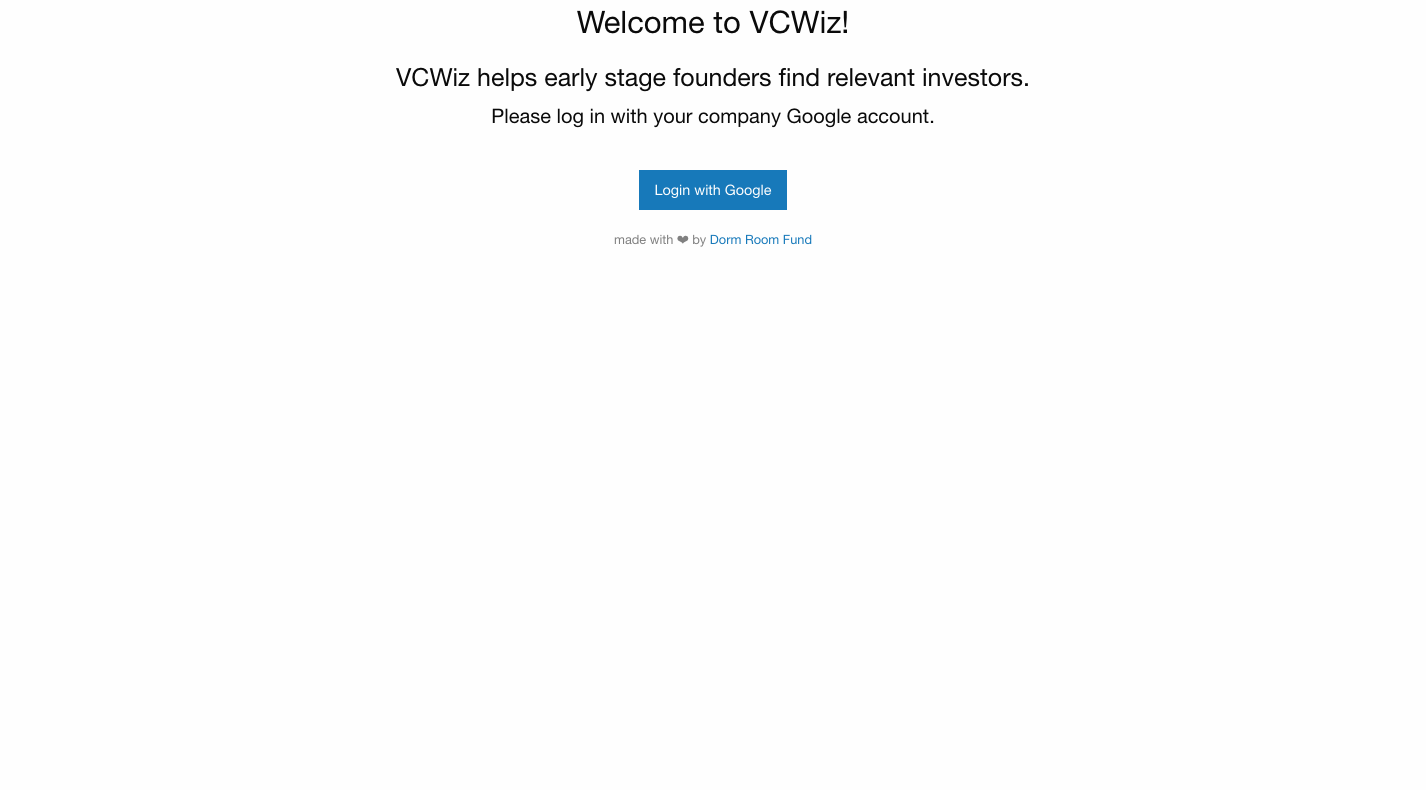}
  \caption{Login}
  \label{screenshots:v1:login}
\end{figure}

\begin{figure}[ht]
  \centering
  \includegraphics[width=\textwidth]{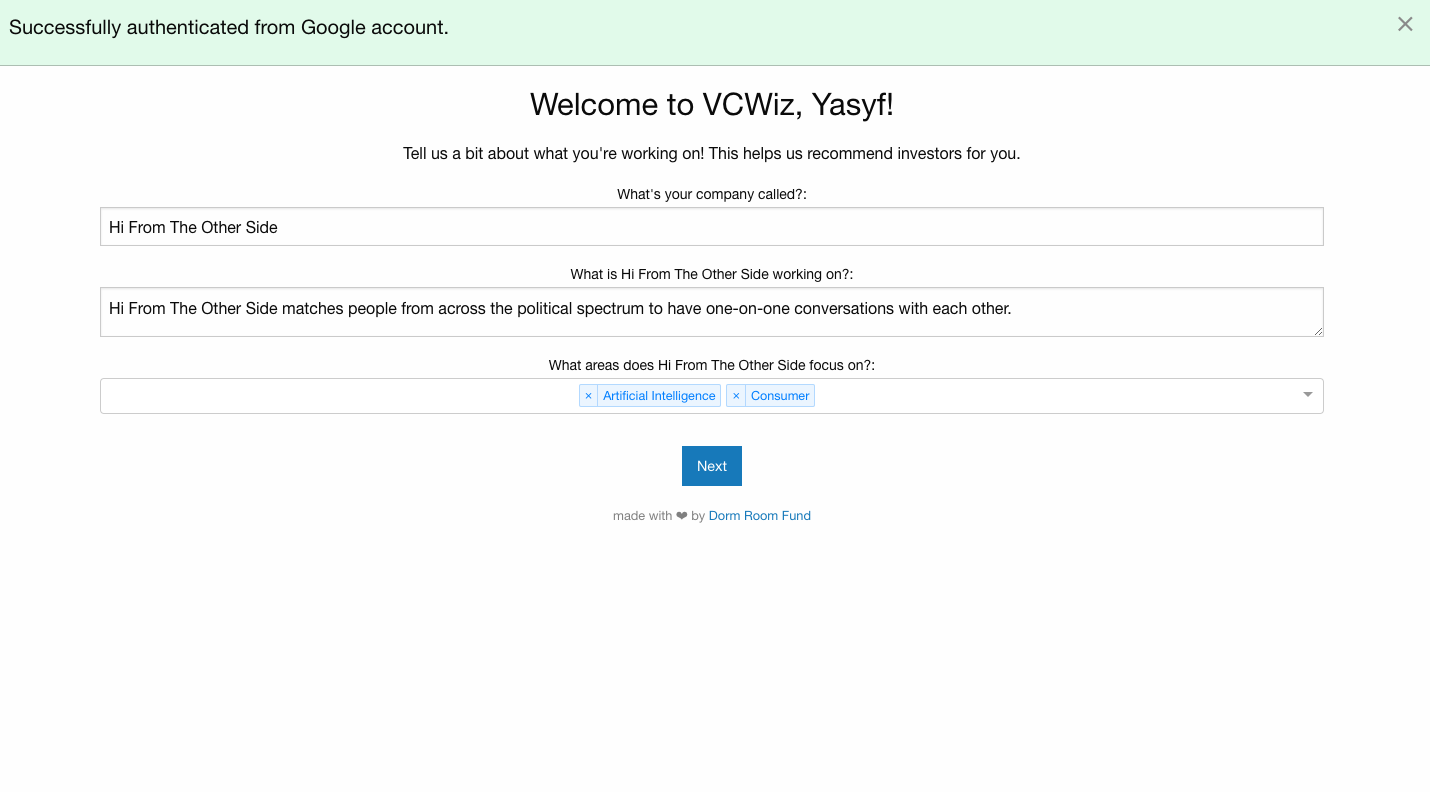}
  \caption{Signup}
  \label{screenshots:v1:signup}
\end{figure}

\begin{figure}[ht]
  \centering
  \includegraphics[width=\textwidth]{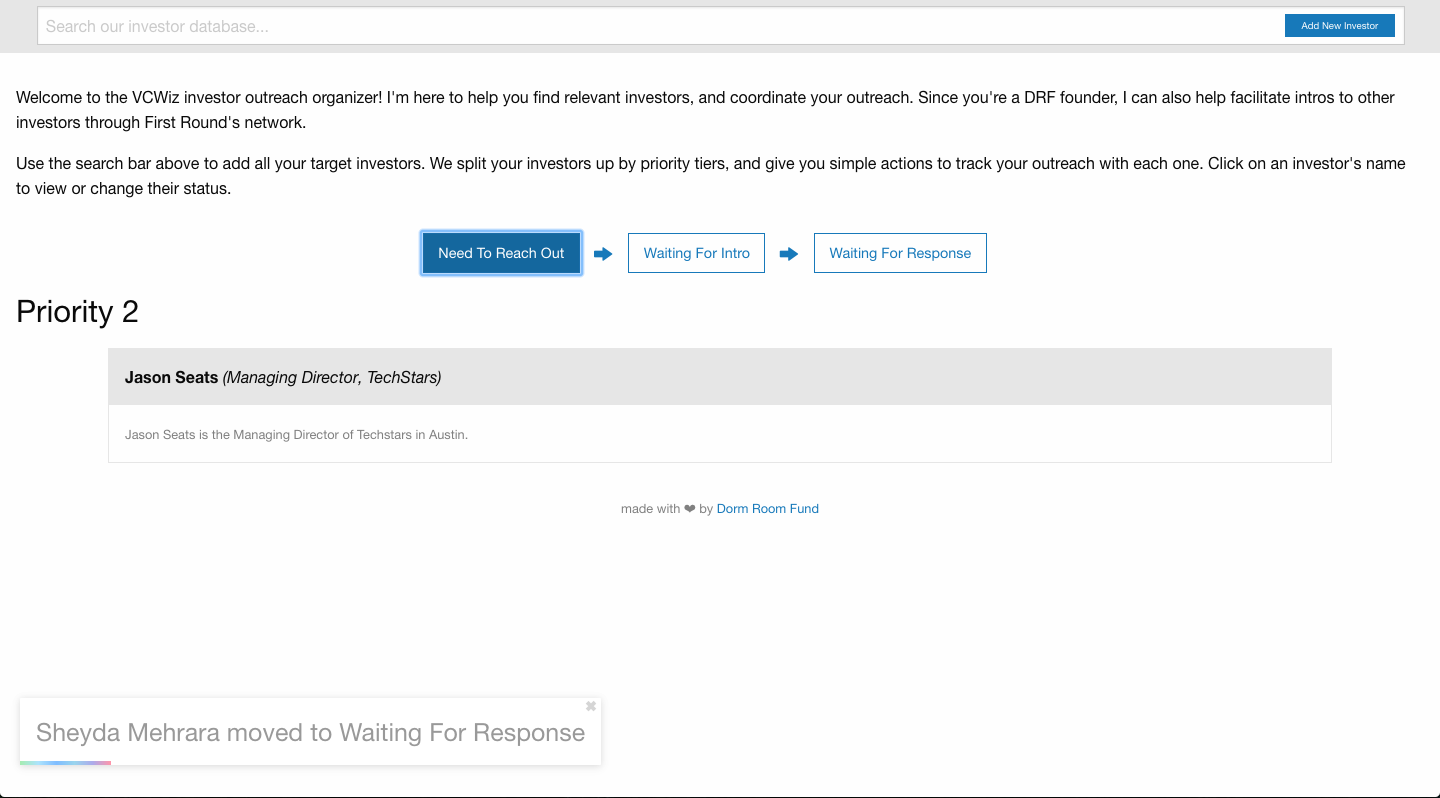}
  \caption{Conversations}
  \label{screenshots:v1:conversations}
\end{figure}

\begin{figure}[ht]
  \centering
  \includegraphics[width=\textwidth]{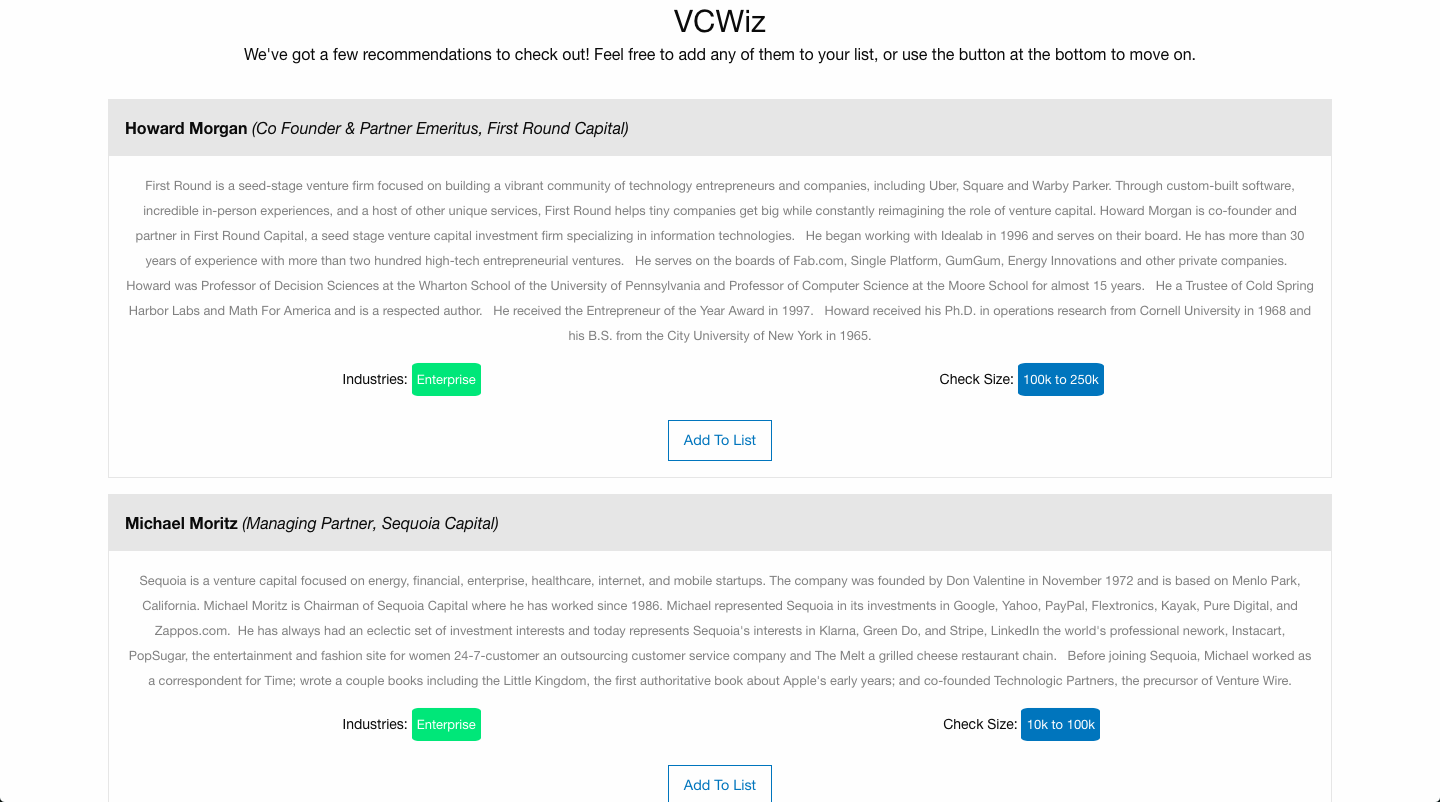}
  \caption{Investor Recommendations}
  \label{screenshots:v2:recs}
\end{figure}

\begin{figure}[ht]
  \centering
  \includegraphics[width=\textwidth]{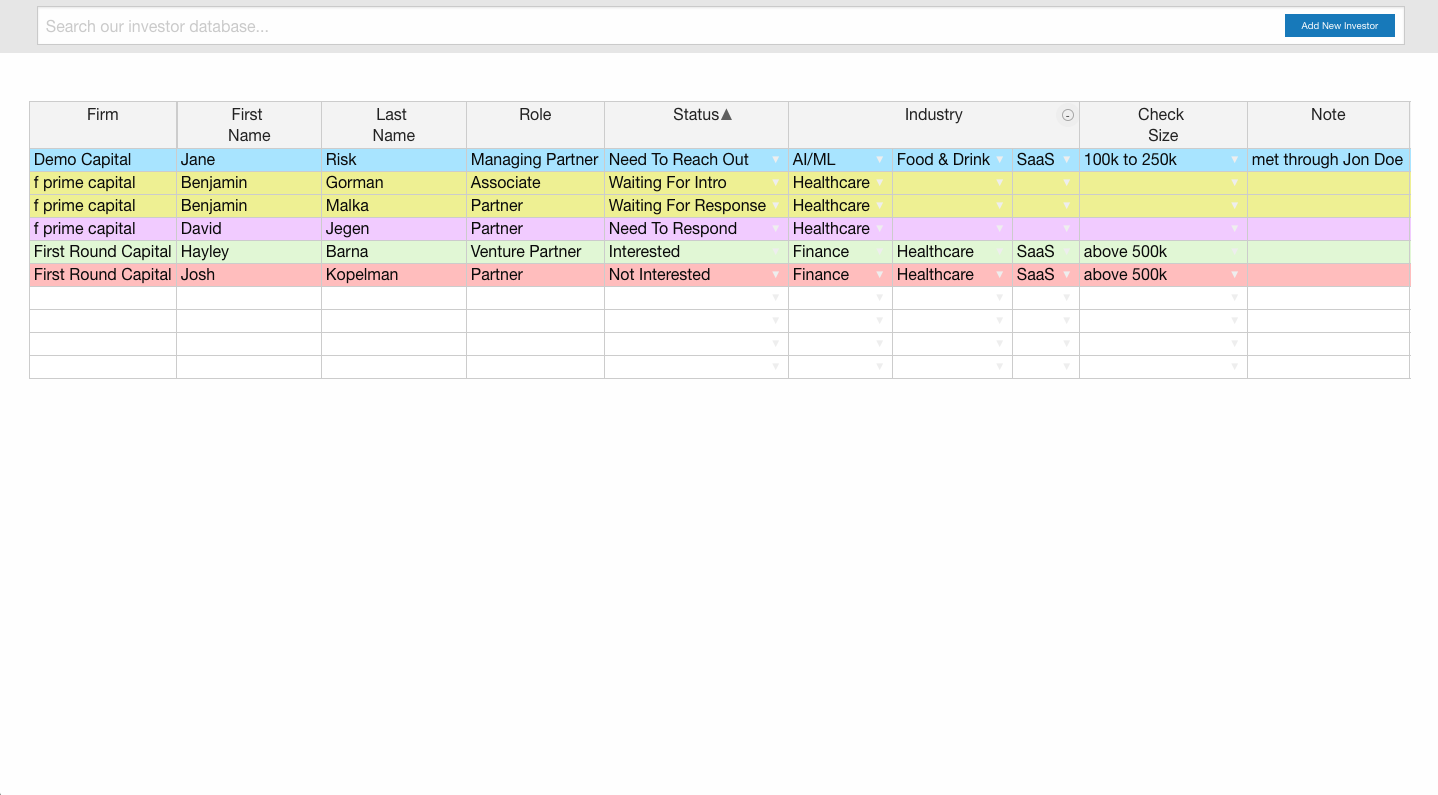}
  \caption{Conversation Tracker}
  \label{screenshots:v2:conversations}
\end{figure}

\begin{figure}[ht]
  \centering
  \includegraphics[width=\textwidth]{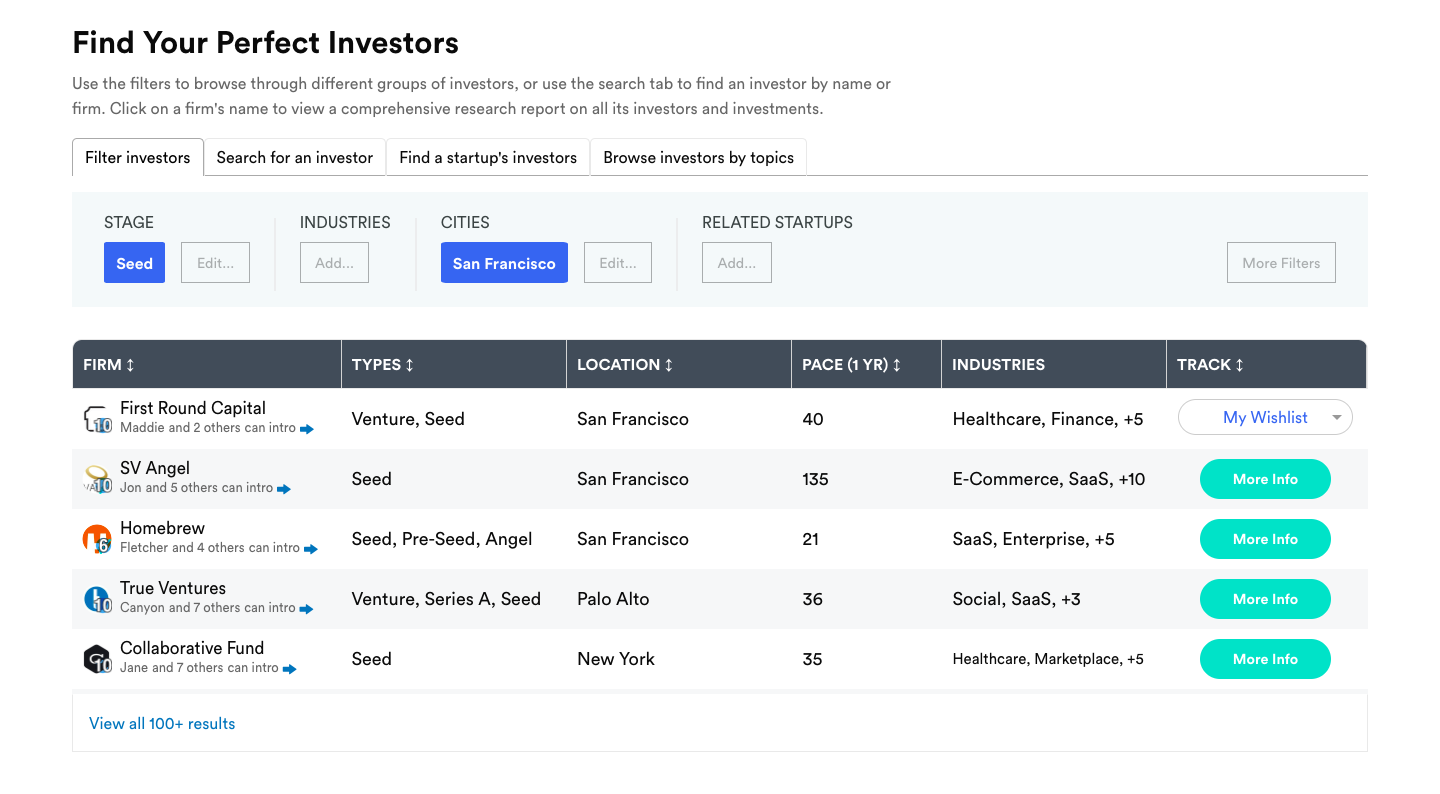}
  \caption{Filter \& Search}
  \label{screenshots:v3:filter}
\end{figure}

\begin{figure}[ht]
  \centering
  \includegraphics[width=\textwidth]{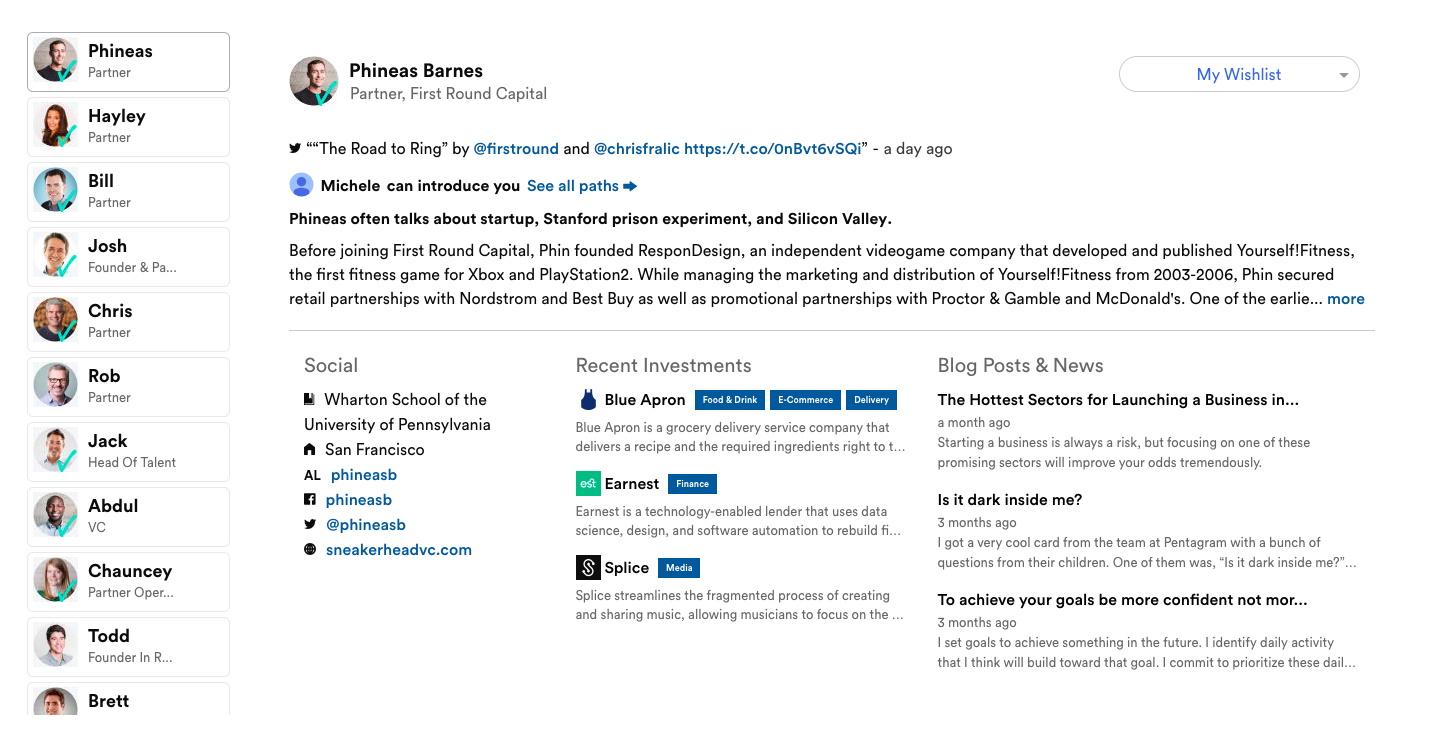}
  \caption{Investor Research}
  \label{screenshots:v3:research}
\end{figure}

\begin{figure}[ht]
  \centering
  \includegraphics[width=\textwidth]{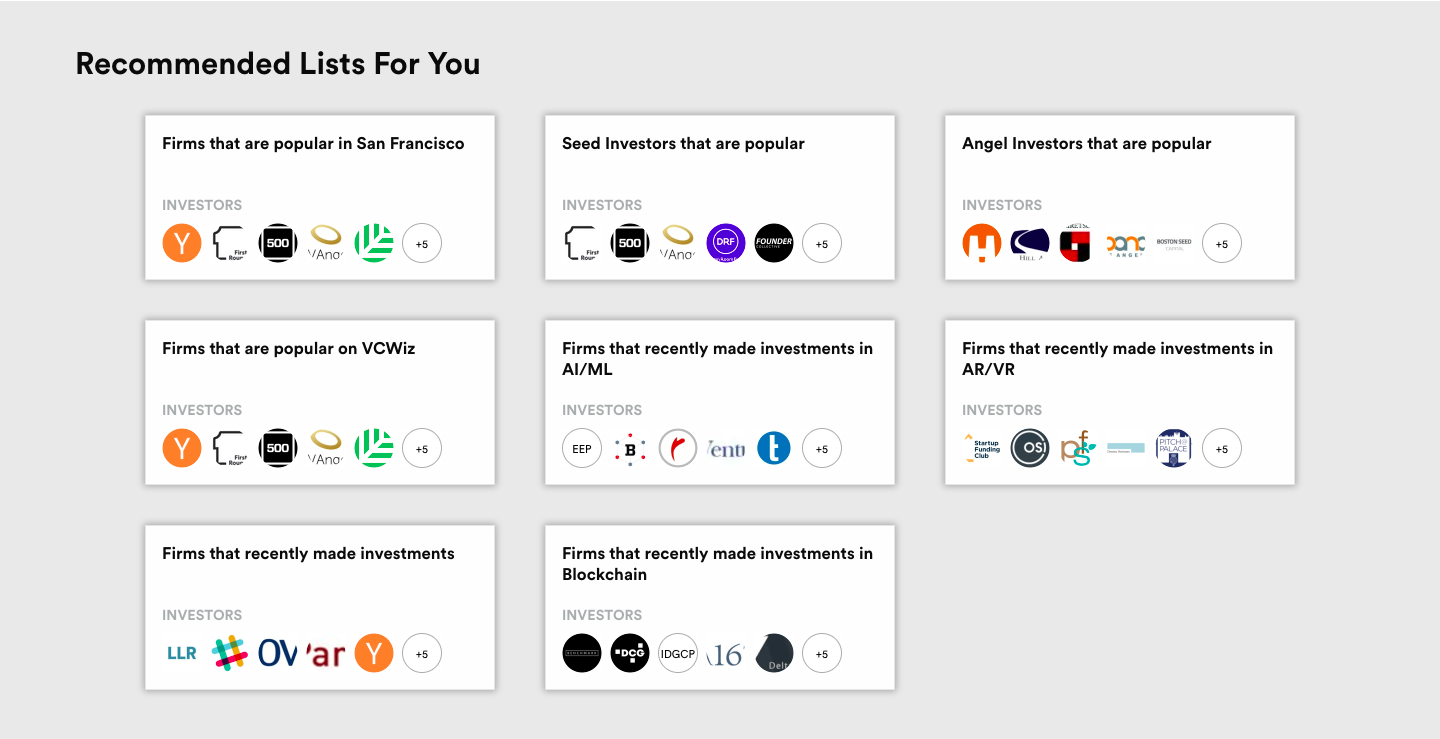}
  \caption{Curated Firm Lists}
  \label{screenshots:v3:lists}
\end{figure}

\begin{figure}[ht]
  \centering
  \includegraphics[width=\textwidth]{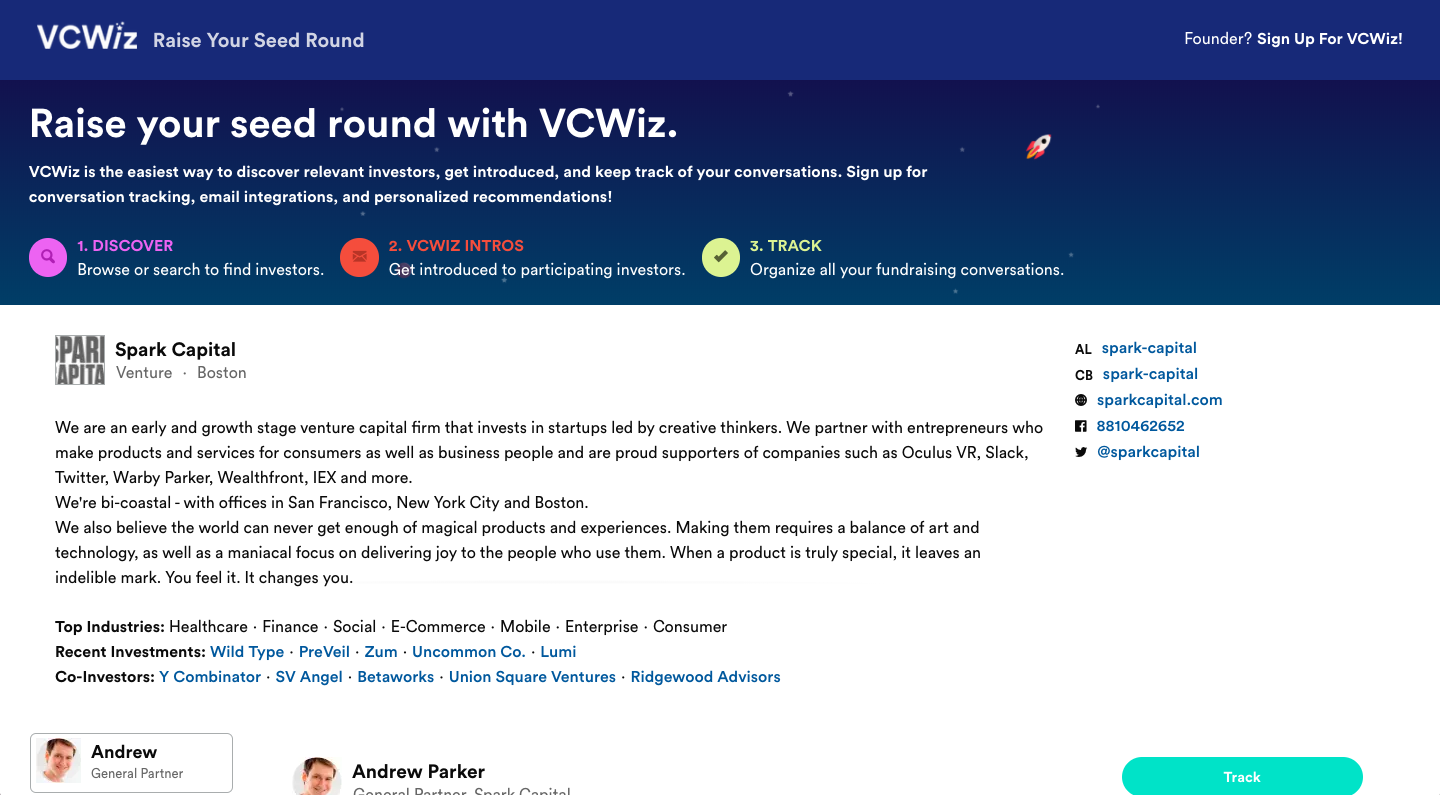}
  \caption{Spark Capital Firm Page}
  \label{screenshots:v3:onboarding}
\end{figure}

\begin{figure}[ht]
  \centering
  \includegraphics[width=\textwidth]{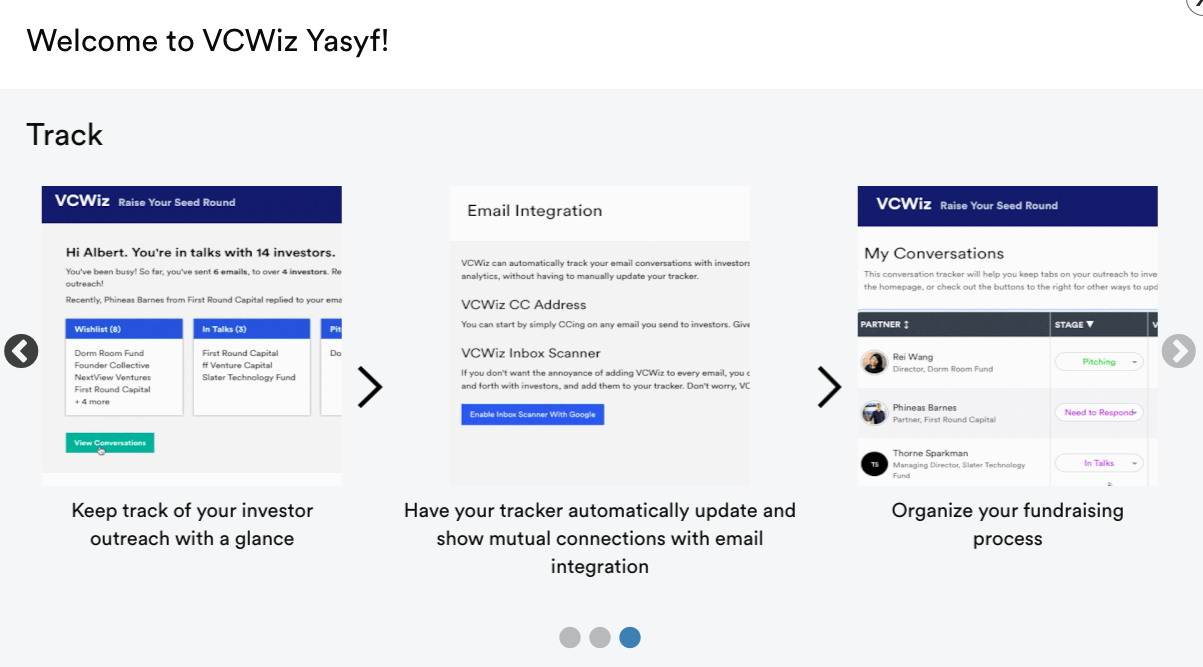}
  \caption{Tutorial}
  \label{screenshots:onboarding:intro}
\end{figure}

\begin{figure}[ht]
  \centering
  \includegraphics[width=\textwidth]{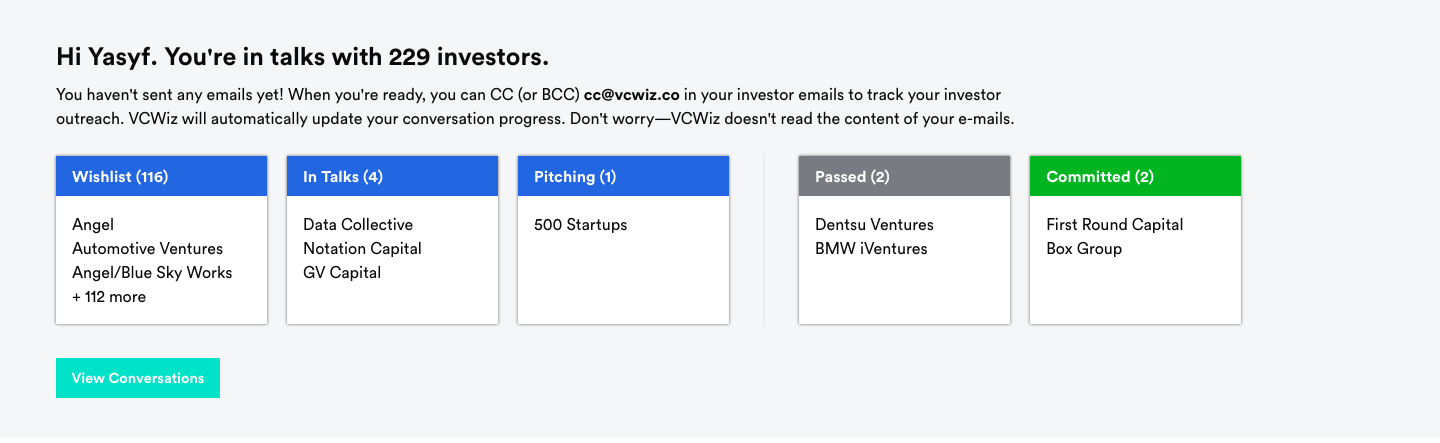}
  \caption{Conversation Summary}
  \label{screenshots:onboarding:summary}
\end{figure}

\begin{figure}[ht]
  \centering
  \includegraphics[width=\textwidth]{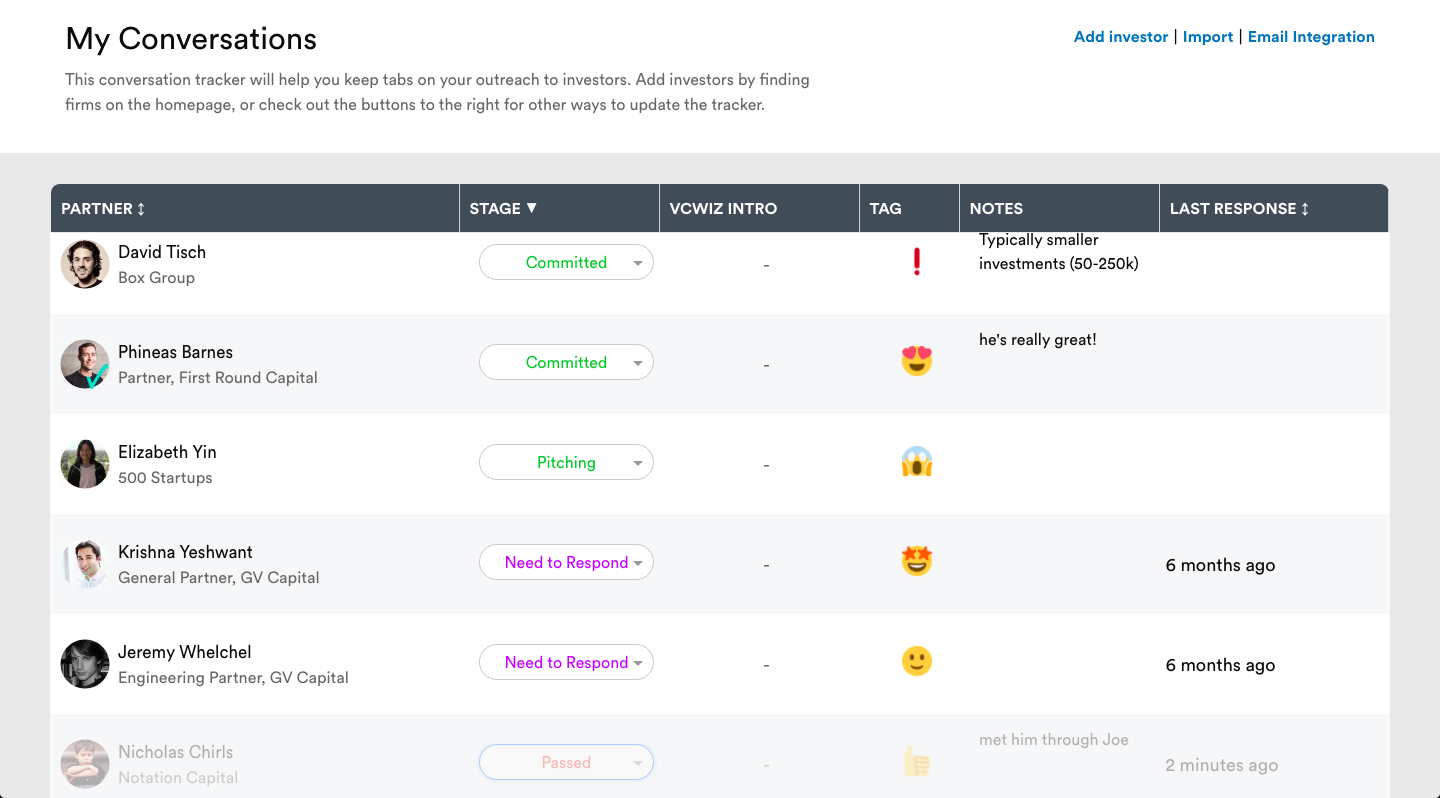}
  \caption{Conversation Tracker}
  \label{screenshots:onboarding:conversations}
\end{figure}

\begin{figure}[ht]
  \centering
  \includegraphics[width=\textwidth]{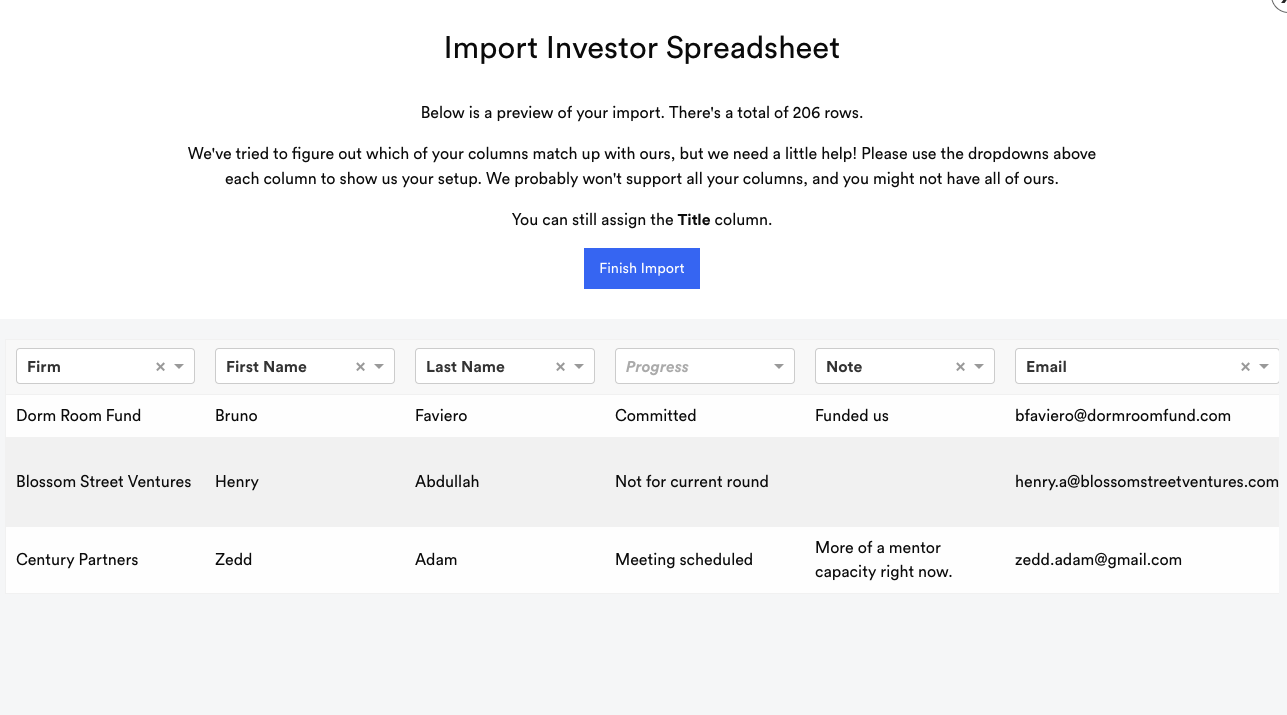}
  \caption{Spreadsheet Import}
  \label{screenshots:import:columns}
\end{figure}

\begin{figure}[ht]
  \centering
  \includegraphics[width=\textwidth]{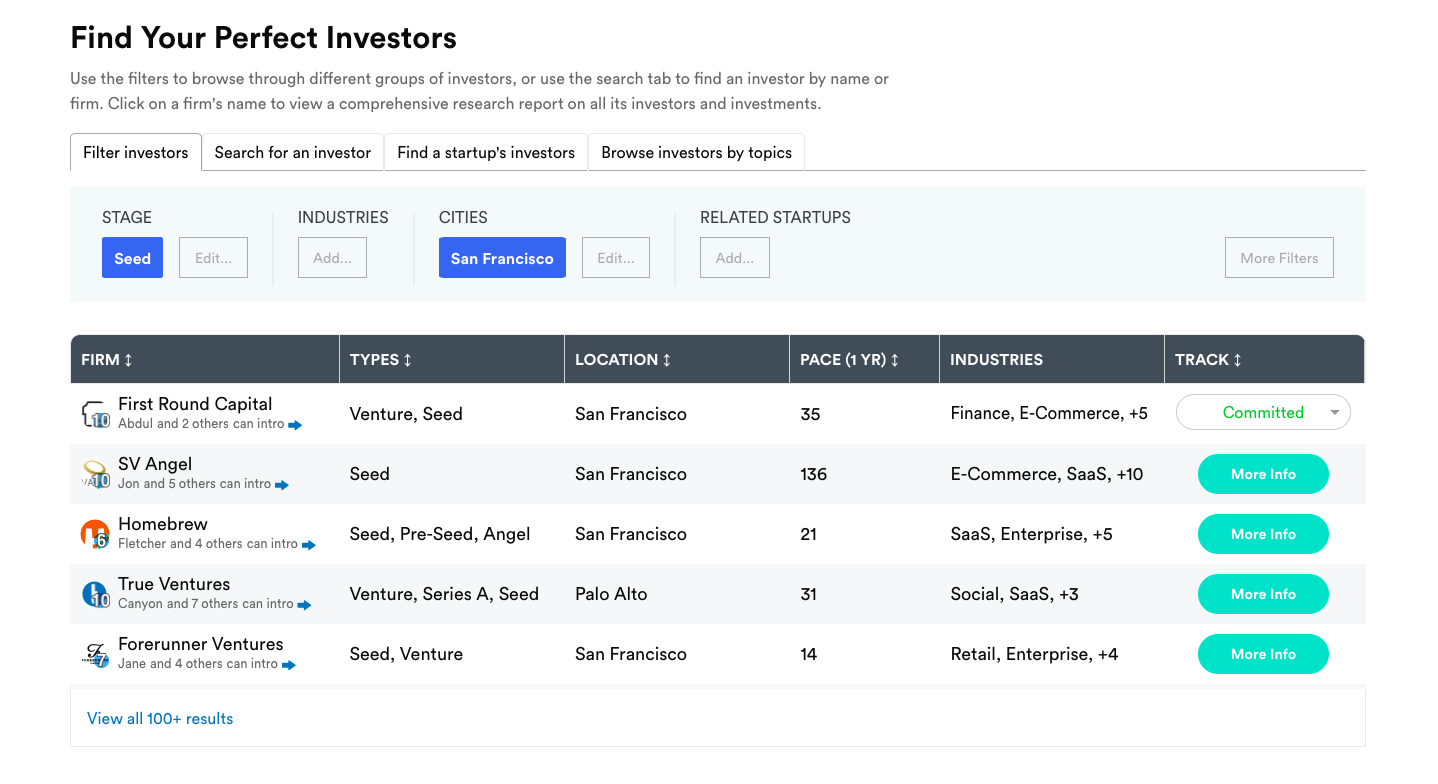}
  \caption{Filter Results}
  \label{screenshots:filtering:results}
\end{figure}

\begin{figure}[ht]
  \centering
  \includegraphics[width=\textwidth]{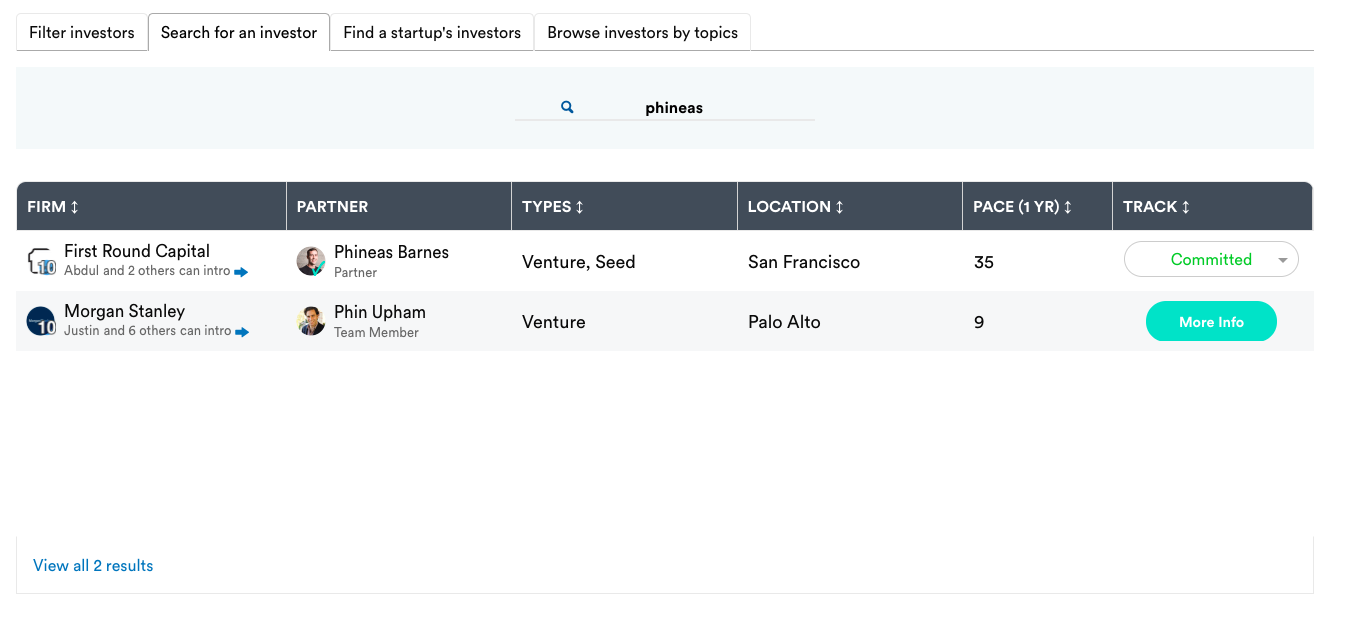}
  \caption{Filter Results with Partner Column}
  \label{screenshots:filtering:partner}
\end{figure}

\begin{figure}[ht]
  \centering
  \includegraphics[width=\textwidth]{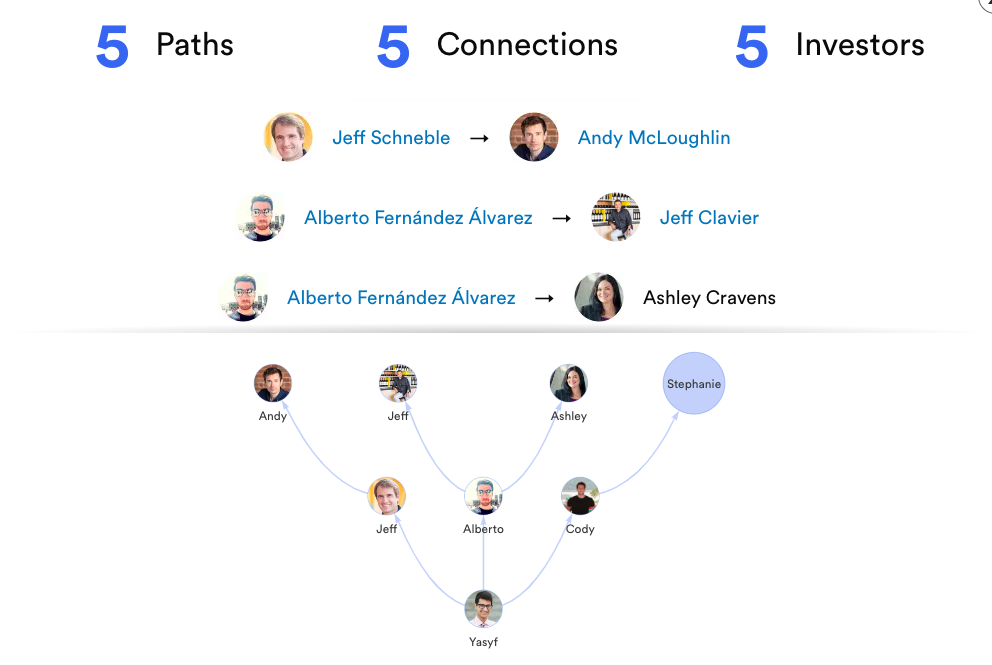}
  \caption{Intro Path Modal}
  \label{screenshots:intro:path}
\end{figure}

\begin{figure}[ht]
  \centering
  \includegraphics[width=\textwidth]{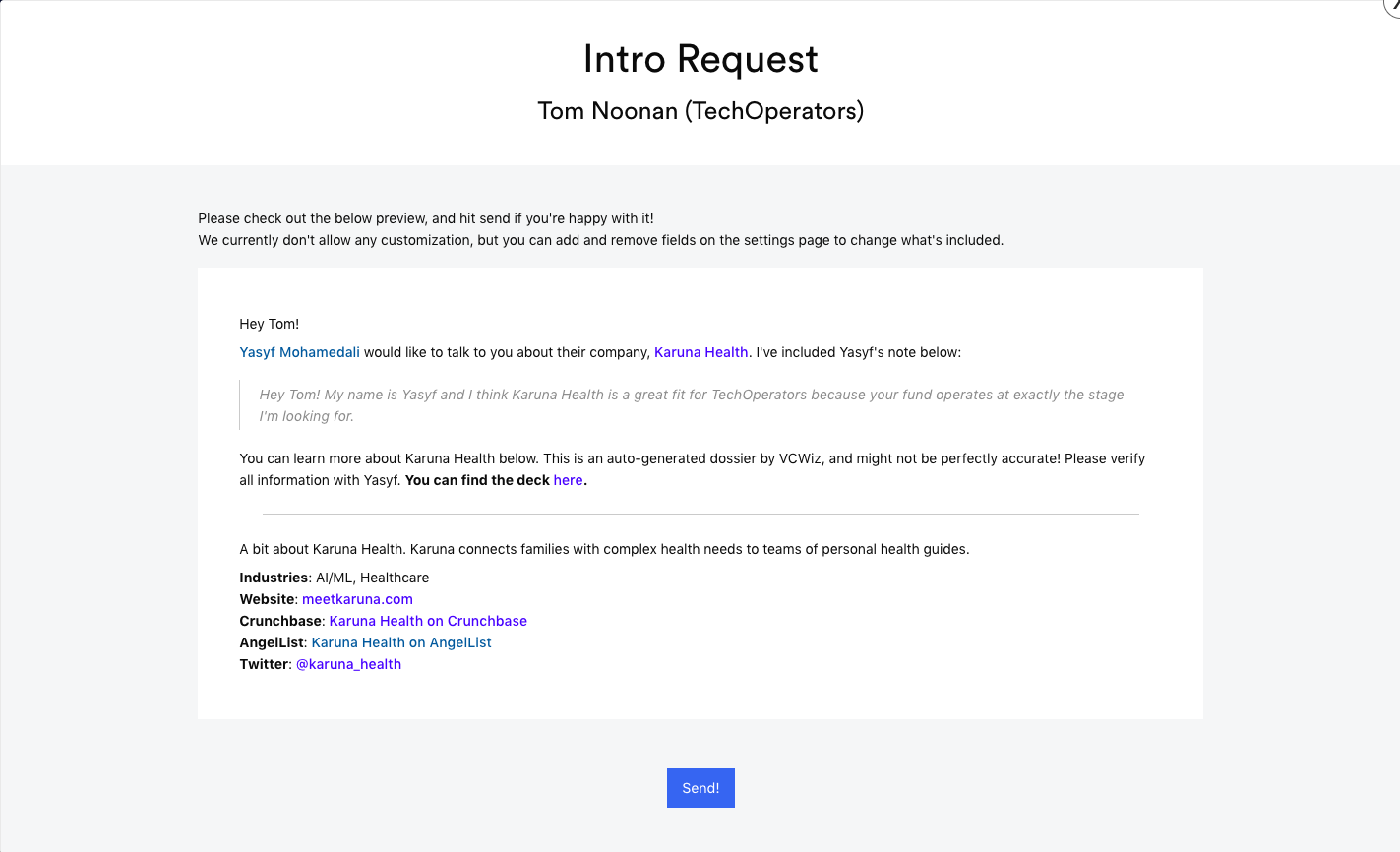}
  \caption{Intro Request}
  \label{screenshots:intro:request}
\end{figure}

\clearpage
\newpage

\chapter{Additional VCWiz Figures}

\begin{figure}[ht]
  \includegraphics[width=\textwidth]{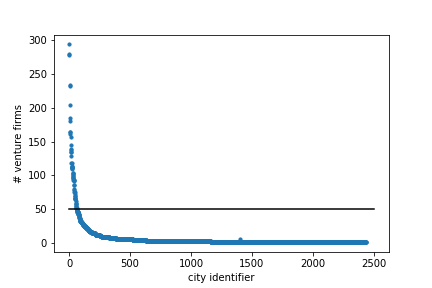}
  \caption{Hub Cities Cutoff}
  \label{vcwiz:fig:hubscuttof}
  \centering
\end{figure}

\begin{figure}[ht]
  \includegraphics[width=\textwidth]{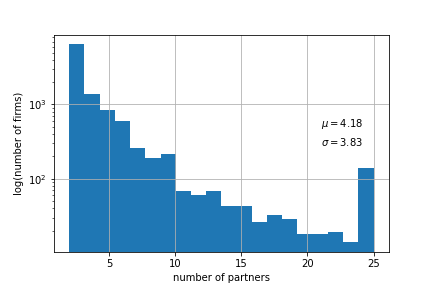}
  \caption{Histogram of General Partners per firm ($\log$)}
  \label{vcwiz:fig:partners}
  \centering
\end{figure}

\begin{figure}[ht]
\begin{multicols}{2}
  \begin{itemize}
    \item San Francisco
    \item New York
    \item London
    \item Boston
    \item Los Angeles
    \item Chicago
    \item Paris
    \item Toronto
    \item Singapore
    \item Austin
    \item Berlin
    \item Seattle
    \item Cambridge
    \item San Diego
    \item Washington, DC
    \item Atlanta
    \item Tokyo
    \item Hong Kong
    \item Sydney
    \item Dallas
    \item Vancouver
    \item Mumbai
    \item Shanghai
    \item Madrid
    \item San Jose
    \item Barcelona
    \item Stockholm
    \item Tel Aviv
    \item Moscow
    \item Santa Monica
    \item Seoul
    \item Beijing
    \item Amsterdam
    \item Redwood City
    \item Sunnyvale
    \item San Mateo
    \item Silicon Valley
    \item Philadelphia
    \item Houston
    \item Denver
    \item Boulder
    \item Santa Clara
    \item Miami
    \item Melbourne
    \item Bengaluru
    \item Portland
    \item Sao Paulo
    \item Dubai
    \item Vienna
    \item Dublin
    \item Munich
    \item Tel Aviv-Yafo
    \item Delhi
    \item United States
    \item Helsinki
    \item Nashville
    \item Minneapolis
    \item Baltimore
    \item Bangalore
    \item Pittsburgh
    \item Montreal
    \item Salt Lake City
    \item Taipei
    \item Charlotte
    \item Montreal
    \item Mexico City
    \item Milan
    \item Cincinnati
    \item Istanbul
  \end{itemize}
\end{multicols}
\caption{Hub Cities}
\label{vcwiz:fig:hubs}
\end{figure}

\begin{figure}[ht]
\begin{multicols}{2}
  \begin{itemize}
    \item AR/VR
    \item Blockchain
    \item Consumer
    \item Enterprise
    \item E-Commerce
    \item Delivery
    \item SaaS
    \item AI/ML
    \item Robotics
    \item Food \& Drink
    \item Mobile
    \item Healthcare
    \item Media
    \item Finance
    \item Education
    \item Life Sci.
    \item Retail
    \item Real Estate
    \item Travel
    \item Automotive
    \item Sports
    \item Clean Tech
    \item IoT
    \item Social
    \item Energy
    \item Hardware
    \item Gaming
    \item Space
    \item Big Data
    \item Transportation
    \item Marketplace
    \item Security
    \item Government
    \item Legal
  \end{itemize}
\end{multicols}
\caption{Company Industries}
\label{vcwiz:fig:industries}
\end{figure}

\begin{figure}[ht]
\begin{multicols}{2}
\begin{verbatim}
/login
/logout
/signup

/discover
/outreach
/filter
/search

/privacy
/terms
/founders/unsubscribe(/:token)

/firm/:id(/:slug)
/investor/:id(/:slug)
/company/:id(/:slug)
/list/:list(/:key/:value)
\end{verbatim}
\end{multicols}
\caption{Frontend Routes}
\label{vcwiz:routes:frontend}
\end{figure}

\begin{figure}[ht]
\begin{multicols}{2}
\begin{verbatim}
/investors/token/:token
/investors/signup
/investors/settings
/investors/contacts
/investors/update_contacts

/intro/opt_in
/intro/decide
/intro/pixel/:token.png
\end{verbatim}
\end{multicols}
\caption{Investor Routes}
\label{vcwiz:routes:investors}
\end{figure}

\begin{figure}[ht]
\begin{multicols}{2}
\begin{verbatim}
/api/intros/:id/preview
/api/intros/:id/confirm
/api/intros
/api/intros/:id
/api/investors/:id/interactions
/api/investors/:id/intro_paths
/api/investors/:id/verify
/api/investors/filter
/api/investors/search
/api/investors/fuzzy_search
/api/investors/entities
/api/investors/recommendations
/api/investors/locations
/api/investors/add
/api/investors
/api/investors/new
/api/investors/:id/edit
/api/investors/:id
/api/companies/search
/api/companies/query
/api/companies/:id
/api/firms/:id/intro_paths
/api/firms/filter
/api/firms/filter_count
/api/firms/locations
/api/firms/intro_path_counts
/api/firms/lists
/api/firms/list/:list(/:key/:value)
/api/firms/:id
/api/message/open
/api/message/click
/api/message/bounce
/api/message/unsubscribe
/api/message/demo
/api/message
/api/pubsub/generation
/api/target_investors/import
/api/target_investors/bulk_import
/api/target_investors/poll/:id
/api/target_investors
/api/target_investors/new
/api/target_investors/:id/edit
/api/target_investors/:id
/api/founder/disable_scanner
/api/founder/event
/api/founder/locations
/api/founder
\end{verbatim}
\end{multicols}
\caption{Internal API Routes}
\label{vcwiz:routes:api}
\end{figure}

\clearpage
\newpage

\begin{singlespace}
\bibliography{main}
\bibliographystyle{IEEEtran}
\end{singlespace}

\end{document}